\documentclass[12pt]{article}
\usepackage{amsmath}
\usepackage{graphicx}
\usepackage{enumerate}
\usepackage{natbib}
\usepackage{url} % not crucial - just used below for the URL 

\usepackage{tikz}
\usepackage{float}
\usepackage{amssymb}
\usepackage{booktabs}
\usepackage[center]{caption}
\usepackage{longtable}
\usepackage[section]{placeins}
\usepackage{multibib}
\newcites{Supp}{References}

\newcommand{\blind}{0}
\newcommand{\independent}{\perp\!\!\!\perp} 
\newcommand{\notindependent}{ \not\!\perp\!\!\!\perp}
\newcommand{\bP}{\mathbb{P}}

\newtheorem{assumption}{Assumption}
\newtheorem{theorem}{Theorem}

\newtheorem{result}{Result}
\newtheorem{proposition}{Proposition}

\begin{document}

\def\spacingset#1{\renewcommand{\baselinestretch}%
{#1}\small\normalsize} \spacingset{1}

%%%%%%%%%%%%%%%%%%%%%%%%%%%%%%%%%%%%%%%%%%%%%%%%%%%%%%%%%%%%%%%%%%%%%%%%%%%%%%

\if0\blind
{
  \title{\bf Mediation analysis with the mediator and outcome missing not at random
  \footnote{Corresponding authors: Peng Ding with e-mail pengdingpku@berkeley.edu and
    Fan Yang with e-mail yangfan1987@tsinghua.edu.cn}
  }
  \author{Shuozhi Zuo$^1$, Debashis Ghosh$^1$, Peng Ding$^{2}$, and Fan Yang$^{3,4}$\\
  \\
    $^1$Department of Biostatistics and Informatics, Colorado School of Public Health\\
    $^2$Department of Statistics, University of California, Berkeley\\
    $^3$Yau Mathematical Sciences Center, Tsinghua University\\
    $^4$Yanqi Lake Beijing Institute of Mathematical Sciences and Applications
   % \\
   %  $^*$Corresponding authors: Peng Ding, pengdingpku@berkeley.edu.\\ 
   %  Fan Yang, yangfan1987@tsinghua.edu.cn.\\
    }
    \date{}
  \maketitle
} \fi

\if1\blind
{
  \bigskip
  \bigskip
  \bigskip
  \begin{center}
    {\LARGE\bf Mediation analysis with the mediator and outcome missing not at random}
\end{center}
  \medskip
} \fi

\bigskip
\begin{abstract}
Mediation analysis is widely used for investigating direct and indirect causal pathways through which an effect arises. However, many mediation analysis studies are challenged by missingness in the mediator and outcome. In general, when the mediator and outcome are missing not at random, the direct and indirect effects are not identifiable without further assumptions. In this work, we study the identifiability of the direct and indirect effects under some interpretable mechanisms that allow for missing not at random in the mediator and outcome. We evaluate the performance of statistical inference under those mechanisms through simulation studies and illustrate the proposed methods via the National Job Corps Study.
\end{abstract}

\noindent%
\vfill

\newpage
\spacingset{1.9} % DON'T change the spacing!
\section{Mediation analysis and the National Job Corps Study}
\label{sec:intro}

Mediation analysis is increasingly adopted by researchers in a variety of fields, including epidemiology \citep{vanderweele2015explanation} and social sciences \citep{imai2010general}, to test specific theories about the underlying mechanism through which an effect arises. In a typical mediation analysis, the average treatment effect on an outcome is decomposed into a natural indirect effect (NIE) that operates through the mediator of interest and a natural direct effect (NDE) that operates through other pathways. The NIE and NDE are identified under the sequential ignorability assumption \citep{Pearl2001,imai2010general,imai2010identification}, that is, when there is no unmeasured pretreatment confounding in the treatment-mediator and treatment-outcome relationships and there is no post-treatment confounding or unmeasured pretreatment confounding in the mediator-outcome relationship. With the identification results, the NIE and NDE can be estimated through various approaches, such as regression \citep{valeri2013mediation}, weighting \citep{hong2010ratio,huber2014identifying}, multiply robust methods \citep{tchetgen2012semiparametric}, imputation-based methods \citep{vansteelandt2012imputation} and simulation-based strategies \citep{imai2010general,imai2010identification}. \citet{vansteelandt2012natural} provided alternative weaker identification assumptions for the NIE and NDE on the treated units. 

However, missingness in the mediator and outcome in mediation analysis are prevalent in empirical research \citep{oneill2020examining,chen2005test, preacher2004spss}. The missingness mechanisms are generally grouped into three categories: (1) missing completely at random (MCAR), meaning that missingness is independent of all study variables; (2) missing at random (MAR), meaning that missingness is independent of the unobservables conditional on all observables; (3) missing not at random (MNAR), meaning that missingness can depend on the unobservables even conditional on all observables. The motivation of our study comes from the well-known National Job Corps Study (NJCS). The NJCS is a multisite randomized nation wide evaluation of the Job Corps program, which is the largest education and vocational training program administered by the U.S. Department of Labor for 16 to 24 years old youths who are unemployed and disconnected from school. Past research showed that Job Corps successfully improves those disadvantaged youths' employment and increases their earnings \citep{schochet2006national, lee2009training, zhang2009likelihood}. Given that education and vocational training are the central elements of the program, \citet{qin2019multisite, qin2021unpacking} further proposed to evaluate how much of
the Job Corps effect on earnings is mediated by educational and vocational attainment across sites and found a significant positive indirect effect. However, the analysis is challenged by missingness in the mediator and outcome. In the NJCS, the assignment to either Job Corps program or the control group was random. The mediator describes whether or not the subject obtained an education credential or vocational certificate after randomization, and the outcome describes subject's weekly earnings in the fourth year after randomization. The mediator and outcome information were collected at the 30-months and 48-months follow-up, respectively, through in-person interviews. The missing rates of the mediator and the outcome are both higher than $20\%$, and less than $70\%$ of the subjects have both the mediator and the outcome observed. Table \ref{tab:Missing Patterns} presents the missingness patterns. \citet{qin2019multisite} applied a non-response weight to samples with both mediator and outcome observed to account for the observed differences in the pretreatment covariates between the subjects with both mediator and outcome measured and the subjects with mediator or outcome missing. As pointed out in their paper, this approach invokes a strong assumption and it is only valid if missingness is MAR. However, we are concerned that missingness is likely MNAR in the NJCS. Conceivably, people who failed to obtain an education credential or vocational certificate may be less likely to provide the information compared to people who successfully obtained a credential or certificate. One could also imagine that people who had no earnings may be less willing to report the amount of earnings in the interview. As a result, the chance of missingness would depend on the unobserved missing value itself, and the data would be MNAR. In such a scenario, the commonly adopted strategies to deal with missing data, such as complete case analysis, multiple imputation under MAR or non-response weighting, may fail to provide valid inference. 

\begin{table}[H]
\centering
\caption{Missingness patterns in the mediator and outcome}
\label{tab:Missing Patterns}
\begin{tabular}{lllllllllll}
\hline 
Mediator & Outcome & Treatment Group $N$ (\%) & Control Group $N$ (\%)\\
\hline 
Missing & Observed & $545~(10.72\%) $ & $361~(9.96\%)$\\
Observed & Missing & $538~(10.58\%)$ & $400~(11.04\%)$\\
Missing & Missing & $497~(9.78\%)$ & $426~(11.76\%)$\\
Observed & Observed & $3504~(68.92\%)$ & $2436~(67.24\%)$\\
\hline  
\multicolumn{2}{l}{Total Number of Subjects} & $5084$ & $3623$\\
\hline 
\end{tabular}
\end{table}

MNAR presents a challenge for causal inference, and is fundamentally more difficult than MAR, because in general, the underlying data distribution can not be identified without further assumptions. This topic has attracted some attention in the literature. For example, \cite{ding2014identifiability} and \cite{yang2019causal} studied the identifiability of subgroup treatment effects and average treatment effects, respectively, with covariates MNAR. In the context of instrumental variable
analysis, methods were developed when missingness in the covariates \citep{yang2014estimation} or in the outcome \citep{frangakis1999addressing,peng2003extended,chen2009identifiability} is MNAR. Previous research also studied various identification problems using graphical models. \cite{fay1986causal} proposed directed acyclic graphs (DAGs) for patterns of non-response and studied the identification of DAGs for categorical outcomes. \cite{glonek1999identifiability} studied the identifiability in models for binary outcome subject to MNAR and \cite{ma2003identification} further studied the identifiability of DAGs with a binary outcome MNAR in longitudinal studies. Using concentration graphical models, \cite{mealli2016identification} studied the identification of principal causal effects by utilizing the conditional independence between two outcomes with one as the focal outcome of interest. Under chain graphs instead of DAGs, \cite{li2022self} studied the identification condition for a self-censoring model for multivariate nonignorable missing data. Previous work also reviewed missing data research in graphical models \citep{mohan2021graphical}. However, limited effort has been made to study the identifiability of causal mediation effects with the mediator and outcome MNAR. Considering missingness in the outcome only, \citet{li2017identifiability} utilized an instrumental variable type of covariate to identify the direct and indirect effects when missingness in the outcome depends on the outcome value itself. To apply their method, we need a covariate that is associated with the outcome, but is conditionally independent of missingness of the outcome. In many studies, such a covariate may not be available. Moreover, \citet{li2017identifiability} did not deal with the issue of missingness in the mediator. 

\subsection{Organization of the paper}

We study the identifiability of causal mediation effects with the mediator and outcome MNAR. We provide conditions for identification under various interpretable MNAR assumptions. The rest of the paper is organized as follows. In section \ref{sec:notation}, we introduce the notation and basic assumptions for causal mediation analysis. In section \ref{sec:mismediator}, we discuss the identification in a simple setup where missingness exists only in the mediator and depends on the missing mediator value itself. In section \ref{sec:misboth}, we extend the results to the more complicated setup where both the mediator and the outcome have missing data. In section \ref{sec:simulation}, we conduct extensive simulation studies to test our theoretical results and to evaluate the performance of the proposed methods. In section \ref{sec:data_analysis}, we apply our methods to the NJCS. We conclude with a discussion and provide proofs of the theorems in the supplementary material.  

\subsection{Notation and some basic definitions}

Let $A\independent B\mid C$ denote that the random variables $A$ and $B$ are conditionally independent given the random variable $C$. Let $\mathcal{X}$, $\mathcal{M}$, and $\mathcal{Y}$ denote the supports of the random variables $X$, $M$, and $Y$, respectively. 
Further, the property of completeness \citep{lehmann1950completeness,basu1955statistics} will play a key role in our nonparametric identification of the mediation effects. Define a function $f(A,B)$ to be complete in $B$ if $\int g(A)f(A,B)d\nu(A)=0$ implies $g(A)=0$ almost surely for any square-integrable function $g$. In the above integral, $\nu(\cdot)$ presents a generic measure, which is the Lebesgue measure for a continuous variable and the counting measure for a discrete variable.

\section{Review of mediation analysis without missing data}
\label{sec:notation}

Consider a sample of size $n$ that are independent and identically distributed samples drawn from an infinite superpopulation. Let $T$ denote the binary treatment assignment, with $t=0$ and $t=1$ representing the control condition and the experimental condition, respectively. Let $  X $ be the vector of measured pre-treatment covariates. We use $M$ and $Y$ to denote the mediator and the outcome, respectively. We adopt the potential outcomes framework to define the causal effects of interest and make the stable unit treatment value assumption (SUTVA) that there is no hidden variations of each treatment condition and the potential outcomes for any unit do not vary with the treatments assigned to other units. We use $M(t)$ to denote the individual's potential mediator value under treatment $t$ for $t=0, 1$, and use $Y(t)$ to denote the individual's potential outcome value under treatment $t$ for $t=0, 1$. The treatment $T$ may affect $Y$ through $M$, hence $Y(t)$ can be written as $Y(t,M(t))$, which is called the ``composition'' assumption \citep{vanderweele2015explanation}. Given the above notation, define the average treatment effect ($\mathrm{ATE}$) as $\mathrm{ATE}=\mathbb{E}\{Y(1)-Y(0)\}$. 

%\citep{neyman1923application, rubin1974estimating} 

Define the nested potential outcome, $Y(1,M(0))$, to describe individual's potential outcome under the experimental condition, however, with the mediator counterfactually taking its value under the control condition \citep{robins1992identifiability}. The $\mathrm{ATE}$ can be decomposed into \citep{Pearl2001}: 
\setlength{\abovedisplayskip}{0pt} 
\setlength{\belowdisplayskip}{0pt}
\begin{align*}
\mathrm{ATE} = \mathrm{NIE}+\mathrm{NDE},
\end{align*} 
where 
\begin{align*}
\mathrm{NIE} = \mathbb{E}\{Y(1,M(1))-Y(1,M(0))\}
\end{align*}
is the natural indirect effect ($\mathrm{NIE}$) and 
\begin{align*}
\mathrm{NDE} = \mathbb{E}\{Y(1,M(0))-Y(0,M(0))\}
\end{align*}
is the natural direct effect ($\mathrm{NDE}$). 

The $\mathrm{NIE}$ quantifies the average treatment effect on the outcome transmitted through the treatment induced change in the mediator from $M(0)$ to $M(1)$. The $\mathrm{NDE}$ quantifies the direct effect of the treatment on the outcome that does not operate through its impact on the mediator. We invoke the standard sequential ignorability assumption \citep{imai2010general,imai2010identification} throughout the paper. Without missing data, we have the following mediation formula \citep{Pearl2001, imai2010general}:
\setlength{\abovedisplayskip}{5pt} 
\setlength{\belowdisplayskip}{5pt}
$$ 
\mathbb{E}\{Y(t, M(t'))\} = \int_\mathcal{X}\int_{\mathcal{M}}\mathbb{E}(Y\mid M=m, T=t, X=x)\,f_M(m\mid T=t', X=x)\text{d}m\,f_X(x)\text{d}x.
$$
With missing data, the key for the identification of the $\mathrm{NIE}$ and $\mathrm{NDE}$ would be to identify the probabilities $\bP(Y=y\mid M=m, T=t, X=x)$ and $\bP(M=m\mid T=t, X=x)$, or equivalently, the joint probability $\bP(Y=y, M=m\mid T=t, X=x)$, from the observable data. 

\section{Missingness only in the mediator}
\label{sec:mismediator}
In this section, we consider a simple setup where the mediator has missing values and the outcome is fully observed. It may happen in studies where the outcome is of primary interest with the mediator being a secondary outcome of interest. This setup serves an a stepstone for later sections because it helps to lay out the basic completeness condition and build intuition for the nonparametric identification under MNAR.

Let $R^M$ be the missingness indicator such that $R^M=1$ if $M$ is observed and $R^M=0$ if $M$ is missing.  When $R^M \independent (Y,M,T,  X )$, the missingness mechanism is MCAR, and complete case analysis is enough to provide consistent estimates of $\bP(Y,M \mid T,  X )$. When $R^M \independent M \mid(Y,T,  X )$, the missingness mechanism is MAR, and $\bP(Y,M \mid T,  X )$ is identifiable given the observed data. However, as we explained, often we have the concern that missingness of $M$ may depend on the value of $M$ itself even conditional on other observed data. In such a case, the missingness mechanism is MNAR. Since the outcome $Y$ occurs after the mediator $M$, it is plausible in many studies to assume that missingness of $M$ is conditionally independent of $Y$. Based on the above discussion, we propose the following MNAR Assumption \ref{ass1}:

\begin{assumption}
$R^M \independent Y\mid(M,T,  X )$ and $Y$ is fully observed.\label{ass1}
\end{assumption}

\begin{figure}[H]
\centering
\begin{tikzpicture}
    \node (t)  at (0,0) {$T$};
    \node (x)  at (2,0) {$M$};
    \node (rm) at (2,2) {$R^M$};
    \node (y)  at (4,0) {$Y$};
    \node (a)  at (2,-1.5) {$(a)$ MCAR};
    
    \path[-latex] (t) edge (x);
    \path[-latex] (x) edge (y);
    \path[-latex] (t) edge [bend right] (y);
    
    \node (t)  at (5,0) {$T$};
    \node (x)  at (7,0) {$M$};
    \node (rm) at (7,2) {$R^M$};
    \node (y)  at (9,0) {$Y$};
    \node (a)  at (7,-1.5) {$(b)$ MAR};
    
    \path[-latex] (t) edge (x);
    \path[-latex] (t) edge (rm);
    \path[-latex] (x) edge (y);
    \path[-latex] (y) edge (rm);
    \path[-latex] (t) edge [bend right] (y);
    
    \node (t)  at (10,0) {$T$};
    \node (x)  at (12,0) {$M$};
    \node (rm) at (12,2) {$R^M$};
    \node (y)  at (14,0) {$Y$};
    \node (a)  at (12,-1.5) {$(c)$ MNAR Assumption \ref{ass1}};
    
    \path[-latex] (t) edge (x);
    \path[-latex] (t) edge (rm);
    \path[-latex] (x) edge (y);
    \path[-latex] (t) edge [bend right] (y);
    \path[-latex] (x) edge (rm);
\end{tikzpicture}
\caption{DAGs describing MCAR, MAR and the MNAR mechanisms when missingness exists only in the mediator (all DAGs condition on $X$ and allow $X$ to have directed arrows to all variables in the DAGs).}
\label{fig:missingness exists only in the mediator}
\end{figure}
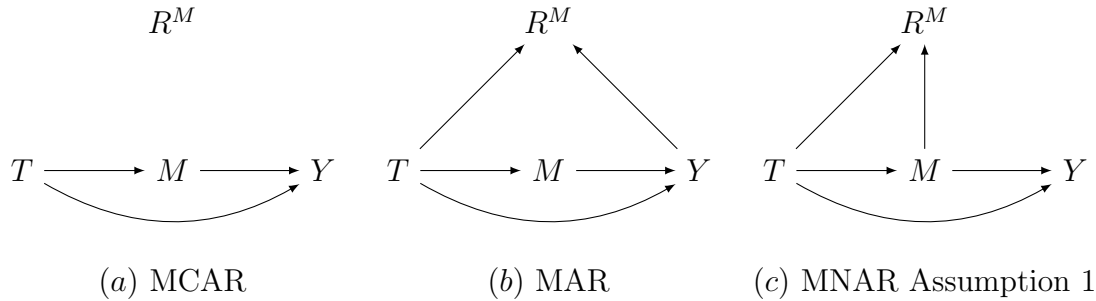

Assumption \ref{ass1} allows $R^M$ to depend on the mediator $M$, $T$ and $  X $. However, we assume $R^M$ to be conditionally independent of the outcome $Y$ given $M$, $T$ and $  X $. The DAGs in Figure \ref{fig:missingness exists only in the mediator} illustrate the different missingness mechanisms under MCAR, MAR and our MNAR Assumption \ref{ass1}, respectively.

The following theorem presents the nonparametric identification results under the MNAR Assumption \ref{ass1}. 

\begin{theorem}\label{the1}
Under Assumption \ref{ass1}, if $\bP(Y,M,R^M=1 \mid T=t,  X=x )$ is complete in $Y$ for all $t$ and $x$, and $\bP(R^M=1\mid M=m,T=t,X=x)>0$ for all $m, t,  x $, then $\bP(Y,M\mid T,  X )$ is identifiable, and therefore, the $\textup{NIE}$ and $\textup{NDE}$ are identifiable.
\end{theorem}

For sufficient conditions on completeness, we refer readers to \citet{d2011completeness} for a comprehensive discussion. The notion of completeness is well-established and consistently applied in the nonparametric identification problems. We provide a list of literature that use the completeness assumption for such problems: the instrumental variable literature \citep{newey2003instrumental, darolles2011nonparametric}, the measurement error literature \citep{an2012well}, the principal stratification literature \citep{jiang2021identification}, the missing data literature \citep{yang2019causal, miao2019identification, li2022self} and the recent proximal inference literature \citep{dukes2023proximal, ghassami2023causal}. Generally speaking, the completeness condition requires that the variables (e.g. instrumental variables, proxies, auxiliary variables, mediators or outcomes) have sufficient dimensions or variability relative to the partly observed or unobserved variable of interest. For discrete $M$ and $Y$, the completeness assumption is equivalent to a rank condition, on which we will provide more details in the next paragraph. In section \ref{sec::counterexamples} of the supplementary material, we present an unidentifiable case when the completeness assumption is violated. Despite of being a general assumption for identification, the concept of completeness is abstract to be put into practice. To facilitate the application of the identification results, parametric models are often adopted. The completeness condition holds under some frequently used parametric models, such as exponential families of distributions \citep{newey2003instrumental} and a class of location-scale distribution families \citep{hu2018nonparametric}. We provide parametric examples that satisfy the corresponding completeness assumptions for each of the theorems in section \ref{sec::parametricexamples} of the supplementary material.

In fact, under Assumption \ref{ass1}, $\bP(Y\mid M, T,  X )$ is identifiable using complete cases, i.e., $\bP(Y\mid M, T,   X ) = \bP(Y\mid M, T, X, R^M=1)$. The completeness assumption is only needed to identify $\bP(M\mid T,  X )$. Here we provide the intuition for the need of the completeness when both $M$ and $Y$ are categorical. Because
\begin{align}
\bP(M=m\mid T=t, X=x) &= \frac{\bP(M=m, R^M=1\mid T=t, X=x)}{\bP(R^M=1\mid M=m, T=t, X=x)},\nonumber
\end{align}
the targeted conditional distribution $\bP(M\mid T,  X )$ is identifiable if $\bP(R^M\mid M, T, X)$ is identifiable. Define 
$$\zeta_{t,x}(m) =  \frac{\bP(R^M=0\mid M=m, T=t, X=x)}{\bP(R^M=1\mid M=m, T=t, X=x)},$$
for each $t$ and $x$, we have the following system of linear equations with $\{\zeta_{t,x}(m): m\in\mathcal{M}\}$ as unknowns: 
\begin{align}
\bP(Y=y, R^M=0\mid T=t, X=x) &= \sum_{m\in \mathcal{M}} \bP(M=m, Y=y, R^M=0\mid T=t, X=x) \nonumber\\&=\sum_{m\in \mathcal{M}}\bP(M=m, Y=y, R^M=1\mid T=t, X=x)  \zeta_{t,x}(m),\nonumber 
\end{align}
for each $y\in\mathcal{Y}$. To ensure the uniqueness of the solutions $\zeta_{t,x}(m)$, we need Rank $(\Theta_{tx})=J$, where $\Theta_{tx}$ is the $J \times K$ matrix with $\bP(M=m, Y=y, R^M=1\mid T=t, X=x)$ as the $(m,y)$th element, $J$ is the number of categories in $M$, and $K$ is the number of categories in $Y$. This is the completeness condition presented in Theorem \ref{the1} in the discrete case. This full rank condition essentially requires that $J\leq K$, and that $M \notindependent  Y\mid (T,  X )$.

%The completeness assumption is equivalent to Rank $(\Theta)=J$, where $\Theta$ is a $J \times K$ matrix with $\bP(M=m, Y=y, R^M=1\mid T=t, X=x)$ as the $(m,y)$th element. This is sufficient to ensure the uniqueness of solutions $\zeta_{t,x}(m)$. We can subsequently identify $\bP(R^M=1\mid M=m, T=t, X=x)$ once $\zeta_{t,x}(m)$ is identified. Then, the identification of $\bP(M=m\mid T=t, X=x)$ follows from
%\begin{align}
%\bP(M=m\mid T=t, X=x) &= \frac{\bP(M=m, R^M=1\mid T=t, X=x)}{\bP(R^M=1\mid M=m, T=t, X=x)}.\nonumber
%\end{align}

%In a special case where $M \independent  Y\mid (T,  X )$ and the completeness assumption is violated, both the $\mathrm{NIE}$ and $\mathrm{NDE}$ remain identifiable since $\bP(Y\mid M, T,  X )$ is identifiable using complete cases. This is because $\bP(Y\mid M, T,  X ) = \bP(Y\mid T,  X )$ when $M \independent  Y\mid (T,  X )$, and therefore, $\mathrm{NIE} = 0$ and $\mathrm{NDE} = \mathrm{ATE} =  \int_{\mathcal{X}}\{\mathbb{E}(Y\mid T=1,    X =  x  )-\mathbb{E}(Y\mid T=0,    X =  x  )\}\,f_X(x)\text{d} x.$ We also illustrate this point using simulation studies.

Since $\bP(Y\mid M, T,  X )$ is identifiable using complete cases, when $M \independent  Y\mid (T,  X )$ and therefore the completeness assumption is violated, both the $\mathrm{NIE}$ and $\mathrm{NDE}$ remain identifiable. This is because $\bP(Y\mid M, T,  X ) = \bP(Y\mid T,  X )$ when $M \independent  Y\mid (T,  X )$, and therefore, $\mathrm{NIE} = 0$ and $\mathrm{NDE} = \mathrm{ATE} =  \int_{\mathcal{X}}\{\mathbb{E}(Y\mid T=1,    X =  x  )-\mathbb{E}(Y\mid T=0,    X =  x  )\}\,f_X(x)\text{d} x.$ We also illustrate this point using simulation studies.

\section{Missingness in both the mediator and outcome}
\label{sec:misboth}

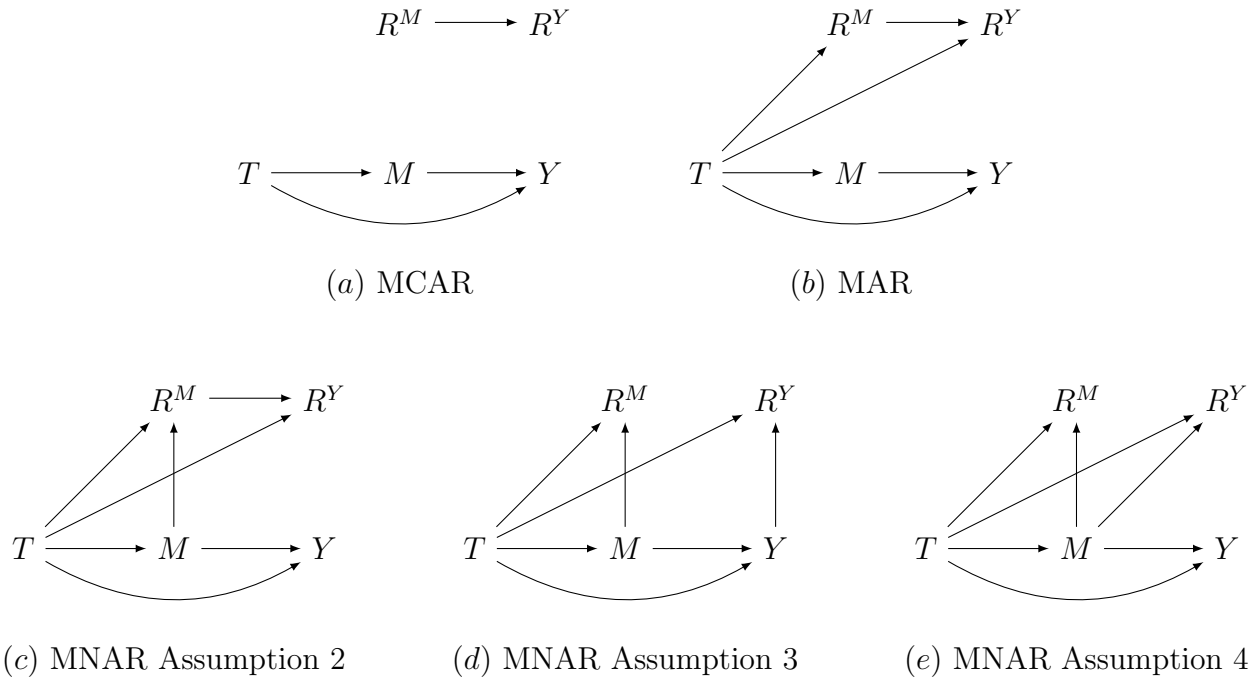
\begin{figure}[ht]
\centering
\begin{tikzpicture}
    \node (t)  at (0,0) {$T$};
    \node (x)  at (2,0) {$M$};
    \node (rm) at (2,2) {$R^M$};
    \node (y)  at (4,0) {$Y$};
    \node (ry) at (4,2) {$R^Y$};
    \node (a)  at (2,-1.5) {$(a)$ MCAR};
 
    \path[-latex] (t) edge (x);
    \path[-latex] (x) edge (y);
    \path[-latex] (t) edge [bend right] (y);
    \path[-latex] (rm) edge (ry);
    
    \node (t)  at (6,0) {$T$};
    \node (x)  at (8,0) {$M$};
    \node (rm) at (8,2) {$R^M$};
    \node (y)  at (10,0) {$Y$};
    \node (ry) at (10,2) {$R^Y$};
    \node (b)  at (8,-1.5) {$(b)$ MAR};

    \path[-latex] (t) edge (x);
    \path[-latex] (t) edge (rm);
    \path[-latex] (t) edge (ry);
    \path[-latex] (x) edge (y);
    \path[-latex] (t) edge [bend right] (y);
    \path[-latex] (rm) edge (ry);
    
    \node (t)  at (-3,-5) {$T$};
    \node (x)  at (-1,-5) {$M$};
    \node (rm) at (-1,-3) {$R^M$};
    \node (y)  at (1,-5) {$Y$};
    \node (ry) at (1,-3) {$R^Y$};
    \node (c)  at (-1,-6.5) {$(c)$ MNAR Assumption \ref{ass2}};

    \path[-latex] (t) edge (x);
    \path[-latex] (t) edge (rm);
    \path[-latex] (t) edge (ry);
    \path[-latex] (x) edge (y);
    \path[-latex] (x) edge (rm);
    \path[-latex] (rm) edge (ry);
    \path[-latex] (t) edge [bend right] (y);
    
    \node (t)  at (3,-5) {$T$};
    \node (x)  at (5,-5) {$M$};
    \node (rm) at (5,-3) {$R^M$};
    \node (y)  at (7,-5) {$Y$};
    \node (ry) at (7,-3) {$R^Y$};
    \node (d)  at (5,-6.5) {$(d)$ MNAR Assumption \ref{ass3}};

    \path[-latex] (t) edge (x);
    \path[-latex] (t) edge (rm);
    \path[-latex] (t) edge (ry);
    \path[-latex] (x) edge (y);
    \path[-latex] (x) edge (rm);
    \path[-latex] (y) edge (ry);
    \path[-latex] (t) edge [bend right] (y);
    
    \node (t)  at (9,-5) {$T$};
    \node (x)  at (11,-5) {$M$};
    \node (rm) at (11,-3) {$R^M$};
    \node (y)  at (13,-5) {$Y$};
    \node (ry) at (13,-3) {$R^Y$};
    \node (e)  at (11,-6.5) {$(e)$ MNAR Assumption \ref{ass4}};
    
    \path[-latex] (t) edge (x);
    \path[-latex] (t) edge (rm);
    \path[-latex] (t) edge (ry);
    \path[-latex] (x) edge (y);
    \path[-latex] (x) edge (rm);
    \path[-latex] (x) edge (ry);
    \path[-latex] (t) edge [bend right] (y);

\end{tikzpicture}
\caption{DAGs describing MCAR, MAR and the MNAR mechanisms when missingness exists in both the mediator and outcome (all DAGs condition on $X$ and allow $X$ to have directed arrows to all variables in the DAGs).}
\label{fig:missingness exists in both the mediator and outcome}
\end{figure}

We now extend the results to the scenario where both the mediator and outcome have missing data. Further let $R^Y$ to denote the missingness indicator for $Y$ such that $R^Y=1$ if $Y$ is observed and $R^Y=0$ otherwise. When $(R^Y, R^M) \independent (Y,M,T,X)$, the missingness mechanism is MCAR. When $(R^Y,R^M) \independent (Y,M)\mid (T,  X )$, the missingness mechanism is MAR. Continuing to allow missingness of $M$ to depend on $M$ itself and assume $R^M \independent Y\mid(M,T,  X )$, we consider the following MNAR mechanisms described in Assumptions \ref{ass2}, \ref{ass3}, and \ref{ass4}, respectively. The DAGs in Figure \ref{fig:missingness exists in both the mediator and outcome} illustrate different missingness mechanisms under MCAR, MAR, and MNAR when missingness exists in both the mediator and outcome. The differences among the MNAR mechanisms under Assumptions \ref{ass2} to \ref{ass4} are in the missingness mechanisms in $Y$.

\subsection{MNAR mechanism under Assumption \ref{ass2}}

\begin{assumption}
$(R^Y, R^M) \independent  Y\mid (M,T,  X )$ and $R^Y \independent M\mid (R^M,T,  X )$.
\label{ass2}
\end{assumption}

The MNAR mechanism under Assumption \ref{ass2} allows missingness in $M$ to depend on $M$ itself in addition to the fully observed variables $T$ and $  X $, and allows missingness in $Y$ to depend on missingness in $M$ in addition to $T$ and $  X $. The MAR mechanism is a special case of the MNAR mechanism under Assumption \ref{ass2} without allowing $M$ to affect $R^M$. In the NJCS, this mechanism suggests that whether or not people successfully obtained a certificate may have an impact on their willingness to report, and their decision on reporting their certificate status may be associated with their decision to report their earnings or not. The following theorem presents the nonparametric identification results under the MNAR Assumption \ref{ass2}.

\begin{theorem}\label{the2}
Under Assumption \ref{ass2}, if $\bP(R^M=1,R^Y=1\mid M=m,T=t,X=x) > 0$ for all $m, t,  x $, we have the following results:\\
(i) If $\bP(R^M=0,R^Y=1\mid M=m,T=t,X=x)>0$ for all $m, t,  x $, and $\bP(Y,M,R^M=1,R^Y=1 \mid T=t,  X =x)$ is complete in $Y$ for all $t$ and $x$, then $\bP(Y,M\mid T,  X )$ is identifiable, and therefore, the \textup{NIE} and \textup{NDE} are identifiable;\\
(ii) If $\bP(R^M=0,R^Y=1\mid M=m,T=t,X=x)=0$ for some $m, t,  x $, and $R^M$ $\independent M\mid (T,  X )$, then $\bP(Y,M\mid T,  X )$ is identifiable, and therefore, the \textup{NIE} and \textup{NDE} are identifiable. %if $R^M \independent M\mid (T,  X )$.
\end{theorem}

The condition $\bP(R^M=0,R^Y=1\mid M=m,T=t,X=x)=0$ for some $m,t,x$ in Theorem \ref{the2} $(ii)$ suggests missingness of $M$ implies missingness of $Y$ for those $m,t,x$, which is commonly referred to as monotone missingness. In such a case, $\bP(Y,M\mid T,  X )$ is not identifiable without additional assumptions. If we further assume $R^M \independent M\mid (T,  X )$, the missingness mechanism becomes MAR as described in Figure \ref{fig:missingness exists in both the mediator and outcome} $(b)$, and $\bP(Y,M\mid T,  X )$ is identifiable using complete cases, i.e., $\bP(Y, M \mid  T, X) = \bP(Y, M \mid T, X, R^Y=1, R^M=1)$. In the NJCS, the percentages of subjects having the outcome $Y$ observed among subjects with missing mediator values are $52.30\%$ and $45.87\%$ in the treatment group and control group, respectively. Therefore, for the rest of the paper, we focus on scenarios described by Theorem \ref{the2} $(i)$ when data are MNAR under Assumption \ref{ass2}.

In Theorem \ref{the2} $(i)$, $Y\independent (R^M, R^Y)\mid (M, T,   X) $, thus $\bP(Y\mid M, T, X )$ is identifiable using complete cases, i.e., $\bP(Y\mid M, T,   X ) = \bP(Y\mid M, T,   X , R^Y=1, R^M=1)$. The completeness assumption is again only used to identify $\bP(M\mid T, X )$. To see the role of the completeness assumption, consider the case where both $M$ and $Y$ are categorical. $\bP(M\mid T,  X )$ is identifiable if $\bP(R^M\mid M, T, X)$ is identifiable as discussed in section \ref{sec:mismediator}. For each $t,x$, we have the following system of linear equations with $\{\zeta_{t,x}(m): m\in\mathcal{M}\}$ as unknowns:
\begin{align}
&~~~~\bP(Y=y, R^M=0, R^Y=1\mid T=t, X=x)\nonumber\\ &=\frac{\bP(R^Y=1\mid R^M=0, T=t, X=x)}{\bP(R^Y=1\mid R^M=1, T=t, X=x)}\sum_{m\in \mathcal{M}}\bP(M=m, Y=y, R^M=1, R^Y=1\mid T=t, X=x) \zeta_{t,x}(m), \nonumber
\end{align}
for each $y\in\mathcal{Y}$. The sufficient condition to ensure the uniqueness of solutions $\zeta_{t,x}(m)$ is the completeness assumption presented in Theorem \ref{the2} $(i)$ that Rank $(\Theta_{tx})=J$, where $\Theta_{tx}$ is a $J \times K$ matrix with $\bP(M=m, Y=y, R^M=1, R^Y=1\mid T=t, X=x)$ as the $(m,y)$th element. This full rank condition again requires that $J\leq K$, and that $M \notindependent  Y\mid (T,  X )$. Since the identification of $\bP(Y\mid M,T, X)$ does not rely on the completeness condition, when $M\independent Y\mid (T,X)$ and therefore the completeness assumption fails, the NIE and NDE are still identifiable as discussed in section \ref{sec:mismediator}.

\subsection{MNAR mechanism under Assumption \ref{ass3}}

\begin{assumption}\label{ass3} 
$R^Y \independent (M,R^M)\mid (Y,T,  X )$ and $R^M \independent (Y,R^Y)\mid (M,T,  X )$.
\end{assumption}

The MNAR mechanism under Assumption \ref{ass3} allows missingness of $Y$ to depend on $Y$ itself instead of $R^M$ as in Assumption \ref{ass2}. In the NJCS, it assumes that the amount of earnings may have an impact on the probability to report earnings, which is also a reasonable concern. Both missingness in $M$ and in $Y$ are MNAR under Assumption \ref{ass3}. Theorem \ref{the3} presents the nonparametric identification results under Assumption \ref{ass3}.

\begin{theorem}\label{the3}
Under Assumption \ref{ass3}, if $\bP(Y,M,R^M=1,R^Y=1 \mid T=t,  X=x )$ is complete in $Y$ for all $t$ and $x$, $\bP(Y,M,R^M=1,R^Y=1 \mid T=t,  X=x )$ is complete in $M$ for all $t$ and $x$, $\bP(R^M=1,R^Y=1\mid Y=y,M=m,T=t,   X =  x  ) > 0$, $\bP(R^M=0,R^Y=1\mid Y=y,M=m,T=t,   X =  x  ) > 0$ and $\bP(R^M=1,R^Y=0\mid Y=y,M=m,T=t,   X =  x  )>0$ for all $y, m,t, x $, then $\bP(Y,M\mid T,  X )$ is identifiable, and therefore, the \textup{NIE} and \textup{NDE} are identifiable.
\end{theorem}

Different from Assumption \ref{ass2}, under the MNAR Assumption \ref{ass3}, $Y$ is not independent of $R^Y$ given $M, T$ and $  X $, and therefore, $\bP(Y\mid M, T,   X )$ is no longer identifiable using complete cases. In fact, the MNAR mechanism under Assumption \ref{ass3} requires the completeness in $Y$ and the completeness in $M$ to be able to identify both $\bP(Y\mid M, T, X )$ and $\bP(M\mid T, X )$, or equivalently $\bP(Y, M\mid T, X )$. We provide the intuition using the discrete case. Because 
\begin{align}
&~~~~\bP(Y=y, M=m\mid T=t, X=x)\nonumber\\&=\frac{\bP(Y=y, M=m, R^M=1, R^Y=1\mid T=t, X=x)}{\bP(R^M=1\mid M=m,T=t, X=x)\bP(R^Y=1\mid Y=y,T=t, X=x)},\nonumber
\end{align}
$\bP(Y=y, M=m\mid T=t, X=x)$ is identifiable if both $\bP(R^M=1\mid M=m,T=t, X=x)$ and $\bP(R^Y=1\mid Y=y,T=t, X=x)$ are identifiable. Define $$\eta_{t,x}(y)=\frac{\bP(R^Y=0\mid Y=y,T=t, X=x)}{\bP(R^Y=1\mid Y=y,T=t, X=x)},$$ for each $t, x$, we have the following two systems of linear equations with $\{\zeta_{t,x}(m): m\in\mathcal{M}\}$ and $\{\eta_{t,x}(y): y\in\mathcal{Y}\}$ as unknowns, respectively:
\begin{align}
&~~~~\bP(Y=y, R^M=0, R^Y=1\mid T=t, X=x)\nonumber\\ &=\sum_{m\in \mathcal{M}}\bP(M=m, Y=y, R^M=1, R^Y=1\mid T=t, X=x) \zeta_{t,x}(m),\nonumber
\end{align}
\text{for each~}$y\in\mathcal{Y}$, and
\begin{align}
&~~~~\bP(M=m, R^M=1, R^Y=0\mid T=t, X=x)\nonumber\\&=\sum_{y\in \mathcal{Y}}\bP(M=m, Y=y, R^M=1, R^Y=1\mid T=t, X=x) \eta_{t,x}(y),\nonumber
\end{align} 
\text{for each~}$m\in\mathcal{M}$. The uniqueness of solutions $\zeta_{t,x}(m)$ requires that $\bP(Y, M, R^M=1, R^Y=1\mid T=t, X=x)$ is complete in $Y$, and the uniqueness of solutions $\eta_{t,x}(y)$ requires that $\bP(Y, M, R^M=1, R^Y=1\mid T=t, X=x)$ is complete in $M$. To satisfy both completeness conditions, $M$ and $Y$ have to share the same numbers of categories, i.e., $J=K$, and that Rank $(\Theta_{tx})=J$, where $\Theta_{tx}$ is a $J \times J$ matrix with $\bP(M=m, Y=y, R^M=1, R^Y=1\mid T=t, X=x)$ as the $(m,y)$th element. The completeness conditions would fail if $J\neq K$ or $M\independent Y\mid (T,   X) $. We present an unidentifiable case when $Y$ has more categories than $M$ in section \ref{sec::counterexamples} of the supplementary material. Different from Assumptions \ref{ass1} and \ref{ass2} where the identification of $\bP(Y\mid M, T, X)$ does not need the completeness conditions, the identification of $\bP(Y\mid M, T, X)$ under Assumption \ref{ass3} relies on the completeness conditions. Therefore, when $M\independent Y\mid (T,   X) $, the NIE and NDE are not identifiable.

%The completeness assumption is equivalent to Rank $(\Theta)=J$, where $\Theta$ is a $J \times J$ matrix with $\bP(M=m, Y=y, R^M=1, R^Y=1)$ as the $(m,y)$th element. This is sufficient to ensure the uniqueness of solutions $\zeta(m)$ and $\eta(y)$. We can subsequently identify $\bP(R^M=1\mid M=m)$ and $\bP(R^Y=1\mid Y=y)$ once $\zeta(m)$ and $\eta(y)$ are identified. Then, the identification of $\bP(Y=y, M=m)$ follows from
%\begin{eqnarray}
%\bP(Y=y, M=m)=\frac{\bP_{my11}}{\bP(R^M=1\mid M=m)\bP(R^Y=1\mid Y=y)}.\nonumber
%\end{eqnarray}

%As a result, when $M\independent Y\mid (T,   X) $, the NIE and NDE are no longer identifiable. Further, since the completeness is required in $Y$ and required in $M$, to achieve identification of $\bP(Y,M\mid T,  X )$ under Assumption \ref{ass3}, the number of elements in the support of $Y$ has to be the same as the number of elements in the support of $M$. We present an unidentifiable case when $Y$ has more categories than $M$ in section \ref{sec::counterexamples} of the supplementary material.

\subsection{MNAR mechanism under Assumption \ref{ass4}}

\begin{assumption}\label{ass4} 
$Y, R^Y$ and $ R^M$ are mutually independent given $(M,T,  X )$.
\end{assumption}

The MNAR mechanism under Assumption \ref{ass4} allows missingness in $Y$ to depend on $M$ instead of $R^M$ or $Y$, another case where both $M$ and $Y$ are MNAR. In the NJCS, it suggests that whether or not people successfully obtained a certificate drives missingness in both $M$ and $Y$ after conditioning on $T$ and $X$. The following theorem presents the nonparametric identification results under the MNAR Assumption \ref{ass4}. Further define a random vector $Y^\dagger=(Y \cdot R^Y, R^Y)$ such that 
$\bP\{Y^\dagger=(y,1)\}=\bP(Y=y,R^Y=1)$ for all $y\in \mathcal{Y}$ and
$\bP\{Y^\dagger=(0,0)\}=\bP(R^Y=0)$.

\begin{theorem}\label{the4}
Under Assumption \ref{ass4}, if $\bP(M,Y^\dagger,R^M=1\mid T=t,  X =x)$ is complete in $Y^\dagger$ for all $t$ and $x$, and $\bP(R^M=1,R^Y=1\mid M=m,T=t,X=x) > 0$ for all $m,t,x$, then $\bP(Y,M\mid T,  X )$ is identifiable, and therefore, the \textup{NIE} and \textup{NDE} are identifiable.
\end{theorem}

Under the MNAR Assumption \ref{ass4}, $\bP(Y\mid M, T, X)$ is identifiable using complete cases, i.e., $\bP(Y\mid M, T, X) = \bP(Y\mid M, T, X, R^Y=1, R^M=1)$. The completeness assumption is again only used to identify $\bP(M\mid T,  X )$. We provide the intuition for the discrete case. $\bP(M\mid T,  X )$ is identifiable if $\bP(R^M\mid M, T, X)$ is identifiable as discussed in section \ref{sec:mismediator}. For each $t,x$, we have the following system of linear equations with $\{\zeta_{t,x}(m): m\in\mathcal{M}\}$ as unknowns:
\begin{align}
&~~~~\bP(Y=y, R^M=0, R^Y=1\mid T=t, X=x) \nonumber\\&=\sum_{m\in \mathcal{M}}\bP(M=m, Y=y, R^M=1, R^Y=1\mid T=t, X=x) \zeta_{t,x}(m), \nonumber
\end{align}
\text{for each~}$y\in\mathcal{Y}$, and
\begin{align}
&~~~~\bP(R^M=0, R^Y=0\mid T=t, X=x)\nonumber\\&=\sum_{m\in \mathcal{M}}\bP(M=m,R^M=1,R^Y=0\mid T=t, X=x) \zeta_{t,x}(m). \nonumber
\end{align}
To ensure the uniqueness of solutions $\zeta_{t,x}(m)$, we need to invoke the following completeness condition that Rank $(\Theta_{tx})=J$, where $\Theta_{tx}$ is the $J \times (K+1)$ matrix with $\bP(M=m, Y=y, R^M=1, R^Y=1 \mid T=t, X=x)$ as the $(m,y)$th element and $\bP(M=m, R^M=1, R^Y=0 \mid T=t, X=x)$ as the $(m,K+1)$th element. The effect of $M$ on $R^Y$, if exists, provides one additional constraint to assist the identification of $\zeta_{t,x}(m)$. The completeness condition would fail if $J>(K+1)$ or $M\independent Y^\dagger\mid (T, X)$. Since the identification of $\bP(Y\mid M, T, X)$ does not rely on the completeness condition, when $M\independent Y\mid (T,X)$, the NIE and NDE are still identifiable as discussed in section \ref{sec:mismediator}.

So far, we have shown the nonparametric identification results of $\bP(Y,M\mid T,  X )$, NIE and NDE, under various MNAR assumptions. However, the nonparametric estimation for these quantities may suffer from the curse of dimensionality in practice, especially with a large number of covariates. Therefore, we adopt a parametric method to obtain likelihood-based inference. The estimation details based on the Expectation-Maximization algorithm \citep{dempster1977maximum} are provided in section \ref{sec::estimation} of the supplementary material.

\section{Simulation}\label{sec:simulation}

We conducted simulation studies to evaluate the performance of the proposed estimators under each of the MNAR assumptions described in sections \ref{sec:mismediator} and \ref{sec:misboth}. In a simple context of a single covariate $X \sim \mathcal{N}(0,1)$ and a randomized $T\sim \mathrm{Bernoulli}(0.5)$, we considered the following four setups representing different relationships in the supports of $M$ and $Y$: (A) binary $M$ and binary $Y$, (B) binary $M$ and continuous $Y$, (C) continuous $M$ and continuous $Y$, and (D) continuous $M$ and binary $Y$. We generated the mediator $M$ from
\setlength{\abovedisplayskip}{0pt} 
\setlength{\belowdisplayskip}{0pt}
$$\mathrm{logit}~\bP(M=1\mid T,X)=\alpha_0+\alpha_t T+\alpha_x X$$
if $M$ is binary; and 
$$M \sim \mathcal{N}(\alpha_0+\alpha_t T+\alpha_x X, 1)$$
if $M$ is continuous. We then generated the outcome $Y$ from
$$\mathrm{logit}~\bP(Y=1\mid M,T,X)=\beta_0+\beta_m M+\beta_t T+\beta_{mt} M \cdot T+\beta_x X$$
if $Y$ is binary; and 
$$Y \sim \mathcal{N}(\beta_0+\beta_m M+\beta_t T+\beta_{mt} M \cdot T+\beta_x X, 1)$$
if $Y$ is continuous.

For each of the four setups described above, we considered the missingness mechanisms under (I) Assumption \ref{ass1}, (II) Assumption \ref{ass2}, (III) Assumption \ref{ass3}, and (IV) Assumption \ref{ass4}, respectively, that is, sixteen simulation scenarios in total. Under MNAR Assumptions \ref{ass1} to \ref{ass4}, $R^M$ is allowed to depends on $M$, $T$ and $X$, and therefore, we generated the binary variable $R^M$ from
$$\mathrm{logit}~\bP(R^M=1\mid M,T,X)=\lambda_0+\lambda_m M+\lambda_t T+\lambda_x X.$$
Under (I) Assumption \ref{ass1}, $Y$ is fully observed. For scenarios (II) to (IV) with missingness in $Y$, the data generating models for $R^Y$ varied according to different MNAR Assumptions:
$$\mathrm{logit}~\bP(R^Y=1\mid R^M,T,X)=\gamma_0+\gamma_{r^M} R^M+\gamma_t T+\gamma_x X$$
if under (II) Assumption \ref{ass2};
$$\mathrm{logit}~\bP(R^Y=1\mid Y,T,X)=\gamma_0+\gamma_y Y+\gamma_t T+\gamma_x X$$
if under (III) Assumption \ref{ass3}; and 
$$\mathrm{logit}~\bP(R^Y=1\mid M,T,X)=\gamma_0+\gamma_m M+\gamma_t T+\gamma_x X$$
if under (IV) Assumption \ref{ass4}.

For each of the sixteen simulation scenarios considered, we tested our theoretical results and evaluated the performance of our methods when $M \not\independent  Y\mid (T,   X )$ and when $M \independent  Y\mid (T,   X )$, respectively. Table \ref{tab:simulation} presents the specifications of parameter values in section \ref{sec::suppsimulation} of the supplementary material. We set parameter values in the $R^M$ and $R^Y$ models to generate missing rates in $M$ and $Y$ both to be around $20\%$ to $25\%$, which are similar to the missing rates in the NJCS. 

\begin{figure}[ht]
\begin{center}
\includegraphics[scale = 0.315]{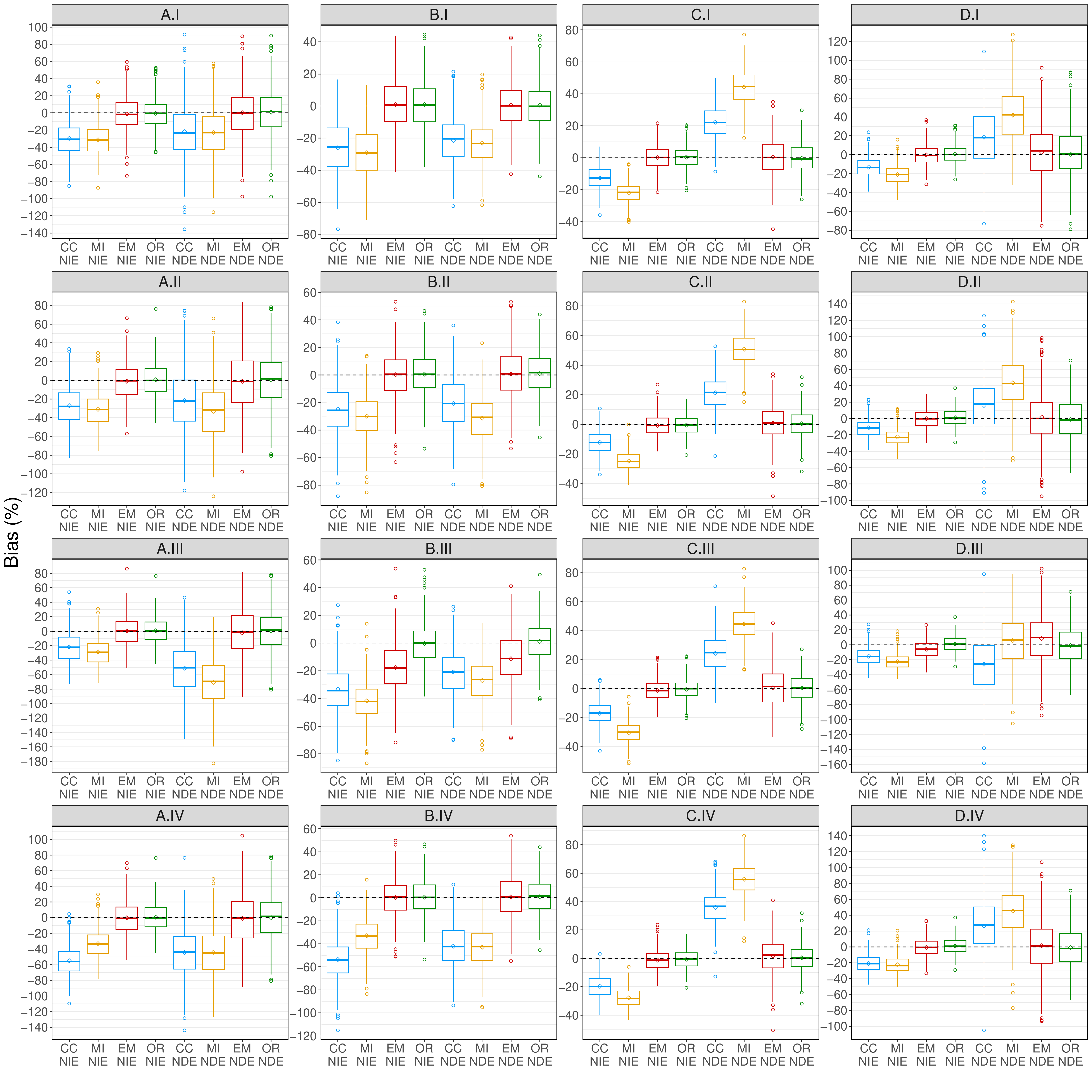}
\end{center}
\caption{Simulation results when $M \not\independent  Y\mid (T,   X )$. A, Binary $M$ and Binary $Y$; B, Binary $M$ and Continuous $Y$; C, Continuous $M$ and Continuous $Y$; D, Continuous $M$ and Binary $Y$; I, Assumption \ref{ass1}; II, Assumption \ref{ass2}; III, Assumption \ref{ass3}; IV, Assumption \ref{ass4}; CC, complete case analysis; MI, multiple imputation estimators; EM, our proposed Expectation-Maximization algorithm; OR, oracle estimators; Bias (\%), \{(estimate-truth)/truth\}*100.}
\label{fig:simulation1}
\end{figure}

We considered a sample size of $1000$, and simulated $500$ data sets for each simulation scenario. We applied the following four methods to compare their results on estimations of the NIE and NDE: 1) complete case analysis, which provides consistent estimates under MCAR; 2) multiple imputation conducted by MICE with default imputation techniques (i.e. predictive mean matching for the numeric scale and logistic regression for the factor with 2 levels) \citep{vanbuuren2011multivariate} assuming MAR; 3) our proposed methods using the Expectation-Maximization algorithm, which are designed to deal with the MNAR assumptions under concern; and 4) oracle estimators, which are obtained by using the true values of the missing data. Figure \ref{fig:simulation1} presents the boxplots of the percentages of bias with respect to the true values for each of the simulation scenarios when $M \not\independent  Y\mid (T,   X )$ across the $500$ replications. When $M\independent Y\mid (T,  X )$, the simulation results are consistent with the theoretical results in sections \ref{sec:mismediator} and \ref{sec:misboth}, and we relegate the details to section \ref{sec::suppsimulation} of the supplementary material due to space limitations.

Under MNAR Assumption \ref{ass1} where missingness only exists in the mediator, the percentages of bias for both the NIE and NDE estimated using our methods are close to zero with slightly larger standard errors than the oracle estimates when $M\not\independent Y\mid (T,  X )$ and the completeness assumption holds (A.I, B.I, C.I), while the estimated NIE and NDE from both complete case analysis and multiple imputation under MAR have substantial biases. It is interesting to observe that when $M\not\independent Y\mid (T,  X )$ and the completeness assumption is violated due to the support of $M$ being larger than the support of $Y$ as in the continuous $M$ and binary $Y$ case (D.I), our methods still recovers the underlying truths with the help of parametric assumptions, while both complete case analysis and multiple imputation under MAR have substantial biases. However, we would like to point out that the parametric assumptions do not always help recover the underlying model parameter values. In section \ref{sec::suppsimulation} of the supplementary material, we present an example where the distribution of model parameters exhibits bimodality, skewness and other irregular patterns, and therefore, the results may not be trustworthy. That is a setup where $M$ follows a multinomial logistic regression model with three categories and $Y$ is binary following a logistic regression model. Under MNAR Assumption \ref{ass2} (A.II to D.II) and MNAR Assumption \ref{ass4} (A.IV to D.IV), we reached the same conclusions as those under MNAR Assumption \ref{ass1}.

Under MNAR Assumption \ref{ass3}, when $M\not\independent Y\mid (T,  X )$, the completeness assumption holds in A.III and C.III. In those cases, the percentages of bias for both the NIE and NDE estimated using our methods are close to zero as expected, while the estimated NIE and NDE from both complete case analysis and multiple imputation under MAR have substantial biases. When the completeness assumption is violated in both the binary $M$ and continuous $Y$ (B.III) and the continuous $M$ and binary $Y$ (D.III) cases, our estimators in general have reduced bias compared to both complete case analysis and multiple imputation under MAR, however, would still fail to recover the underlying truths even with the help of the parametric models. Furthermore, we find that the performance of our estimators is highly sensitive to the specifications of the parameter values when the completeness assumption is violated in B.III and D.III, while our methods always recovers the underlying truths when the completeness assumption holds in A.III and C.III. 

As pointed out by \citet{cox2011principles}, ``If an issue can be addressed nonparametrically then it will often be better to tackle it parametrically; however, if it cannot be resolved nonparametrically then it is usually dangerous to resolve it parametrically". Our simulation results echo this point. When the corresponding completeness assumption holds and we can achieve nonparametric identification, the underlying data distribution can be consistently estimated if the missingness mechanism is correctly accounted for. However, when the corresponding completeness assumption does not hold and we cannot achieve nonparametric identification, the degree to which parametric assumptions can help vary from case to case. 

\section{Application to the National Job Corps Study}\label{sec:data_analysis}

\subsection{Data}

The data describes $8707$ eligible applicants in the mid-1990s who lived in the areas selected for in-person interviews at the baseline. The subjects were randomized either to the experimental group $(T=1)$ where they could join the Job Corps program soon after randomization, or to the control group $(T=0)$ where they were not provided the Job Corps program for three years \citep{schochet2001national}. The mediator ($M$) was collected at the 30-months follow-up describing subject's educational and vocational attainment, measured by whether or not the subject obtained an education credential or vocational certificate after randomization. We use $M=1$ to denote that an education credential or vocational certificate was obtained, and $M=0$ otherwise. The outcome ($Y$) was collected at the 48-months follow-up describing the subject's weekly earnings in the fourth year after randomization. The covariates $  X $ include information on gender, age, race, education level, earnings in the year before participating in the study, whether the subject had a child or not, and whether the subject had ever been arrested or not. There are some small portions of missingness in covariates $X$, including education level $(0.65\%)$, earnings levels in the year before participating in the study $(9.59\%)$, and whether the subject had ever been arrested or not $(6.74\%)$. Since covariates $X$ are all discrete, we treat missingness as another category for each covariate in the analysis. We provide details on the distribution of the covariates in section \ref{sec::xtable} of the supplementary material.

The number of subjects with missing information in the mediator or outcome is nontrivial. The missingness patterns in the mediator and outcome are described in Table \ref{tab:Missing Patterns} for the experimental group and the control group. We suspect that missingness may be MNAR in the data. Besides the potential impacts of $T$ and $X$ on missingness, we have the following concerns: (1) conceivably, people who failed to obtain an education credential or vocational certificate ($M=0$) may be less likely to report compared to people who successfully obtained an educational credential or vocational certificate ($M=1$), that is, $M$ may have a direct effect on $R^M$; (2) we are concerned that people who were unwilling to respond to questions at the 30-months follow-up may also be unwilling to respond at the 48-months follow-up, that is, $R^M$ may have a direct impact on $R^Y$; (3) people who had no earnings $(Y=0)$ may be less likely to report their earnings compared to people who had earnings $(Y>0)$, which results in a direct effect of $Y$ on $R^Y$; (4) in addition, $M$ occurs before $R^Y$, and therefore, may potentially have an impact on people's probability of reporting earnings through channels other than $R^M$ and $Y$. Concerns $(1)$ and $(2)$ can be addressed by the MNAR Assumption \ref{ass2}, concerns $(1)$ and $(3)$ can be addressed by the MNAR Assumption \ref{ass3}, and concerns $(1)$ and $(4)$ can be addressed by the MNAR Assumption \ref{ass4}. When incorporating the weekly earnings $Y$ into the prediction of $R^Y$ under MNAR Assumption \ref{ass3}, we assume that it is the binary indicator $\mathrm{I}(Y>0)$ describing whether $Y$ is positive or not that predicts $R^Y$. This is because of the following two considerations: first, there are excessive zero values of the earnings $(14.85\%)$ in the data; second, the identification under MNAR Assumption \ref{ass3} requires $Y$ to be binary given that our mediator $M$ in this study is binary (see the completeness assumption in Theorem \ref{the3}). 

\subsection{Models}
Our outcome $Y$, weekly earnings, contains many zero values as well as heavily right skewed positive values. To address those complications, we adopted two-part models. Let $H_i$ be the binary indicator describing whether the earning is greater than $0$ or not, i.e., $H_i=1$ if $Y_i>0$ and $H_i=0$ if $Y_i=0$.
We used the following logistic regression to model $H_{i}$:
$$\mathrm{logit}~\bP(H_i=1\mid M_i=m,T_i=t,   X _i=  x   ) = \delta_0+\delta_m m+\delta_t t+\delta_{mt} m \cdot t+\delta_x^ \textsc{t}  x .$$
Conditioning on $H_{i}=1$, we considered two commonly adopted models, Gamma and log-normal models to fit the positively skewed values of earnings. As an illustration, we describe the Gamma model here: $Y_i \mid (H_i=1, M_i=m, T_i=t,   X _i=  x   ) \sim \mathrm{Gamma}\{\nu,\nu/\mu_i(m, t, x)\}$, where $\nu$ denotes the shape parameter, $\nu / \mu_i(m, t, x)$ is the rate parameter, and the function $\mu_i(m, t, x)$ is parameterized as $\mathrm{exp}(\beta_0+\beta_m m+\beta_t t + \beta_{mt} m\cdot t+\beta_x^ \textsc{t}  x ).$

We used a logistic model for $M$: 
$$\mathrm{logit}~\bP(M_i=1\mid T_i=t,   X _i=  x   )=\alpha_0+\alpha_{t}t+\alpha_{x}^ \textsc{t}  x ,$$
and a logistic model for $R^M$ allowing an effect from $M$:
$$\mathrm{logit} ~\bP(R^M_i=1\mid M_i=m, T_i=t,    X _i=  x   ) = \lambda_0+\lambda_m m+\lambda_t t +\lambda_x^ \textsc{t}  x .$$
The model for $R^Y$ varied according to different MNAR assumptions. Under MNAR Assumption \ref{ass2}, we specified that 
$$\mathrm{logit} ~\bP(R^Y_i=1\mid R^M_i=r^M, T_i=t,    X _i=  x   ) = \gamma_0+\gamma_{r^M}r^M+\gamma_t t +\gamma_x^ \textsc{t}  x ;$$
under MNAR Assumption \ref{ass3}, we specified that 
$$\mathrm{logit} ~\bP(R^Y_i=1\mid H_i=h, T_i=t,    X _i=  x   ) = \gamma_0+\gamma_{h} h+\gamma_t t +\gamma_x^ \textsc{t}  x ;$$
and under MNAR Assumption \ref{ass4}, we adopted that
$$\mathrm{logit} ~\bP(R^Y_i=1\mid M_i=m, T_i=t,    X _i=  x   ) = \gamma_0+\gamma_{m} m+\gamma_t t +\gamma_x^ \textsc{t}  x .$$

\subsection{Results}

We compared the performance and results using two-part Gamma and two-part log-normal models for the outcome under MNAR Assumptions \ref{ass2}, \ref{ass3}, and \ref{ass4}. Table \ref{tab:Model comparison} presents the log-likelihoods evaluated at the corresponding Maximum Likelihood Estimates (MLEs) and results on the NIE and NDE for those six models. The causal conclusions on the NIE and NDE are consistent across those six models. Since all six models have the same numbers of parameters, we compared likelihoods. The model that stands out is the two-part Gamma model with MNAR mechanism under Assumption \ref{ass2}, which allows an impact on $R^M$ from $M$ and an impact on $R^Y$ from $R^M$, besides the impacts from $T$ and $  X $. Below we describe the results based on the two-part Gamma model under MNAR Assumption \ref{ass2}.

\begin{table}[H]
\begin{center}
\caption{Model comparison among models under MNAR Assumptions \ref{ass2}, \ref{ass3}, and \ref{ass4} using two-part Gamma and two-part log-normal models for the outcome. The log-likelihoods are evaluated at the corresponding MLEs; Est, estimate; CI, confidence interval based on $500$ bootstrap samples.} 
\label{tab:Model comparison}
\begin{tabular}{lllllllllll}
\hline 
\multicolumn{1}{l}{Assumption}&\multicolumn{1}{l}{Model}&\multicolumn{1}{l}{Log-likelihood}&\multicolumn{2}{l}{NIE}&\multicolumn{2}{l}{NDE}\\
& & & Est & $95\%$ CI & Est & $95\%$ CI\\ \hline
\ref{ass2} & Gamma & $-53131.35$\checkmark & $10.94$ & $(7.94, ~14.29)$ & $12.93$ & $(-1.95, ~27.64)$\\
\ref{ass3} & Gamma & $-53488.54$ & $14.87$ & $(11.59, ~18.35)$ & $9.99$ & $(-2.99,~22.73)$\\
\ref{ass4} & Gamma & $-53475.01$ & $10.14$ & $(7.26, ~13.25)$ & $11.29$ & $(-3.85,~26.21)$\\
\ref{ass2} & Log-normal & $-53799.79$ & $15.50$ & $(11.06, ~20.16)$ & $4.14$ & $(-18.14,~26.42)$\\
\ref{ass3} & Log-normal & $-54159.23$ & $19.23$ & $(14.79, ~23.81)$ & $3.21$ & $(-15.37,~21.74)$\\
\ref{ass4} & Log-normal & $-54145.92$ & $14.36$ & $(10.16,~18.71)$ & $1.68$ & $(-20.82,~24.21)$\\
\hline
\end{tabular}
\end{center}
\end{table}

The coefficient of $M$ in the $R^M$ model is estimated to be $1.73$ with $95\%$ CI $(0.34,~3.33)$ - see $\lambda_m$ in Table \ref{tab:Data Analysis}, which suggests that people who had acquired certificates were more likely to report their certificate status. The coefficient of $R^M$ in the $R^Y$ model is estimated to be $1.87$ with $95\%$ CI $(1.76,~2.00)$ - see $\gamma_{r^M}$ in Table \ref{tab:Data Analysis}, which suggests that people who were willing to report their certificate status were also more likely to report their earnings. The strong association between $R^M$ and $R^Y$ might be a result of both being affected by the subjects' unmeasured tendency or willingness to respond to interview questions. The natural indirect effect is estimated to be $10.94$ with $95\%$ CI $(7.94,~14.29)$ - see $\mathrm{NIE}$ in Table \ref{tab:Data Analysis}, which indicates that there was a significant indirect effect of the program assignment on weekly earnings through an educational credential or vocational certificate at the $0.05$ significance level. The natural direct effect is estimated to be $12.93$ with $95\%$ CI $(-1.95,~27.64)$ - see NDE in Table \ref{tab:Data Analysis}, which indicates that there was no significant direct effect of the program assignment on weekly earnings at the $0.05$ significance level. The causal conclusions regarding the NIE and NDE are the same among complete case analysis, multiple imputation under MAR and our proposed
Expectation-Maximization algorithm, in spite of the significant effect of $M$ on $R^M$ measured by $\lambda_m$.  

% and the significant effect of $R^M \rightarrow R^Y$ ($\gamma_{r^M}$)

\begin{table}[H]
\centering
\caption{Data analysis results from the Gamma model under MNAR Assumption \ref{ass2}. Est, estimate; CI, confidence interval based on $500$ bootstrap samples; $\lambda_m$, coefficient of $M$ in the $R^M$ model; $\gamma_{r^M}$, coefficient of $R^M$ in the $R^Y$ model.}
\label{tab:Data Analysis}
\begin{tabular}{lllllllllll}
\hline 
\multicolumn{1}{l}{}&\multicolumn{2}{l}{Complete Case Analysis}&\multicolumn{2}{l}{Multiple Imputation}&\multicolumn{2}{l}{EM Algorithm}\\
Parameters & Est & $95\%$ CI & Est & $95\%$ CI & Est & $95\%$ CI\\ \hline
$\lambda_m$    & NA & NA & NA & NA & $1.73$ & $(0.34, ~3.33)$\\ 
$\gamma_{r^M}$ & NA & NA & NA & NA & $1.87$ & $(1.76, ~2.00)$\\ 
NIE            & $12.00$ & $(8.65, ~15.57)$ & $12.04$ & $(8.25, ~14.60)$ & $10.94$ & $(7.94,~14.29)$\\ 
NDE            & $14.75$ & $(-0.05, ~29.50)$ & $9.22$ &  $(-5.85,~23.23)$ & $12.93$ & $(-1.95,~27.64)$\\  \hline
\end{tabular}
\end{table}

\section{Discussion}\label{sec:disc}

We relegate several extensions to the supplementary material due to space limitations. First, we provide examples of missingness mechanisms where identification cannot be achieved without further assumptions, and we present conditions for identification under those complex missingness mechanisms by utilizing information on a future outcome in section \ref{sec::counterexamples}. We do not have such information available in our application data, but it offers guidance to data collection in study designs where certain unidentifiable missingness mechanisms are expected. Second, to address the concern that missingness of $Y$ may also depends on $Y$ itself and $M$ in addition to $R^M$ in the NJCS, we develop a sensitivity analysis approach to study the robustness of conclusions on the causal mediation effects if the underlying missingness mechanism is beyond the mechanisms that allow identification in section \ref{sec::suppsens} and found that our conclusions on the NIE and NDE are not sensitive to some large impacts on $R^Y$ from $Y$ and $M$. Third, due to the similarity, we extend our nonparametric identification results to the instrumental variable analysis with treatment and outcome MNAR in section \ref{sec::suppiv}. 

Several limitations of the work are to be addressed in future research. First, prevalent missingness may exist in both the mediator/outcome and the covariates, and missingness of covariates may also be MNAR. Identifiability in this more complicated scenario is to be explored. Second, the NIE and NDE evaluated in the NJCS should be interpreted as the effects of the program assignment instead of the effects of the program. According to \cite{schochet2006national}, although control group subjects were barred from participating in the Job Corps program, there were about $30\%$ of the program group subjects who failed to participate in the program. This one-sided noncompliance issue is to be addressed. Third, we considered a setting where there is a single mediator. In practice, cases with multiple mediators may arise, and the potential missingness mechanisms may vary depending on whether those mediators have a sequential relationship or parallel relationship, and etc. Fourth, the sequential ignorability assumption could be violated in the NJCS. One approach to partly address this issue is to conduct sensitivity analysis to quantify the potential bias due to the departure from the sequential ignorability assumption \citep{vanderweele2015explanation,Hong2022posttreatment}. Fifth, we ignore the site membership in the NJCS and focus on the
single-level setting. In other words, we only considered the population-average causal effects without investigating the between-site heterogeneity of causal effects in the NJCS. By viewing the site as a discrete covariate, our nonparametric identification results can be extended to multilevel settings \citep{qin2019multisite, qin2021unpacking}. However, in finite-sample, one needs more sophisticated multi-level models to estimate the site-specific NIE and NDE.

\section*{Supplementary material}\label{sec:supp}
The supplementary material contains proofs of the theorems, parametric examples for the completeness conditions, counterexamples and further discussion for the unidentifiable cases, details for the parametric estimation, details on the simulation studies, details on the distribution of the covariates in the NJCS, sensitivity analysis results on the NJCS and extension of our identification results to the instrumental variable analysis.

\section*{Acknowledgements}\label{sec:ack}
We thank William Lippitt, Xu Qin, Guanglei Hong, and Wang Miao for helpful discussions. Shuozhi Zuo was supported by NIH R01GM108711, NSF DMS 1914937 and NSF SES 2149492. Debashis Ghosh was partially supported by NSF DMS 1914937 and NSF SES 2149492. Peng Ding was partially supported by the NSF DMS 1945136. Fan Yang was partially supported by NIH R01GM108711 and IES R305D200031. 

%\section*{Additional information}\label{sec:add}
%\subsection*{Funding}\label{sec:fund}

\bibliographystyle{apalike}
\bibliography{ref}

\newpage
\bigskip
\begin{center}
{\title\bf Supplementary material for ``Mediation analysis with the mediator and outcome missing not at random"}\\
\end{center}
\if0\blind
{}\fi

\setcounter{page}{1}
\setcounter{equation}{0}
\setcounter{section}{0}
\setcounter{figure}{0}
\setcounter{table}{0}
\renewcommand {\thepage} {S\arabic{page}}
\renewcommand {\theequation} {S\arabic{equation}}
\renewcommand {\thesection} {S\arabic{section}}
\renewcommand{\thefigure}{S\arabic{figure}}
\renewcommand{\thetable}{S\arabic{table}}

Section \ref{sec::proofs} gives proofs of the theorems.

Section \ref{sec::parametricexamples} gives parametric examples for the completeness conditions.

Section \ref{sec::counterexamples} gives counterexamples and further discussion for the unidentifiable cases.

Section \ref{sec::estimation} provides details for the parametric estimation.

Section \ref{sec::suppsimulation} provides details on the simulation studies.

Section \ref{sec::xtable} provides details on the distribution of the covariates in the NJCS.

Section \ref{sec::suppsens} provides sensitivity analysis results on the NJCS.

Section \ref{sec::suppiv} extends our identification results to the instrumental variable analysis.

\section{Proofs}\label{sec::proofs}
\subsection{Proof of Theorem \ref{the1}}
\setlength{\abovedisplayskip}{5pt} 
\setlength{\belowdisplayskip}{5pt}
The identification of $\bP(Y=y\mid M=m, T=t,    X =   x  )$ follows from
\begin{align}
\bP(Y=y\mid M=m, T=t,    X =   x  ) = \bP(Y=y\mid R^M = 1, M=m, T=t,    X =   x  )\nonumber.
\end{align}
We now focus on the identification of $\bP(M=m\mid T=t,    X =   x  )$. Define 
\begin{eqnarray*}
    \bP_{my1\mid t, x} &=& \bP(M=m, Y=y, R^M=1\mid T=t,    X =   x  ), \\
    \bP_{+y0\mid t,x} &=& \bP(Y=y, R^M=0\mid T=t,    X =   x  ), \\
    \zeta_{t,x}(m) &=&  \frac{\bP(R^M=0\mid M=m, T=t,    X =   x  )}{\bP(R^M=1\mid M=m, T=t,    X =   x  )}.
\end{eqnarray*}
Since
$$
\bP_{my1\mid t, x} = \bP(M=m, Y=y\mid T=t,    X =   x  )\bP(R^M=1\mid M=m,T=t,   X =   x  ),
$$
we have:
\begin{align}
\bP_{+y0\mid t,x} &= \int_{m\in \mathcal{M}} \bP(M=m, Y=y, R^M=0\mid T=t,    X =   x  )\textup{d}m \nonumber\\&=\int_{m\in \mathcal{M}} \bP(M=m, Y=y\mid T=t,    X =   x  ) \bP(R^M=0\mid M=m, T=t,    X =   x  )\textup{d}m \nonumber\\&=\int_{m\in \mathcal{M}}\bP_{my1\mid t, x}  \frac{\bP(R^M=0\mid M=m, T=t,    X =   x  )}{\bP(R^M=1\mid M=m, T=t,    X =   x  )}\textup{d}m \nonumber\\&=\int_{m\in \mathcal{M}}\bP_{my1\mid t, x}  \zeta_{t,x}(m)\textup{d}m,\nonumber
\end{align}
\text{for each} $y\in\mathcal{Y}$. The uniqueness of solutions $\zeta_{t,x}(m)$ requires that $\bP(Y,M,R^M=1 \mid T=t,  X=x )$ is complete in $Y$ for all $t$ and $x$. For discrete $M$ and discrete $Y$, the completeness assumption is equivalent to Rank $(\Theta_{tx})=J$, where $\Theta_{tx}$ is a $J \times K$ matrix with $\bP_{my1\mid t, x}$ as the $(m,y)$th element. For binary $M$, the rank condition further reduces to $M \not\independent  Y\mid (T,  X )$, which is equivalent to the testable condition $M \not\independent  Y\mid (T,  X ,R^M=1)$. For continuous $M$ and continuous $Y$, the dimension of $Y$ needs to be no smaller than the dimension of $M$ in general as required by the completeness assumption.

We can subsequently identify $\bP(R^M=1\mid M=m, T=t,    X =   x  )$ once $\zeta_{t,x}(m)$ is identified. Then, the identification of $\bP(M=m\mid T=t, X = x)$ follows from 
\begin{align}
\bP(M=m\mid T=t, X = x) &= \frac{\bP(M=m, R^M=1\mid T=t, X = x)}{\bP(R^M=1\mid M=m, T=t, X = x)}.\nonumber
\end{align}

\subsection{Proof of Theorem \ref{the2}}
\subsubsection*{Theorem \ref{the2} $(i)$} 

The identification of $\bP(Y=y\mid M=m, T=t,    X =   x  )$ follows from
\begin{align}
\bP(Y=y\mid M=m, T=t,    X =   x  ) = \bP(Y=y\mid R^M=1, R^Y=1, M=m, T=t,    X =   x  ).\nonumber
\end{align}
We now focus on the identification of $\bP(M=m\mid T=t,    X =   x  )$. Define 
\begin{eqnarray*}
    \bP_{my11\mid t,x} &=& \bP(M=m, Y=y, R^M=1, R^Y=1\mid T=t,    X =   x  ), \\
    \bP_{+y01\mid t,x} &=& \bP(Y=y, R^M=0, R^Y=1\mid T=t,    X =   x  ), \\
    \zeta_{t,x}(m)&=&\frac{\bP(R^M=0\mid M=m,T=t,   X =   x  )}{\bP(R^M=1\mid M=m,T=t,   X =   x  )}.
\end{eqnarray*}
Since
\begin{eqnarray*}
\bP_{my11\mid t,x} &=& \bP(M=m, Y=y\mid T=t,    X =   x  ) \\
&& \cdot \bP(R^Y=1\mid R^M=1,T=t,   X =   x )\bP(R^M=1\mid M=m,T=t,   X =   x  ),
\end{eqnarray*}
we have:
\begin{align}
\bP_{+y01\mid t,x} &= \int_{m\in \mathcal{M}} \bP(M=m, Y=y, R^M=0, R^Y=1\mid T=t,    X =   x  )\textup{d}m\nonumber\\&=\int_{m\in \mathcal{M}}\bP_{my11\mid t,x}  \frac{\bP(R^Y=1\mid R^M=0,T=t,   X =   x  )\bP(R^M=0\mid M=m,T=t,   X =   x  )}{\bP(R^Y=1\mid R^M=1,T=t,   X =   x  )\bP(R^M=1\mid M=m,T=t,   X =   x  )}\textup{d}m\nonumber\\&=\frac{\bP(R^Y=1\mid R^M=0,T=t,   X =   x  )}{\bP(R^Y=1\mid R^M=1,T=t,   X =   x  )}\int_{m\in \mathcal{M}}\bP_{my11\mid t,x} \zeta_{t,x}(m)\textup{d}m,\nonumber
\end{align}
\text{for each} $y\in\mathcal{Y}$. The uniqueness of solutions $\zeta_{t,x}(m)$ requires that $\bP(Y,M,R^M=1,R^Y=1 \mid T=t,  X=x )$ is complete in $Y$ for all $t$ and $x$. For discrete $M$ and discrete $Y$, the completeness assumption is equivalent to Rank $(\Theta_{tx})=J$, where $\Theta_{tx}$ is a $J \times K$ matrix with $\bP_{my11\mid t,x}$ as the $(m,y)$th element. For binary $M$, the rank condition further reduces to $M \not\independent  Y\mid (T,  X )$, which is equivalent to the testable condition $M \not\independent  Y\mid (T,  X ,R^M=1,R^Y=1)$. For continuous $M$ and continuous $Y$, the dimension of $Y$ needs to be no smaller than the dimension of $M$ in general as required by the completeness assumption.

We can subsequently identify $\bP(R^M=1\mid M=m, T=t,    X =   x  )$ once $\zeta_{t,x}(m)$ is identified. Then, the identification of $\bP(M=m\mid T=t, X = x)$ follows from
\begin{align}
\bP(M=m\mid T=t, X = x) &= \frac{\bP(M=m, R^M=1\mid T=t, X = x)}{\bP(R^M=1\mid M=m, T=t, X = x)}.\nonumber
\end{align}

\subsubsection*{Theorem \ref{the2} $(ii)$} 
The identification of $\bP(Y=y, M=m \mid T=t, X=x)$ follows from
\begin{align}
\bP(Y=y, M=m \mid T=t, X=x) = \bP(Y=y, M=m \mid R^M=1, R^Y=1, T=t, X = x).\nonumber
\end{align}

\subsection{Proof of Theorem \ref{the3}}

We discuss the identification of $\bP(M=m,Y=y\mid T=t,    X =   x  )$. Define 
\begin{eqnarray*}
    \bP_{my11\mid t,x} &=& \bP(M=m, Y=y, R^M=1, R^Y=1\mid T=t,    X =   x  ), \\
    \bP_{+y01\mid t,x} &=& \bP(Y=y, R^M=0, R^Y=1\mid T=t,    X =   x  ), \\
    \bP_{m+10\mid t,x} &=& \bP(M=m, R^M=1, R^Y=0\mid T=t,    X =   x  ), \\
    \zeta_{t,x}(m)&=&\frac{\bP(R^M=0\mid M=m,T=t,   X =   x  )}{\bP(R^M=1\mid M=m,T=t,   X =   x  )},\\
    \eta_{t,x}(y)&=&\frac{\bP(R^Y=0\mid Y=y,T=t,   X =   x  )}{\bP(R^Y=1\mid Y=y,T=t,   X =   x  )}.
\end{eqnarray*}
Since 
\begin{eqnarray*}
\bP_{my11|t,x} 
&=& \bP(M=m, Y=y\mid T=t,    X =   x  ) \\
&& \cdot\bP(R^Y=1\mid Y=y,T=t,   X =   x  )\bP(R^M=1\mid M=m,T=t,   X =   x  ),
\end{eqnarray*}
We have:
\begin{align}
\bP_{+y01\mid t,x} &= \int_{m\in \mathcal{M}} \bP(M=m, Y=y, R^M=0, R^Y=1\mid T=t,    X =   x  )\textup{d}m\nonumber\\&=\int_{m\in \mathcal{M}}\bP_{my11\mid t,x}\frac{\bP(R^M=0\mid M=m,T=t,   X =   x  )}{\bP(R^M=1\mid M=m,T=t,   X =   x  )}\textup{d}m\nonumber\\&=\int_{m\in \mathcal{M}}\bP_{my11\mid t,x} \zeta_{t,x}(m)\textup{d}m,\nonumber
\end{align} 
\text{for each} $y\in\mathcal{Y}$, and
\begin{align}
\bP_{m+10\mid t,x} &= \int_{y\in \mathcal{Y}}\bP(M=m, Y=y, R^M=1, R^Y=0\mid T=t,    X =   x  )\textup{d}y\nonumber\\&=\int_{y\in \mathcal{Y}}\bP_{my11\mid t,x}  \frac{\bP(R^Y=0\mid Y=y,T=t,   X =   x  )}{\bP(R^Y=1\mid Y=y,T=t,   X =   x  )}\textup{d}y\nonumber\\&=\int_{y\in \mathcal{Y}}\bP_{my11\mid t,x} \eta_{t,x}(y)\textup{d}y,\nonumber
\end{align}
\text{for each} $m\in\mathcal{M}$. The uniqueness of solutions $\zeta_{t,x}(m)$ requires that $\bP(Y,M,R^M=1,R^Y=1 \mid T=t,  X=x )$ is complete in $Y$ for all $t$ and $x$, and the uniqueness of solutions $\eta_{t,x}(y)$ require that $\bP(Y,M,R^M=1,R^Y=1 \mid T=t,  X=x )$ is complete in $M$ for all $t$ and $x$. For discrete $M$ and discrete $Y$, the above completeness assumptions are equivalent to $J=K$ and Rank $(\Theta_{tx})=J$, where $\Theta_{tx}$ is a $J \times J$ matrix with $\bP_{my11\mid t,x}$ as the $(m,y)$th element. For binary $M$ and binary $Y$, the rank condition reduces to $M \not\independent  Y\mid (T,  X )$. For continuous $M$ and continuous $Y$, the dimension of $Y$ needs to be the same as the dimension of $M$ in general as required by $\bP(Y,M,R^Y=1,R^M=1\mid T,  X )$ being complete in $M$ and being complete in $Y$.

We can subsequently identify $\bP(R^M=1\mid M=m, T=t,    X =   x  )$ and $\bP(R^Y=1\mid Y=y, T=t,    X =   x  )$ once $\zeta_{t,x}(m)$ and $\eta_{t,x}(y)$ are identified. Then, the identification of $\bP(Y=y, M=m\mid T=t, X = x)$ follows from
\begin{eqnarray}
&&\bP(Y=y, M=m\mid T=t,    X =   x  )\nonumber\\&=&\frac{\bP_{my11\mid t,x}}{\bP(R^M=1\mid M=m, T=t,    X =   x  )\bP(R^Y=1\mid Y=y, T=t,    X =   x  )}.\nonumber
\end{eqnarray}

\subsection{Proof of Theorem \ref{the4}}

The identification of $\bP(Y=y\mid M=m, T=t,    X =   x  )$ follows from
\begin{align}
\bP(Y=y\mid M=m, T=t,    X =   x  ) = \bP(Y=y\mid R^M = 1, R^Y=1, M=m, T=t,    X =   x  )\nonumber.
\end{align}
We now focus on the identification of $\bP(M=m\mid T=t,    X =   x  )$. Define 
\begin{eqnarray*}
    \bP_{my^\dagger1\mid t,x} &=& \bP(M=m, Y^\dagger=y^\dagger, R^M=1\mid T=t, X = x), \\
    \bP_{my11\mid t,x} &=& \bP(M=m, Y=y, R^M=1, R^Y=1\mid T=t,    X =   x  ), \\
    \bP_{+y01\mid t,x} &=& \bP(Y=y, R^M=0, R^Y=1\mid T=t,    X =   x  ), \\
    \bP_{m+10|t,x} &=& \bP(M=m,R^M=1,R^Y=0\mid T=t, X = x), \\
    \bP_{++00\mid t,x} &=& \bP(R^M=0, R^Y=0\mid T=t,    X =   x  ), \\ 
    \zeta_{t,x}(m)&=&\frac{\bP(R^M=0\mid M=m,T=t,   X =   x  )}{\bP(R^M=1\mid M=m,T=t,   X =   x  )}.
\end{eqnarray*}
Since
\begin{eqnarray*}
\bP_{my11\mid t,x} &=& \bP(M=m, Y=y \mid T=t, X = x)\\
&&\cdot\bP(R^M=1\mid M=m, T=t, X = x)\bP(R^Y=1\mid M=m, T=t,    X =   x  ),
\end{eqnarray*}
and
\begin{eqnarray*}
\bP_{m+10\mid t,x} &=& \bP(M=m \mid T=t, X = x)\\
&&\cdot\bP(R^M=1\mid M=m, T=t, X = x)\bP(R^Y=0\mid M=m, T=t,    X =   x  ),
\end{eqnarray*}
We have:
\begin{align}
\bP_{+y01\mid t,x} &= \int_{m\in \mathcal{M}} \bP(M=m, Y=y, R^M=0, R^Y=1\mid T=t,    X =   x  )\textup{d}m\nonumber\\&=\int_{m\in \mathcal{M}}\bP_{my11\mid t,x} \frac{\bP(R^M=0\mid M=m, T=t,    X =   x  )}{\bP(R^M=1\mid M=m, T=t,    X =   x  )}\textup{d}m\nonumber\\&=\int_{m\in \mathcal{M}}\bP_{my11\mid t,x} \zeta_{t,x}(m)\textup{d}m,\nonumber
\end{align}
\text{for each} $y\in\mathcal{Y}$, and 
\begin{align}
\bP_{++00\mid t,x} &= \int_{m\in \mathcal{M}} \bP(M=m, R^M=0, R^Y=0\mid T=t,X=x)\textup{d}m \nonumber\\&=\int_{m\in \mathcal{M}} \bP_{m+10|t,x}  \frac{\bP(R^M=0\mid M=m, T=t,    X =   x  )}{\bP(R^M=1\mid M=m, T=t,    X =   x  )}\textup{d}m \nonumber\\&=\int_{m\in \mathcal{M}}\bP_{m+10|t,x}  \zeta_{t,x}(m)\textup{d}m. \nonumber
\end{align}
The uniqueness of solutions $\zeta_{t,x}(m)$ requires that $\bP(M,Y^\dagger,R^M=1 \mid T=t,  X=x )$ is complete in $Y^\dagger$ for all $t$ and $x$. For discrete $M$ and discrete $Y$, the completeness assumption is equivalent to Rank $(\Theta_{tx})=J$, where $\Theta_{tx}$ is a $J \times (K+1)$ matrix with $\bP_{my11\mid t,x}$ as the $(m,y)$th element and $\bP_{m+10\mid t,x}$ as the $(m,K+1)$th element. The effect of $M$ on $R^Y$, if exists, provides one additional constraint to assist the identification of $\zeta_{t,x}(m)$. For binary $M$, the rank condition further reduces to $M \not\independent  Y^\dagger \mid (T,  X )$, that is $M \not\independent  Y\mid (T,  X )$ or $M \not\independent  R^Y\mid (T,  X )$, which is equivalent to the testable condition $M \not\independent  Y\mid (T,  X ,R^M=1,R^Y=1)$ or $M \not\independent  R^Y\mid (T,  X ,R^M=1)$. For continuous $M$ and continuous $Y$, the dimension of $Y^\dagger$ needs to be no smaller than the dimension of $M$ in general as required by the completeness assumption.

We can subsequently identify $\bP(R^M=1\mid M=m, T=t,    X =   x  )$ once $\zeta_{t,x}(m)$ is identified. Then, the identification of $\bP(M=m\mid T=t, X = x)$ follows from
\begin{align}
\bP(M=m\mid T=t, X = x) &= \frac{\bP(M=m, R^M=1\mid T=t, X = x)}{\bP(R^M=1\mid M=m, T=t, X = x)}.\nonumber
\end{align}

\section{Parametric examples}\label{sec::parametricexamples}

Theorem 2.2 in \citeSupp{suppnewey2003instrumental} presents the following result on the completeness of distributions of data from an exponential family. 

\begin{result}\label{res1}
The distribution $\bP(Y,M)=\psi(M)h(Y)\exp\{\lambda(Y)^\textsc{t}\eta(M)\}$ is complete in $Y$ if (i) $\psi(M)>0$, (ii) the support of $\lambda(Y)$ is an open set, and (iii) the mapping $M \rightarrow \eta(M)$ is one to one.
\end{result}

For illustration, we present examples of parametric models below that satisfy the corresponding completeness assumption for each of Theorems \ref{the1} to \ref{the4}.

\subsection{An example for Theorem \ref{the1}}
\begin{proposition}\label{prop1}
For continuous $Y$, under a linear model $~Y\mid (M, T, X)\sim \mathcal{N}(\beta_0+\beta_m M+\beta_t T+\beta_{mt} M \cdot T+\beta_x X, \sigma^2)$ with $\beta_m\neq 0$ and $\beta_m+\beta_{mt}\neq 0$, the distribution
\begin{eqnarray*}
&&\bP(Y,M,R^M=1\mid T=t,X=x)\\&=&\bP(Y\mid M,T=t,X=x)\bP(M,R^M=1\mid T=t,X=x)\\&=&\frac{1}{(2\pi\sigma^2)^{1/2}}\exp \left\{-\frac{(Y-\beta_0-\beta_m M-\beta_t t-\beta_{mt} M\cdot t-\beta_x x)^2}{2\sigma^2}\right\}\bP(M,R^M=1\mid T=t,X=x)
\end{eqnarray*}
is complete in $Y$ for all $t$ and $x$.
\end{proposition}
Proposition \ref{prop1} follows from Result \ref{res1} with $\lambda(Y)=\sigma^{-2}(\beta_m+\beta_{mt}t)Y$ and $\eta(M)=M$.

\subsection{An example for Theorem \ref{the2}}
\begin{proposition}\label{prop2}
For continuous $Y$, under a linear model $~Y\mid (M, T, X)\sim \mathcal{N}(\beta_0+\beta_m M+\beta_t T+\beta_{mt} M \cdot T+\beta_x X, \sigma^2)$ with $\beta_m\neq 0$ and $\beta_m+\beta_{mt}\neq 0$, the distribution
\begin{eqnarray*}
&&\bP(Y,M,R^M=1,R^Y=1\mid T=t,X=x)\\&=&\bP(Y\mid M,T=t,X=x)\bP(M,R^M=1,R^Y=1\mid T=t,X=x)\\&=&\frac{1}{(2\pi\sigma^2)^{1/2}}\exp \left\{-\frac{(Y-\beta_0-\beta_m M-\beta_t t-\beta_{mt} M\cdot t-\beta_x x)^2}{2\sigma^2}\right\}\\
&& \cdot \bP(M,R^M=1,R^Y=1\mid T=t,X=x)
\end{eqnarray*}
is complete in $Y$ for all $t$ and $x$.
\end{proposition}
Proposition \ref{prop2} follows from Result \ref{res1} with $\lambda(Y)=\sigma^{-2}(\beta_m+\beta_{mt}t)Y$ and $\eta(M)=M$.

\subsection{An example for Theorem \ref{the3}}
\begin{proposition}\label{prop3}
For continuous $Y$ and continuous $M$, under a linear model $~Y\mid (M, T, X)\sim \mathcal{N}(\beta_0+\beta_m M+\beta_t T+\beta_{mt} M \cdot T+\beta_x X, \sigma^2)$ with $\beta_m\neq 0$ and $\beta_m+\beta_{mt}\neq 0$, the distribution
\begin{eqnarray*}
&&\bP(Y,M,R^M=1,R^Y=1\mid T=t,X=x)\\&=&\bP(Y\mid M,T=t,X=x)\bP(M,R^M=1\mid T=t,X=x)\bP(R^Y=1\mid Y,T=t,X=x)\\&=&\frac{1}{(2\pi\sigma^2)^{1/2}}\exp \left\{-\frac{(Y-\beta_0-\beta_m M-\beta_t t-\beta_{mt} M\cdot t-\beta_x x)^2}{2\sigma^2}\right\}\\
&& \cdot \bP(M,R^M=1\mid T=t,X=x)\bP(R^Y=1\mid Y,T=t,X=x)
\end{eqnarray*}
is complete in $Y$ and is complete in $M$ for all $t$ and $x$.
\end{proposition}
Proposition \ref{prop3} follows from Result \ref{res1} with $\lambda(Y)=\sigma^{-2}(\beta_m+\beta_{mt}t)Y$ and $\eta(M)=M$ and with $\lambda(M)=\sigma^{-2}(\beta_m+\beta_{mt}t)M$ and $\eta(Y)=Y$.

\subsection{Examples for Theorem \ref{the4}}

\begin{proposition}\label{prop4}
For continuous $Y$, under a linear model $~Y\mid (M, T, X)\sim \mathcal{N}(\beta_0+\beta_m M+\beta_t T+\beta_{mt} M \cdot T+\beta_x X, \sigma^2)$ with $\beta_m\neq 0$ and $\beta_m+\beta_{mt}\neq 0$, the distribution
\begin{eqnarray*}
&&\bP(Y,M,R^M=1,R^Y=1\mid T=t,X=x)\\&=&\bP(Y\mid M,T=t,X=x)\bP(M,R^M=1,R^Y=1\mid T=t,X=x)\\&=&\frac{1}{(2\pi\sigma^2)^{1/2}}\exp \left\{-\frac{(Y-\beta_0-\beta_m M-\beta_t t-\beta_{mt} M\cdot t-\beta_x x)^2}{2\sigma^2}\right\}\\
&& \cdot \bP(M,R^M=1,R^Y=1\mid T=t,X=x)
\end{eqnarray*}
is complete in $Y$ for all $t$ and $x$.
\end{proposition}
Proposition \ref{prop4} follows from Result \ref{res1} with $\lambda(Y)=\sigma^{-2}(\beta_m+\beta_{mt}t)Y$ and $\eta(M)=M$.

\begin{proposition}\label{prop5}
For binary $M$ and binary $R^Y$, under a logistic regression model $~\mathrm{logit}~\bP(R^Y=1\mid M,T,X)=\beta_0+\beta_m M+\beta_t T+\beta_{mt} M \cdot T+\beta_x X$ with $\beta_m\neq 0$ and $\beta_m+\beta_{mt}\neq 0$, the distribution
\begin{eqnarray*}
&&\bP(M,R^M=1,R^Y\mid T=t,X=x)\\&=&\bP(R^Y\mid M,T=t,X=x)\bP(M,R^M=1\mid T=t,X=x)\\&=&\frac{\exp\{R^Y(\beta_0+\beta_m M+\beta_t t+\beta_{mt} M\cdot t+\beta_x x)\}}{1+\exp(\beta_0+\beta_m M+\beta_t t+\beta_{mt} M\cdot t+\beta_x x)} \bP(M,R^M=1\mid T=t,X=x)
\end{eqnarray*}
is complete in $R^Y$ for all $t$ and $x$.
\end{proposition}
For binary $M$ and binary $R^Y$, the completeness condition reduces to $M \notindependent R^Y \mid (T=t,X=x)$ for all $t$ and $x$, and therefore, Proposition \ref{prop5} follows when $\beta_m\neq 0$ and $\beta_m+\beta_{mt}\neq 0$.

\section{The unidentifiable cases: counterexamples and conditions for identification}\label{sec::counterexamples}

Assuming $Y \independent R^M\mid (M,T,X)$, and allowing $M$ to have a impact on $R^M$, we have shown that the joint distribution $\bP(Y,M\mid T,  X )$ is identifiable under some completeness assumptions when $R^Y$ only depends on one of $(R^M,Y,M)$ given $T$ and $  X $. When $R^Y$ depends on more than one of $(R^M,Y,M)$ given $T$ and $X$ as in those missingness mechanisms described in Figure \ref{fig:unidentifiable missingness mechanisms}, the joint distribution $\bP(Y,M\mid T,X)$ is no longer identifiable without further assumptions. 

\begin{figure}[ht]
\centering
\begin{tikzpicture}
    \node (t)  at (0,0) {$T$};
    \node (x)  at (2,0) {$M$};
    \node (rm) at (2,2) {$R^M$};
    \node (y)  at (4,0) {$Y$};
    \node (ry) at (4,2) {$R^Y$};
    \node (c)  at (2,-1.5) {$(i)$ unidentifiable case};

    \path[-latex] (t) edge (x);
    \path[-latex] (t) edge (rm);
    \path[-latex] (t) edge (ry);
    \path[-latex] (x) edge (y);
    \path[-latex] (t) edge [bend right] (y);
    \path[-latex] (rm) edge (ry);
    \path[-latex] (x) edge (rm);
    \path[-latex] (y) edge (ry);

    \node (t)  at (6,0) {$T$};
    \node (x)  at (8,0) {$M$};
    \node (rm) at (8,2) {$R^M$};
    \node (y)  at (10,0) {$Y$};
    \node (ry) at (10,2) {$R^Y$};
    \node (c)  at (8,-1.5) {$(ii)$ unidentifiable case};

    \path[-latex] (t) edge (x);
    \path[-latex] (t) edge (rm);
    \path[-latex] (t) edge (ry);
    \path[-latex] (x) edge (y);
    \path[-latex] (t) edge [bend right] (y);
    \path[-latex] (rm) edge (ry);
    \path[-latex] (x) edge (rm);
    \path[-latex] (x) edge (ry);

    \node (t)  at (0,-5) {$T$};
    \node (x)  at (2,-5) {$M$};
    \node (rm) at (2,-3) {$R^M$};
    \node (y)  at (4,-5) {$Y$};
    \node (ry) at (4,-3) {$R^Y$};
    \node (c)  at (2,-6.5) {$(iii)$ unidentifiable case};

    \path[-latex] (t) edge (x);
    \path[-latex] (t) edge (rm);
    \path[-latex] (t) edge (ry);
    \path[-latex] (x) edge (y);
    \path[-latex] (t) edge [bend right] (y);
    \path[-latex] (y) edge (ry);
    \path[-latex] (x) edge (rm);
    \path[-latex] (x) edge (ry);

    \node (t)  at (6,-5) {$T$};
    \node (x)  at (8,-5) {$M$};
    \node (rm) at (8,-3) {$R^M$};
    \node (y)  at (10,-5) {$Y$};
    \node (ry) at (10,-3) {$R^Y$};
    \node (c)  at (8,-6.5) {$(iv)$ unidentifiable case};

    \path[-latex] (t) edge (x);
    \path[-latex] (t) edge (rm);
    \path[-latex] (t) edge (ry);
    \path[-latex] (x) edge (y);
    \path[-latex] (t) edge [bend right] (y);
    \path[-latex] (y) edge (ry);
    \path[-latex] (x) edge (rm);
    \path[-latex] (rm) edge (ry);
    \path[-latex] (x) edge (ry);
    
\end{tikzpicture}
\caption{DAGs describing the unidentifiable missingness mechanisms when missingness exists in both the mediator and outcome (all DAGs condition on $X$ and allow $X$ to have directed arrows to all variables in the DAGs).}
\label{fig:unidentifiable missingness mechanisms}
\end{figure}

As discussed in section \ref{sec:notation}, the identification of the NIE and NDE relies on the identification of the joint distribution $P(Y,M\mid T,X)$. Below, we first show that this joint distribution of $Y$ and $M$ cannot always be uniquely determined by the observable data probabilities without further assumptions if missingness of $Y$ depends on more than one of $(R^M, Y, M)$ or the completeness assumption is violated. We explain the reasons and provide concrete examples in subsection \ref{subsec::counterexamples}. We then show that the identification is plausible by exploiting the information on a future outcome under the complex MNAR mechanism where missingness of $Y$ depends on more than one of $(R^M, Y, M)$ in subsection \ref{subsec::identifiability}. To simplify the notation, all DAGs and probabilities below are conditioning on $T$ and $X$.

\subsection{Counterexamples}\label{subsec::counterexamples}
Define 
\begin{eqnarray*}
    \bP_{my11} &=& \bP(M=m, Y=y, R^M=1, R^Y=1), \\
    \bP_{+y01} &=& \bP(Y=y, R^M=0, R^Y=1),\\
    \bP_{m+10} &=& \bP(M=m, R^M=1, R^Y=0), \\
    \bP_{++00} &=& \bP(R^M=0, R^Y=0).
\end{eqnarray*}

In $(i)$ to $(iv)$, we present examples where the identification cannot be achieved without further assumptions if missingness of $Y$ depends on more than one of $(R^M, Y, M)$ in a simple setup of a binary
mediator $M$ and a binary outcome $Y$. Based on the observable data probabilities, we can directly identify $\bP_{my11}$, $\bP_{+y01}$, $\bP_{m+10}$ and $\bP_{++00}$, and $\sum^{1}_{m=0}\sum^{1}_{y=0}\bP_{my11}+\sum_{y=0}^1 \bP_{+y01}+\sum_{m=0}^1 \bP_{m+10}+ \bP_{++00}=1$. In $(v)$, we present an unidentifiable case when $M$ has more categories than $Y$ under MNAR Assumptions \ref{ass1} to \ref{ass4} . In $(vi)$, we present an unidentifiable case when $Y$ has more categories than $M$ under MNAR Assumption \ref{ass3}. 

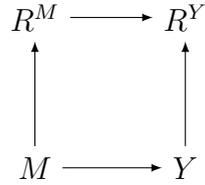
\begin{figure}[H]
\centering
\begin{tikzpicture}
    \node (x)  at (12,0) {$M$};
    \node (rm) at (12,2) {$R^M$};
    \node (y)  at (14,0) {$Y$};
    \node (ry) at (14,2) {$R^Y$};
    
    \path[-latex] (x) edge (y);
    \path[-latex] (x) edge (rm);
    \path[-latex] (y) edge (ry);
    \path[-latex] (rm) edge (ry);
\end{tikzpicture}
\caption{$R^Y$ depends on both $Y$ and $R^M$}
\label{suppfig1}
\end{figure}

\subsubsection*{$(i)$ We present below an unidentified case when $R^Y$ depends on both $Y$ and $R^M$ as described by Figure \ref{suppfig1}}

Consider the following observable data probabilities:
\begin{align}
(\bP_{1111},\bP_{0111},\bP_{1011},\bP_{0011},\bP_{+101},\bP_{+001},\bP_{1+10},\bP_{0+10},\bP_{++00})
=\left(\frac{6}{20},\frac{2}{20},\frac{1}{20},\frac{1}{20},\frac{2}{20},\frac{1}{20},\frac{2}{20},\frac{1}{20},\frac{4}{20} \right)\nonumber.
\end{align}
The key to identify $\bP(Y=y, M=m)$ is to identify both $\bP(R^M=1\mid M=m)$ and $\bP(R^Y=1\mid Y=y,R^M=1)$ in the following formula
\begin{align}
\bP(Y=y, M=m)=\frac{\bP_{my11}}{\bP(R^M=1\mid M=m)\bP(R^Y=1\mid Y=y,R^M=1)}.\nonumber
\end{align}

We now show that the identification of $\bP(R^Y=1\mid Y=y,R^M=1)$ can be achieved. This is because 
\begin{align}
\bP_{m+10} &= \sum_{y\in \mathcal{Y}}\bP(M=m, Y=y, R^M=1, R^Y=0)\nonumber\\&=\sum_{y\in \mathcal{Y}}\bP_{my11}\frac{\bP(R^Y=0\mid Y=y,R^M=1)}{\bP(R^Y=1\mid Y=y,R^M=1)}.\nonumber
\end{align}
By plugging in the two possible values for $m$ and $y$ in the above formula, we have
\begin{align}
\bP_{1+10} &= \bP_{1111}\frac{\bP(R^Y=0\mid Y=1,R^M=1)}{\bP(R^Y=1\mid Y=1,R^M=1)}+\bP_{1011}\frac{\bP(R^Y=0\mid Y=0,R^M=1)}{\bP(R^Y=1\mid Y=0,R^M=1)},\label{eqS1}
\end{align}
\begin{align}
\bP_{0+10} &= \bP_{0111}\frac{\bP(R^Y=0\mid Y=1,R^M=1)}{\bP(R^Y=1\mid Y=1,R^M=1)}+\bP_{0011}\frac{\bP(R^Y=0\mid Y=0,R^M=1)}{\bP(R^Y=1\mid Y=0,R^M=1)}.\label{eqS2}
\end{align}
Therefore, $\bP(R^Y=1\mid Y=y,R^M=1)$ can be identified by solving the linear equations (\ref{eqS1}) and (\ref{eqS2}). Based on the observable data probabilities, $\bP(R^Y=1\mid Y=1,R^M=1)=\frac{4}{5}$ and $\bP(R^Y=1\mid Y=0,R^M=1)=\frac{2}{3}$.

We now focus on the identifiability of $\bP(R^M=1\mid M=m)$ and show that $\bP(R^M=1\mid M=m)$ cannot be identified without further assumptions. We have 
\begin{align}
\bP_{+y01} &= \sum_{m\in \mathcal{M}} \bP(M=m, Y=y, R^M=0, R^Y=1)\nonumber\\&=\sum_{m\in \mathcal{M}}\bP_{my11}\frac{\bP(R^Y=1\mid Y=y,R^M=0)\bP(R^M=0\mid M=m)}{\bP(R^Y=1\mid Y=y,R^M=1)\bP(R^M=1\mid M=m)}\nonumber\\&= \frac{\bP(R^Y=1\mid Y=y,R^M=0)}{\bP(R^Y=1\mid Y=y,R^M=1)} \sum_{m\in \mathcal{M}}\bP_{my11}\frac{\bP(R^M=0\mid M=m)}{\bP(R^M=1\mid M=m)},\nonumber
\end{align}
and as a result,
\begin{align}
\bP_{+y01}\frac{\bP(R^Y=1\mid Y=y,R^M=1)}{\bP(R^Y=1\mid Y=y,R^M=0)} &= \sum_{m\in \mathcal{M}}\bP_{my11}\frac{\bP(R^M=0\mid M=m)}{\bP(R^M=1\mid M=m)}.\nonumber
\end{align}
By plugging in the two possible values for $m$ and $y$ in the above formula, we have 
\begin{align}
\bP_{+101}\frac{\bP(R^Y=1\mid Y=1,R^M=1)}{\bP(R^Y=1\mid Y=1,R^M=0)}&= \bP_{1111}\frac{\bP(R^M=0\mid M=1)}{\bP(R^M=1\mid M=1)}\nonumber+\bP_{0111}\frac{\bP(R^M=0\mid M=0)}{\bP(R^M=1\mid M=0)},\nonumber
\end{align}
\begin{align}
\bP_{+001}\frac{\bP(R^Y=1\mid Y=0,R^M=1)}{\bP(R^Y=1\mid Y=0,R^M=0)}&= \bP_{1011}\frac{\bP(R^M=0\mid M=1)}{\bP(R^M=1\mid M=1)}\nonumber+\bP_{0011}\frac{\bP(R^M=0\mid M=0)}{\bP(R^M=1\mid M=0)}.\nonumber
\end{align}

Since $\bP(R^Y=1\mid Y=y,R^M=1)$ are identified from the previous step, the identifiability of $\bP(R^M=1 \mid M=m)$ depends on the identifiability of $\bP(R^Y=1\mid Y=y,R^M=0)$. We have
\begin{eqnarray}
&&\bP(R^Y=1\mid Y=y,R^M=0)\nonumber \\&=& \frac{\bP(Y=y,R^M=0,R^Y=1)}{\bP(Y=y,R^M=0)}\nonumber\\&=& \frac{\bP(Y=y,R^M=0,R^Y=1)}{\bP(Y=y,R^M=0,R^Y=0)+\bP(Y=y,R^M=0,R^Y=1)}\nonumber\\&=&\frac{\bP(Y=y,R^M=0,R^Y=1)}{\bP(Y=y \mid R^M=0, R^Y=0)\bP(R^M=0,R^Y=0)+\bP(Y=y, R^M=0,R^Y=1)}\nonumber\\&=&\frac{\bP_{+y01}}{\bP(Y=y \mid R^M=0, R^Y=0)\bP_{++00}+\bP_{+y01}}.\nonumber
\end{eqnarray}

In the above expression, $\bP_{+y01}$ and $\bP_{++00}$ are known, but $\bP(Y=y \mid R^Y=0,R^M=0)$ is not observable or identifiable. Different values of $\bP(Y=y \mid R^Y=0,R^M=0)$ will result in different values of $\bP(R^M=1 \mid M=m)$, which in turn will give different values of $\bP(Y=y, M=m)$. For example, let $\bP(Y=1 \mid R^Y=0,R^M=0)=\frac{5}{6}$,  and the corresponding $\bP(R^Y=1\mid Y=1,R^M=0)$ and $\bP(R^Y=1\mid Y=0,R^M=0)$ equal $\frac{3}{8}$ and $\frac{3}{5}$, respectively. As a result, we have $\bP(R^M=1 \mid M=1)=\frac{45}{68}$ and $\bP(R^M=1 \mid M=0)=\frac{5}{8}$. Subsequently, we have $\bP(Y=1,M=1)=\frac{17}{30}$, $\bP(Y=0,M=1)=\frac{17}{150}$,  $\bP(Y=1,M=0)=\frac{1}{5}$ and $\bP(Y=0,M=0)=\frac{3}{25}$. Alternatively, let $\bP(Y=1 \mid R^Y=0,R^M=0)=\frac{7}{8}$, and the corresponding $\bP(R^Y=1\mid Y=1,R^M=0)$ and $\bP(R^Y=1\mid Y=0,R^M=0)$ equal $\frac{4}{11}$ and $\frac{2}{3}$, respectively. As a result, we have $\bP(R^M=1 \mid M=1)=\frac{5}{8}$ and $\bP(R^M=1 \mid M=0)=\frac{5}{7}$. Subsequently, we have $\bP(Y=1,M=1)=\frac{3}{5}$, $\bP(Y=0,M=1)=\frac{3}{25}$,  $\bP(Y=1,M=0)=\frac{7}{40}$ and $\bP(Y=0,M=0)=\frac{21}{200}$. 

The two sets of values of $\bP(Y=y, M=m)$ correspond to the same observable data probabilities: $(\bP_{1111},\bP_{0111},\bP_{1011},\bP_{0011},\bP_{+101},\bP_{+001},\bP_{1+10},\bP_{0+10},\bP_{++00})%=\left(\frac{3}{10},\frac{1}{10},\frac{1}{20},\frac{1}{20},\frac{1}{10},\frac{1}{20},\frac{1}{10},\frac{1}{20},\frac{1}{5}\right),
$, and therefore, $\bP(Y=y, M=m)$ can not be uniquely identified without further assumptions.

This unidentifiable result is not contradictory to the conclusion in \citetSupp{li2022selfsupp} discussing the identifiability of self-censoring model under assumptions imposed by chain graphs instead of DAGs. 

\begin{figure}[H]
\centering
\begin{tikzpicture}
    \node (x)  at (12,0) {$M$};
    \node (rm) at (12,2) {$R^M$};
    \node (y)  at (14,0) {$Y$};
    \node (ry) at (14,2) {$R^Y$};
    
    \path[-latex] (x) edge (y);
    \path[-latex] (x) edge (rm);
    \path[-latex] (x) edge (ry);
    \path[-latex] (rm) edge (ry);
\end{tikzpicture}
\caption{$R^Y$ depends on both $M$ and $R^M$}
\label{suppfig2}
\end{figure}
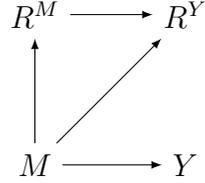

\subsubsection*{$(ii)$ We present below an unidentified case when $R^Y$ depends on both $M$ and $R^M$ as described by Figure \ref{suppfig2}}

Consider the following observable data probabilities:
\begin{align}
(\bP_{1111},\bP_{0111},\bP_{1011},\bP_{0011},\bP_{+101},\bP_{+001},\bP_{1+10},\bP_{0+10},\bP_{++00})=\left(\frac{12}{40},\frac{4}{40},\frac{2}{40},\frac{4}{40},\frac{4}{40},\frac{1}{40},\frac{4}{40},\frac{4}{40},\frac{5}{40}\right)\nonumber.
\end{align}
Define 
\begingroup
\allowdisplaybreaks
\begin{align*}
&\bP_M=\bP(M=1),\\
&\bP_{Y1}=\bP(Y=1\mid M=1),\\
&\bP_{Y0}=\bP(Y=1\mid M=0),\\
&\bP_{R^M1}=\bP(R^M=1\mid M=1),\\
&\bP_{R^M0}=\bP(R^M=1\mid M=0),\\
&\bP_{R^Y00}=\bP(R^Y=1\mid M=0,R^M=0),\\
&\bP_{R^Y01}=\bP(R^Y=1\mid M=0,R^M=1),\\
&\bP_{R^Y10}=\bP(R^Y=1\mid M=1,R^M=0),\\
&\bP_{R^Y11}=\bP(R^Y=1\mid M=1,R^M=1).
\end{align*}
\endgroup
Below we study the identifiablity of the above $9$ parameters describing the graphical model in Figure \ref{suppfig2} based on the observable data probabilities. Although there are $9$ observable data probabilities, the degree of freedom in the probabilities is only $8$ given that they sum up to $1$.

The following relationships between the observable data probabilities and the parameters hold,
\begingroup
\allowdisplaybreaks
\begin{align}
&\bP_{1111}=\bP_M\bP_{Y1}\bP_{R^M1}\bP_{R^Y11}, \label{eqS3}\\
&\bP_{1011}=\bP_M(1-\bP_{Y1})\bP_{R^M1}\bP_{R^Y11},\label{eqS4}\\
&\bP_{0111}=(1-\bP_M)\bP_{Y0}\bP_{R^M0}\bP_{R^Y01},\label{eqS5}\\
&\bP_{0011}=(1-\bP_M)(1-\bP_{Y0})\bP_{R^M0}\bP_{R^Y01},\label{eqS6}\\
&\bP_{1+10}=\bP_M\bP_{R^M1}(1-\bP_{R^Y11}),\label{eqS7}\\
&\bP_{0+10}=(1-\bP_M)\bP_{R^M0}(1-\bP_{R^Y01}),\label{eqS8}\\
&\bP_{+101}=\bP_M\bP_{Y1}(1-\bP_{R^M1})\bP_{R^Y10}+(1-\bP_M)\bP_{Y0}(1-\bP_{R^M0})\bP_{R^Y00},\label{eqS9}\\
&\bP_{+001}=\bP_M(1-\bP_{Y1})(1-\bP_{R^M1})\bP_{R^Y10}+(1-\bP_M)(1-\bP_{Y0})(1-\bP_{R^M0})\bP_{R^Y00}.\label{eqS10}
\end{align}
\endgroup

By solving the equations (\ref{eqS3}) to (\ref{eqS8}), we can identify the parameters $\bP_{Y1}$, $\bP_{Y0}$, $\bP_{R^Y11}$ and $\bP_{R^Y01}$:
\begin{align*}
&\bP_{Y1}=\frac{\bP_{1111}}{\bP_{1111}+\bP_{1011}},\\
&\bP_{Y0}=\frac{\bP_{0111}}{\bP_{0111}+\bP_{0011}},\\
&\bP_{R^Y11}=\frac{\bP_{1111}+\bP_{1011}}{\bP_{1111}+\bP_{1011}+\bP_{1+10}},\\
&\bP_{R^Y01}=\frac{\bP_{0111}+\bP_{0011}}{\bP_{0111}+\bP_{0011}+\bP_{0+10}}.
\end{align*}
Based on the observable data probabilities, $\bP_{Y1}=\frac{6}{7}$, $\bP_{Y0}=\frac{1}{2}$, $\bP_{R^Y11} = \frac{7}{9}$ and $\bP_{R^Y01} = \frac{2}{3}$. In addition, we can identify the following products of parameters based on equations (\ref{eqS7}) to (\ref{eqS10}):
$\bP_M\bP_{R^M1}$, $(1-\bP_M)\bP_{R^M0}$, $\bP_M(1-\bP_{R^M1})\bP_{R^Y10}$ and $(1-\bP_M)(1-\bP_{R^M0})\bP_{R^Y00}$. As a result, when $\bP_M$ is known, one can solve for $\bP_{R^M1}$, $\bP_{R^M0}$, $\bP_{R^Y10}$ and $\bP_{R^Y00}$.

For example, let $\bP_M=\frac{3}{5}$, we have $\bP_{R^M1}=\frac{3}{4}$, $\bP_{R^M0}=\frac{3}{4}$, $\bP_{R^Y10}=\frac{7}{10}$ and $\bP_{R^Y00}=\frac{1}{5}$. This set of parameter values give us the following joint probabilities of $M$ and $Y$ as $\bP(Y=1,M=1) =\frac{18}{35} $, $\bP(Y=0,M=1) = \frac{3}{35}$, $\bP(Y=1,M=0) = \frac{1}{5}$ and $\bP(Y=0,M=0) = \frac{1}{5}$. Alternatively, let $\bP_M=\frac{13}{20}$, we have $\bP_{R^M1}=\frac{9}{13}$, $\bP_{R^M0}=\frac{6}{7}$, $\bP_{R^Y10}=\frac{21}{40}$ and $\bP_{R^Y00}=\frac{2}{5}$. This alternative set of parameter values give us the following joint probabilities of $M$ and $Y$ as $\bP(Y=1,M=1) = \frac{39}{70}$, $\bP(Y=0,M=1) =\frac{13}{140} $, $\bP(Y=1,M=0) = \frac{7}{40}$ and $\bP(Y=0,M=0) = \frac{7}{40}$.

The two sets of values of $\bP(Y=y, M=m)$ correspond to the same observable data probabilities:
$
(\bP_{1111},\bP_{0111},\bP_{1011},\bP_{0011},\bP_{+101},\bP_{+001},\bP_{1+10},\bP_{0+10},\bP_{++00})$,
%=\left(\frac{3}{10},\frac{1}{10},\frac{1}{20},\frac{1}{10},\frac{1}{10},\frac{1}{40},\frac{1}{10},\frac{1}{10},\frac{5}{40}\right)\nonumber.
and therefore, $\bP(Y=y, M=m)$ can not be uniquely identified without further assumptions.

\begin{figure}[H]
\centering
\begin{tikzpicture}
    \node (x)  at (12,0) {$M$};
    \node (rm) at (12,2) {$R^M$};
    \node (y)  at (14,0) {$Y$};
    \node (ry) at (14,2) {$R^Y$};
    
    \path[-latex] (x) edge (y);
    \path[-latex] (x) edge (rm);
    \path[-latex] (x) edge (ry);
    \path[-latex] (y) edge (ry);
\end{tikzpicture}
\caption{$R^Y$ depends on both $Y$ and $M$}
\label{suppfig3}
\end{figure}
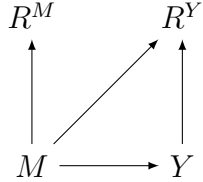

\subsubsection*{$(iii)$ We present below an unidentified case when $R^Y$ depends on both $Y$ and $M$ as described by Figure \ref{suppfig3}}

Consider the following probabilities from the observable data:
\begin{align}
(\bP_{1111},\bP_{0111},\bP_{1011},\bP_{0011},\bP_{+101},\bP_{+001},\bP_{1+10},\bP_{0+10},\bP_{++00})=\left(\frac{18}{96},\frac{6}{96},\frac{3}{96},\frac{2}{96},\frac{12}{96},\frac{3}{96},\frac{15}{96},\frac{16}{96},\frac{21}{96}\right)\nonumber.
\end{align}
Define 
\begin{align*}
&\bP_M=\bP(M=1),\\
&\bP_{Y1}=\bP(Y=1\mid M=1),\\
&\bP_{Y0}=\bP(Y=1\mid M=0),\\
&\bP_{R^M1}=\bP(R^M=1\mid M=1),\\
&\bP_{R^M0}=\bP(R^M=1\mid M=0),\\
&\bP_{R^Y00}=\bP(R^Y=1\mid M=0,Y=0),\\
&\bP_{R^Y01}=\bP(R^Y=1\mid M=0,Y=1),\\
&\bP_{R^Y10}=\bP(R^Y=1\mid M=1,Y=0),\\
&\bP_{R^Y11}=\bP(R^Y=1\mid M=1,Y=1).
\end{align*}
Below we study the identifiablity of the above $9$ parameters describing the graphical model in Figure \ref{suppfig3} based on the observable data probabilities. Although there are $9$ observable data probabilities, the degree of freedom in the probabilities is only $8$ given that they sum up to $1$.

The following relationships between the observable data probabilities and the parameters hold,
\begingroup
\allowdisplaybreaks
\begin{align}
&\bP_{1111}=\bP_M\bP_{Y1}\bP_{R^M1}\bP_{R^Y11},\label{eqS11}\\
&\bP_{1011}=\bP_M(1-\bP_{Y1})\bP_{R^M1}\bP_{R^Y10},\label{eqS12}\\
&\bP_{0111}=(1-\bP_M)\bP_{Y0}\bP_{R^M0}\bP_{R^Y01},\label{eqS13}\\
&\bP_{0011}=(1-\bP_M)(1-\bP_{Y0})\bP_{R^M0}\bP_{R^Y00},\label{eqS14}\\
&\bP_{1+10}=\bP_M\bP_{Y1}\bP_{R^M1}(1-\bP_{R^Y11})+\bP_M(1-\bP_{Y1})\bP_{R^M1}(1-\bP_{R^Y10}),\label{eqS15}\\
&\bP_{0+10}=(1-\bP_M)\bP_{Y0}\bP_{R^M0}(1-\bP_{R^Y01})+(1-\bP_M)(1-\bP_{Y0})\bP_{R^M0}(1-\bP_{R^Y00}),\label{eqS16}\\
&\bP_{+101}=\bP_M\bP_{Y1}(1-\bP_{R^M1})\bP_{R^Y11}+(1-\bP_M)\bP_{Y0}(1-\bP_{R^M0})\bP_{R^Y01},\label{eqS17}\\
&\bP_{+001}=\bP_M(1-\bP_{Y1})(1-\bP_{R^M1})\bP_{R^Y10}+(1-\bP_M)(1-\bP_{Y0})(1-\bP_{R^M0})\bP_{R^Y00}.\label{eqS18}
\end{align}
\endgroup

By solving the equations (\ref{eqS11}) to (\ref{eqS18}), we can identify the parameters $\bP_{M}$, $\bP_{R^M1}$ and $\bP_{R^M0}$. Given the observable data probabilities, $\bP_{M} = \frac{1}{2}$, $\bP_{R^M1} = \frac{3}{4}$ and $\bP_{R^M0} =\frac{1}{2} $.  However, $\bP_{Y1}$, $\bP_{Y0}$, $\bP_{R^Y11}$, $\bP_{R^Y10}$, $\bP_{R^Y01}$ and $\bP_{R^Y00}$ are not identifiable. For example, we can have $\bP_{Y1}=\frac{3}{4}$, $\bP_{Y0}=\frac{1}{2}$, $\bP_{R^Y11}=\frac{2}{3}$, $\bP_{R^Y10}=\frac{1}{3}$, $\bP_{R^Y01}=\frac{1}{2}$ and $\bP_{R^Y00}=\frac{1}{6}$, which in turn give us $\bP(Y=1, M=1)=\frac{3}{8}$, $\bP(Y=1, M=0)=\frac{1}{4}$, $\bP(Y=0, M=1)=\frac{1}{8}$ and $\bP(Y=0, M=0)=\frac{1}{4}$. Alternatively, we can have $\bP_{Y1}=\frac{2}{3}$, $\bP_{Y0}=\frac{3}{4}$, $\bP_{R^Y11}=\frac{3}{4}$, $\bP_{R^Y10}=\frac{1}{4}$, $\bP_{R^Y01}=\frac{1}{3}$ and $\bP_{R^Y00}=\frac{1}{3}$, which in turn give us $\bP(Y=1, M=1)=\frac{1}{3}$, $\bP(Y=1, M=0)=\frac{3}{8}$, $\bP(Y=0, M=1)=\frac{1}{6}$ and $\bP(Y=0, M=0)=\frac{1}{8}$.

The two sets of values of $\bP(Y=y, M=m)$ correspond to the same observable data probabilities: $(\bP_{1111},\bP_{0111},\bP_{1011},\bP_{0011},\bP_{+101},\bP_{+001},\bP_{1+10},\bP_{0+10},\bP_{++00})$, and therefore, $\bP(Y=y, M=m)$ can not be uniquely identified without further assumptions.

\begin{figure}[H]
\centering
\begin{tikzpicture}
    \node (x)  at (12,0) {$M$};
    \node (rm) at (12,2) {$R^M$};
    \node (y)  at (14,0) {$Y$};
    \node (ry) at (14,2) {$R^Y$};
    
    \path[-latex] (x) edge (y);
    \path[-latex] (x) edge (rm);
    \path[-latex] (x) edge (ry);
    \path[-latex] (y) edge (ry);
    \path[-latex] (rm) edge (ry);
\end{tikzpicture}
\caption{$R^Y$ depends on $Y$, $M$ and $R^M$}
\label{suppfig4}
\end{figure}
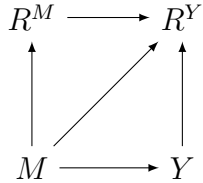

\subsubsection*{$(iv)$ We present below an unidentified case when $R^Y$ depends on $Y$, $M$ and $R^M$ as described by Figure \ref{suppfig4}}

The counterexamples presented in $(i)$ to $(iii)$ can all be viewed as special cases of the missingness mechanism described by Figure \ref{suppfig4}.

\begin{figure}[H]
\centering
\begin{tikzpicture}
    \node (x)  at (0,0) {$M$};
    \node (rm) at (0,2) {$R^M$};
    \node (y)  at (2,0) {$Y$};
    \node (a)  at (1,-1) {$(a)$ Assumption \ref{ass1}};
    \path[-latex] (x) edge (y);
    \path[-latex] (x) edge (rm);

    \node (x)  at (4,0) {$M$};
    \node (rm) at (4,2) {$R^M$};
    \node (y)  at (6,0) {$Y$};
    \node (ry) at (6,2) {$R^Y$};
    \node (b)  at (5,-1) {$(b)$ Assumption \ref{ass2}};
    \path[-latex] (x) edge (y);
    \path[-latex] (x) edge (rm);
    \path[-latex] (rm) edge (ry);

    \node (x)  at (8,0) {$M$};
    \node (rm) at (8,2) {$R^M$};
    \node (y)  at (10,0) {$Y$};
    \node (ry) at (10,2) {$R^Y$};
    \node (b)  at (9,-1) {$(c)$ Assumption \ref{ass3}};
    \path[-latex] (x) edge (y);
    \path[-latex] (x) edge (rm);
    \path[-latex] (y) edge (ry);

    \node (x)  at (12,0) {$M$};
    \node (rm) at (12,2) {$R^M$};
    \node (y)  at (14,0) {$Y$};
    \node (ry) at (14,2) {$R^Y$};
    \node (b)  at (13,-1) {$(d)$ Assumption \ref{ass4}};
    
    \path[-latex] (x) edge (y);
    \path[-latex] (x) edge (rm);
    \path[-latex] (x) edge (ry);
    
\end{tikzpicture}
\caption{$M$ has more categories than $Y$ under MNAR Assumptions \ref{ass1} to \ref{ass4}}
\label{suppfig5}
\end{figure}
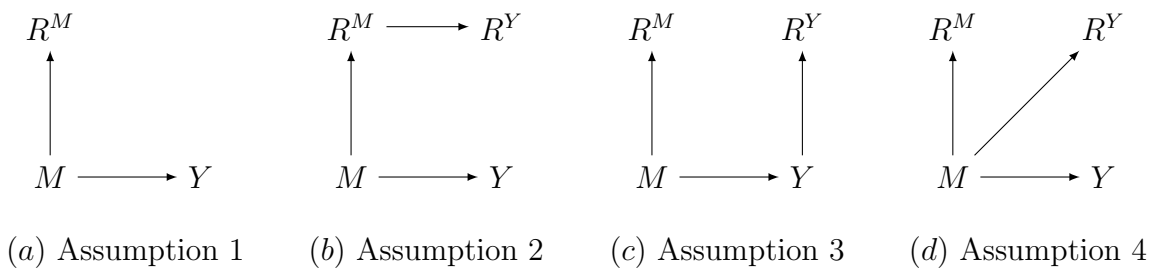

\subsubsection*{$(v)$ We present below an unidentified case when $M$ has more categories than $Y$ under MNAR Assumptions \ref{ass1} to \ref{ass4} as described by Figure \ref{suppfig5}}

As an illustration, we present a counterexample for the missingness mechanism under MNAR Assumption \ref{ass1} $(a)$. The counterexample for $(a)$ can be viewed as a special case for the missingness mechanism under MNAR Assumptions \ref{ass2}, \ref{ass3}, and \ref{ass4} with $\bP(R^Y=1)=1$. 

For $(d)$, when $M$ has more categories than $Y$, we can still achieve identification if the rank condition holds as illustrated in Theorem \ref{the4}. We provide a simulation result in section \ref{sec::suppsimulation} showing that the identifiability of model parameters is improved under MNAR Assumption \ref{ass4} compare to MNAR Assumption \ref{ass1} when $M$ has more categories than $Y$.

Define 
\begin{eqnarray*}
    \bP_{my1} &=& \bP(M=m, Y=y, R^M=1), \\
    \bP_{+y0} &=& \bP(Y=y, R^M=0).
\end{eqnarray*}

Consider a binary outcome $Y$ and a categorical $M$ with three categories, denoted as $0$, $1$, and $2$, respectively. Consider the following probabilities from the observable data:
\begin{eqnarray*}(\bP_{211},\bP_{111},\bP_{011},\bP_{201},\bP_{101},\bP_{001},\bP_{+10},\bP_{+00})=\left(\frac{4}{96},\frac{4}{96},\frac{6}{96},\frac{8}{96},\frac{8}{96},\frac{6}{96},\frac{22}{96},\frac{38}{96}\right)\nonumber.
\end{eqnarray*}
Define 
\begin{align*}
&\bP_{M^2}=\bP(M=2),\\
&\bP_{M^1}=\bP(M=1),\\
&\bP_{Y2}=\bP(Y=1\mid M=2),\\
&\bP_{Y1}=\bP(Y=1\mid M=1),\\
&\bP_{Y0}=\bP(Y=1\mid M=0),\\
&\bP_{R^M2}=\bP(R^M=1\mid M=2),\\
&\bP_{R^M1}=\bP(R^M=1\mid M=1),\\
&\bP_{R^M0}=\bP(R^M=1\mid M=0).
\end{align*}

The following relationships between the observable data probabilities and the parameters hold,
\begingroup
\allowdisplaybreaks
\begin{align}
&\bP_{211}=\bP_{M^2}\bP_{Y2}\bP_{R^M2},\label{eqS19}\\
&\bP_{111}=\bP_{M^1}\bP_{Y1}\bP_{R^M1},\label{eqS20}\\
&\bP_{011}=(1-\bP_{M^2}-\bP_{M^1})\bP_{Y0}\bP_{R^M0},\label{eqS21}\\
&\bP_{201}=\bP_{M^2}(1-\bP_{Y2})\bP_{R^M2},\label{eqS22}\\
&\bP_{101}=\bP_{M^1}(1-\bP_{Y1})\bP_{R^M1},\label{eqS23}\\
&\bP_{001}=(1-\bP_{M^2}-\bP_{M^1})(1-\bP_{Y0})\bP_{R^M0},\label{eqS24}\\
&\bP_{+10}=\bP_{M^2}\bP_{Y2}(1-\bP_{R^M2})+\bP_{M^1}\bP_{Y1}(1-\bP_{R^M1})\nonumber\\&\hspace{1.5cm}+(1-\bP_{M^2}-\bP_{M^1})\bP_{Y0}(1-\bP_{R^M0}),\label{eqS25}\\
&\bP_{+00}=\bP_{M^2}(1-\bP_{Y2})(1-\bP_{R^M2})+\bP_{M^1}(1-\bP_{Y1})(1-\bP_{R^M1})\nonumber\\&\hspace{1.5cm}+(1-\bP_{M^2}-\bP_{M^1})(1-\bP_{Y0})(1-\bP_{R^M0}).\label{eqS26}
\end{align}
\endgroup

By solving the equations (\ref{eqS19}) to (\ref{eqS26}), we can identify the parameters $\bP_{Y2}$, $\bP_{Y1}$ and $\bP_{Y0}$. Given the observable data probabilities, $\bP_{Y2}=\frac{1}{3}$, $\bP_{Y1}=\frac{1}{3}$ and $\bP_{Y0}=\frac{1}{2}$. However, $\bP_{M^2}$, $\bP_{M^1}$, $\bP_{R^M2}$, $\bP_{R^M1}$ and $\bP_{R^M0}$ are not identifiable. For example, we can have $\bP_{M^2}=\frac{1}{4}$, $\bP_{M^1}=\frac{1}{2}$, $\bP_{R^M2}=\frac{1}{2}$, $\bP_{R^M1}=\frac{1}{4}$ and $\bP_{R^M0}=\frac{1}{2}$, which in turn give us $\bP(Y=1, M=2)=\frac{1}{12}$, $\bP(Y=1, M=1)=\frac{1}{6}$, $\bP(Y=1, M=0)=\frac{1}{8}$, $\bP(Y=0, M=2)=\frac{1}{6}$, $\bP(Y=0, M=1)=\frac{1}{3}$ and $\bP(Y=0, M=0)=\frac{1}{8}$. Alternatively, we can have $\bP_{M^2}=\frac{3}{8}$, $\bP_{M^1}=\frac{3}{8}$, $\bP_{R^M2}=\frac{1}{3}$, $\bP_{R^M1}=\frac{1}{3}$ and $\bP_{R^M0}=\frac{1}{2}$, which in turn give us $\bP(Y=1, M=2)=\frac{1}{8}$, $\bP(Y=1, M=1)=\frac{1}{8}$, $\bP(Y=1, M=0)=\frac{1}{8}$, $\bP(Y=0, M=2)=\frac{1}{4}$, $\bP(Y=0, M=1)=\frac{1}{4}$ and $\bP(Y=0, M=0)=\frac{1}{8}$.

The two sets of values of $\bP(Y=y, M=m)$ correspond to the same observable data probabilities: $(\bP_{211},\bP_{111},\bP_{011},\bP_{201},\bP_{101},\bP_{001},\bP_{+10},\bP_{+00})$, and therefore, $\bP(Y=y, M=m)$ can not be uniquely identified without further assumptions.

\begin{figure}[H]
\centering
\begin{tikzpicture}
    \node (x)  at (12,0) {$M$};
    \node (rm) at (12,2) {$R^M$};
    \node (y)  at (14,0) {$Y$};
    \node (ry) at (14,2) {$R^Y$};
    
    \path[-latex] (x) edge (y);
    \path[-latex] (x) edge (rm);
    \path[-latex] (y) edge (ry);
\end{tikzpicture}
\caption{$Y$ has more categories than $M$ under MNAR Assumption \ref{ass3}}
\label{suppfig7}
\end{figure}
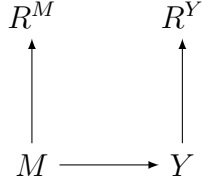

\subsubsection*{$(vi)$ We present below an unidentified case when $Y$ has more categories than $M$ under MNAR Assumption \ref{ass3} as described by Figure \ref{suppfig7}}

Consider a binary $M$ and an outcome $Y$ with three categories, denoted as $0$, $1$, and $2$, respectively. Consider the following probabilities from the observable data:
\begin{eqnarray*}
&&(\bP_{1211},\bP_{1111},\bP_{1011},\bP_{0211},\bP_{0111},\bP_{0011},\bP_{1+10},\bP_{0+10},\bP_{+201},\bP_{+101},\bP_{+001},\bP_{++00})\nonumber\\&=&\left(\frac{180}{1440},\frac{60}{1440},\frac{15}{1440},\frac{120}{1440},\frac{30}{1440},\frac{15}{1440},\frac{285}{1440},\frac{195}{1440},\frac{180}{1440},\frac{50}{1440},\frac{20}{1440},\frac{290}{1440}\right)\nonumber.
\end{eqnarray*}
Define 
\begin{align*}
&\bP_M=\bP(M=1),\\
&\bP_{Y^21}=\bP(Y=2\mid M=1),\\
&\bP_{Y^11}=\bP(Y=1\mid M=1),\\
&\bP_{Y^20}=\bP(Y=2\mid M=0),\\
&\bP_{Y^10}=\bP(Y=1\mid M=0),\\
&\bP_{R^M1}=\bP(R^M=1\mid M=1),\\
&\bP_{R^M0}=\bP(R^M=1\mid M=0),\\
&\bP_{R^Y2}=\bP(R^Y=1\mid Y=2),\\
&\bP_{R^Y1}=\bP(R^Y=1\mid Y=1),\\
&\bP_{R^Y0}=\bP(R^Y=1\mid Y=0).
\end{align*}

The following relationships between the observable data probabilities and the parameters hold,
\begingroup
\allowdisplaybreaks
\begin{align}
&\bP_{1211}=\bP_M\bP_{Y^21}\bP_{R^M1}\bP_{R^Y2},\label{eqS27}\\
&\bP_{1111}=\bP_M\bP_{Y^11}\bP_{R^M1}\bP_{R^Y1},\label{eqS28}\\
&\bP_{1011}=\bP_M(1-\bP_{Y^21}-\bP_{Y^11})\bP_{R^M1}\bP_{R^Y0},\label{eqS29}\\
&\bP_{0211}=(1-\bP_M)\bP_{Y^20}\bP_{R^M0}\bP_{R^Y2},\label{eqS30}\\
&\bP_{0111}=(1-\bP_M)\bP_{Y^10}\bP_{R^M0}\bP_{R^Y1},\label{eqS31}\\
&\bP_{0011}=(1-\bP_M)(1-\bP_{Y^20}-\bP_{Y^10})\bP_{R^M0}\bP_{R^Y0},\label{eqS32}\\
&\bP_{+201}=\bP_M\bP_{Y^21}(1-\bP_{R^M1})\bP_{R^Y2}+(1-\bP_M)\bP_{Y^20}(1-\bP_{R^M0})\bP_{R^Y2},\label{eqS33}\\
&\bP_{+101}=\bP_M\bP_{Y^11}(1-\bP_{R^M1})\bP_{R^Y1}+(1-\bP_M)\bP_{Y^10}(1-\bP_{R^M0})\bP_{R^Y1},\label{eqS34}\\
&\bP_{+001}=\bP_M(1-\bP_{Y^21}-\bP_{Y^11})(1-\bP_{R^M1})\bP_{R^Y0}\nonumber\\&\hspace{1.5cm}+(1-\bP_M)(1-\bP_{Y^20}-\bP_{Y^10})(1-\bP_{R^M0})\bP_{R^Y0},\label{eqS35}\\
&\bP_{1+10}=\bP_M\bP_{Y^21}\bP_{R^M1}(1-\bP_{R^Y2})+\bP_M\bP_{Y^11}\bP_{R^M1}(1-\bP_{R^Y1})\nonumber\\&\hspace{1.5cm}+\bP_M(1-\bP_{Y^21}-\bP_{Y^11})\bP_{R^M1}(1-\bP_{R^Y0}),\label{eqS36}\\
&\bP_{0+10}=(1-\bP_M)\bP_{Y^20}\bP_{R^M0}(1-\bP_{R^Y2})+(1-\bP_M)\bP_{Y^10}\bP_{R^M0}(1-\bP_{R^Y1})\nonumber\\&\hspace{1.5cm}+(1-\bP_M)(1-\bP_{Y^20}-\bP_{Y^10})\bP_{R^M0}(1-\bP_{R^Y0}).\label{eqS37}
\end{align}
\endgroup

By solving the equations (\ref{eqS27}) to (\ref{eqS37}), we can identify the parameters $\bP_{M}$, $\bP_{R^M1}$ and $\bP_{R^M0}$. Given the observable data probabilities, $\bP_{M}=\frac{1}{2}$, $\bP_{R^M1}=\frac{3}{4}$ and $\bP_{R^M0}=\frac{1}{2}$. However, $\bP_{Y^21}$, $\bP_{Y^11}$, $\bP_{Y^20}$, $\bP_{Y^10}$, $\bP_{R^Y2}$, $\bP_{R^Y1}$ and $\bP_{R^Y0}$ are not identifiable. For example, we can have $\bP_{Y^21}=\frac{1}{2}$, $\bP_{Y^11}=\frac{1}{3}$, $\bP_{Y^20}=\frac{1}{2}$,  $\bP_{Y^10}=\frac{1}{4}$, $\bP_{R^Y2}=\frac{2}{3}$, $\bP_{R^Y1}=\frac{1}{3}$ and $\bP_{R^Y0}=\frac{1}{6}$, which in turn give us $\bP(Y=2, M=1)=\frac{1}{4}$, $\bP(Y=1, M=1)=\frac{1}{6}$, $\bP(Y=0, M=1)=\frac{1}{12}$, $\bP(Y=2, M=0)=\frac{1}{4}$, $\bP(Y=1, M=0)=\frac{1}{8}$ and $\bP(Y=0, M=0)=\frac{1}{8}$. Alternatively, we can have $\bP_{Y^21}=\frac{5}{8}$, $\bP_{Y^11}=\frac{1}{4}$, $\bP_{Y^20}=\frac{5}{8}$,  $\bP_{Y^10}=\frac{3}{16}$, $\bP_{R^Y2}=\frac{8}{15}$, $\bP_{R^Y1}=\frac{4}{9}$ and $\bP_{R^Y0}=\frac{2}{9}$, which in turn give us $\bP(Y=2, M=1)=\frac{5}{16}$, $\bP(Y=1, M=1)=\frac{1}{8}$, $\bP(Y=0, M=1)=\frac{1}{16}$, $\bP(Y=2, M=0)=\frac{5}{16}$, $\bP(Y=1, M=0)=\frac{3}{32}$ and $\bP(Y=0, M=0)=\frac{3}{32}$.

The two sets of values of $\bP(Y=y, M=m)$ correspond to the same observable data probabilities: $(\bP_{1211},\bP_{1111},\bP_{1011},\bP_{0211},\bP_{0111},\bP_{0011},\bP_{1+10},\bP_{0+10},\bP_{+201},\bP_{+101},\bP_{+001},\bP_{++00})$, and therefore, $\bP(Y=y, M=m)$ can not be uniquely identified without further assumptions.

\subsection{Improved identifiability with a future outcome}\label{subsec::identifiability}

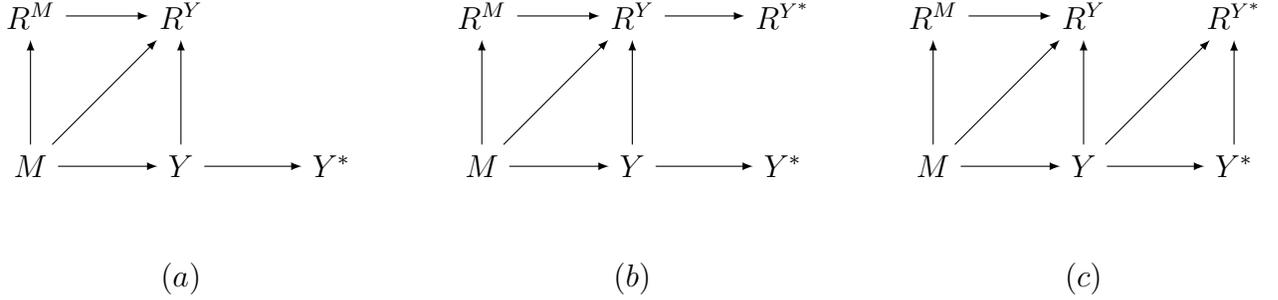
\begin{figure}[ht]
\centering
\begin{tikzpicture}

    \node (x)  at (-3,0) {$M$};
    \node (rm) at (-3,2) {$R^M$};
    \node (y)  at (-1,0) {$Y$};
    \node (ry) at (-1,2) {$R^Y$};
    \node (y1)  at (1,0) {$Y^\ast$};
    \node (c)  at (-1,-1.5) {$(a)$};

    \path[-latex] (x) edge (y);
    \path[-latex] (y) edge (y1);
    \path[-latex] (x) edge (rm);
    \path[-latex] (x) edge (ry);
    \path[-latex] (y) edge (ry);
    \path[-latex] (rm) edge (ry);
    
    \node (x)  at (3,0) {$M$};
    \node (rm) at (3,2) {$R^M$};
    \node (y)  at (5,0) {$Y$};
    \node (ry) at (5,2) {$R^Y$};
    \node (y1)  at (7,0) {$Y^\ast$};
    \node (ry1) at (7,2) {$R^{Y^\ast}$};
    \node (c)  at (5,-1.5) {$(b)$};

    \path[-latex] (x) edge (y);
    \path[-latex] (y) edge (y1);
    \path[-latex] (x) edge (rm);
    \path[-latex] (x) edge (ry);
    \path[-latex] (y) edge (ry);
    \path[-latex] (rm) edge (ry);
    \path[-latex] (ry) edge (ry1);

    \node (x)  at (9,0) {$M$};
    \node (rm) at (9,2) {$R^M$};
    \node (y)  at (11,0) {$Y$};
    \node (ry) at (11,2) {$R^Y$};
    \node (y1)  at (13,0) {$Y^\ast$};
    \node (ry1) at (13,2) {$R^{Y^\ast}$};
    \node (c)  at (11,-1.5) {$(c)$};

    \path[-latex] (x) edge (y);
    \path[-latex] (y) edge (y1);
    \path[-latex] (x) edge (rm);
    \path[-latex] (x) edge (ry);
    \path[-latex] (y) edge (ry);
    \path[-latex] (rm) edge (ry);
    \path[-latex] (y1) edge (ry1);
    \path[-latex] (y) edge (ry1);

\end{tikzpicture}
\caption{DAGs describing the unidentifiable case $(iv)$ that can become identifiable with a fully observed $Y^\ast$ or $Y^\ast$ subject to missingness.}
\label{suppfig8}
\end{figure}

We use $Y^\ast$ to denote the future outcome with $\mathcal{Y}^*$ denoting its support, and let $R^{Y^\ast}$ be the missingness indicator for $Y^\ast$ such that $R^{Y^\ast}=1$ if $Y^\ast$ is observed and $R^{Y^\ast}=0$ otherwise. Under the assumption that the future variables are independent of the past variables conditional on the present variables, we provide some scenarios where the identification of $\bP(Y=y,M=m)$ is plausible under the unidentifiable case $(iv)$ by exploiting the information on a future outcome as described in Figure \ref{suppfig8} $(a)$ to $(c)$. The same results can apply to the reduced unidentifiable cases $(i)$ to $(iii)$. To simplify notation, all DAGs and discussions in this subsection condition on $T$ and $X$ and allow $T$ and $X$ to have directed arrows to all variables in the DAGs.

According to the structures of the DAGs,  the identification of $\bP(Y=y)$ in $(a)$ to $(c)$ can be established on the basis of the theoretical results presented in the main paper under some completeness assumptions. Specifically, let $Y$, $Y^\ast$, $R^Y$, $R^{Y^\ast}$ play the roles as $M$, $Y$, $R^M$, $R^Y$, respectively, the identification of $\bP(Y=y)$ in $(a)$ to $(c)$ can be achieved following the identification of $\bP(M=m)$ in the proof of Theorems \ref{the1} to \ref{the4}. 

In all of the DAGs in Figure \ref{suppfig8}, $R^M \independent Y \mid M$, we can identify $\bP(Y=y \mid M=m)$ if $\bP(R^Y=1 \mid Y=y, M=m, R^M=1)$ is identifiable. This is because 
\begin{align}
\bP(Y=y \mid M=m)&=\bP(Y=y \mid M=m, R^M=1)\nonumber\\&=\frac{\bP(Y=y,R^Y=1 \mid M=m, R^M=1)}{\bP(R^Y=1 \mid Y=y, M=m, R^M=1)}.\label{eq::S38}
\end{align}

In the expression (\ref{eq::S38}), $\bP(Y=y,R^Y=1 \mid M=m, R^M=1)$ is observable. Below we show that the identification of $\bP(R^Y=1\mid Y=y,M=m,R^M=1)$ can be achieved with a fully observed $Y^\ast$ or $Y^\ast$ subject to missingness according to $(a)$ to $(c)$. Define 
\begin{eqnarray*}
    \bP_{{y^\ast}y1|m1} &=& \bP(Y^\ast=y^\ast, Y=y, R^Y=1 \mid M=m, R^M=1),\\
    \bP_{{y^\ast}+0|m1} &=& \bP(Y^\ast=y^\ast, R^Y=0 \mid M=m, R^M=1),\\
    \bP_{{y^\ast}y11|m1} &=& \bP(Y^\ast=y^\ast, Y=y, R^{Y^\ast}=1, R^Y=1 \mid M=m,R^M=1),\\
    \bP_{{y^\ast}+10|m1} &=& \bP(Y^\ast=y^\ast, R^{Y^\ast}=1, R^Y=0 \mid M=m,R^M=1),\\
    \bP_{+y01|m1} &=& \bP(Y=y, R^{Y^\ast}=0, R^Y=1 \mid M=m, R^M=1),\\
    \bP_{++00|m1} &=& \bP(R^{Y^\ast}=0, R^Y=0 \mid M=m, R^M=1).
\end{eqnarray*}

In $(a)$, we have 
\begin{align*}
  \bP_{{y^\ast}y1|m1}=\bP(Y^\ast=y^\ast \mid Y=y)\bP(Y=y \mid M=m)\bP(R^Y=1 \mid Y=y, M=m, R^M=1),
\end{align*}
and therefore, for each $y^*\in\mathcal{Y^*}$,
\begin{align*}
\bP_{{y^\ast}+0|m1}&=\int_{y\in \mathcal{Y}}\bP(Y^\ast=y^\ast, Y=y, R^Y=0 \mid M=m, R^M=1)\textup{d}y\\&=\int_{y\in \mathcal{Y}}\bP_{{y^\ast}y1|m1}\frac{\bP(R^Y=0\mid Y=y,M=m,R^M=1)}{\bP(R^Y=1\mid Y=y,M=m,R^M=1)}\textup{d}y.
\end{align*}
The uniqueness of solutions $\bP(R^Y=1\mid Y=y,M=m,R^M=1)$ in $(a)$ requires that $\bP(Y^\ast,Y,R^Y=1\mid M=m,R^M=1)$ is complete in $Y^\ast$ for all $m$.

In $(b)$, we have 
\begin{eqnarray*}
\bP_{{y^\ast}y11|m1}&=&\bP(Y^\ast=y^\ast \mid Y=y)\bP(Y=y \mid M=m)\\
&&\cdot\bP(R^{Y^\ast}=1 \mid R^Y=1)\bP(R^Y=1 \mid Y=y, M=m, R^M=1),
\end{eqnarray*}
and therefore, for each $y^*\in\mathcal{Y^*}$,
\begin{align*}
\bP_{{y^\ast}+10|m1}&=\int_{y\in \mathcal{Y}}\bP(Y^\ast=y^\ast, Y=y, R^{Y^\ast}=1, R^Y=0 \mid M=m, R^M=1)\textup{d}y\\&=\int_{y\in \mathcal{Y}}\bP_{{y^\ast}y11|m1}\frac{\bP(R^Y=0\mid Y=y,M=m,R^M=1)\bP(R^{Y^\ast}=1\mid R^Y=0)}{\bP(R^Y=1\mid Y=y,M=m,R^M=1)\bP(R^{Y^\ast}=1\mid R^Y=1)}\textup{d}y.
\end{align*}
The uniqueness of solutions $\bP(R^Y=1\mid Y=y,M=m,R^M=1)$ in $(b)$ requires that $\bP(Y^\ast,Y,R^{Y^\ast}=1,R^Y=1\mid M=m,R^M=1)$ is complete in $Y^\ast$ for all $m$.

In $(c)$, we have 
\begin{eqnarray*}
\bP_{{y^\ast}y11|m1}&=&\bP(Y^\ast=y^\ast \mid Y=y)\bP(Y=y \mid M=m)\\
&&\cdot\bP(R^{Y^\ast}=1 \mid Y=y, Y^\ast=y^\ast)\bP(R^Y=1 \mid Y=y, M=m, R^M=1),
\end{eqnarray*}
and 
\begin{align*}
\bP_{+y01|m1}=\bP(Y=y \mid M=m)\bP(R^{Y^\ast}=0 \mid Y=y)\bP(R^Y=1 \mid Y=y, M=m, R^M=1).
\end{align*}
Therefore, we have 
\begin{align*}
\bP_{{y^\ast}+10|m1}&=\int_{y\in \mathcal{Y}}\bP(Y^\ast=y^\ast, Y=y, R^{Y^\ast}=1, R^Y=0 \mid M=m, R^M=1)\textup{d}y\\&=
\int_{y\in \mathcal{Y}}\bP_{{y^\ast}y11|m1}\frac{\bP(R^Y=0\mid Y=y,M=m,R^M=1)}{\bP(R^Y=1\mid Y=y,M=m,R^M=1)}\textup{d}y,
\end{align*}
\text{for each} $y^*\in\mathcal{Y^*}$, and 
\begin{align*}
\bP_{++00|m1}&=\int_{y\in \mathcal{Y}}\bP(Y=y, R^{Y^\ast}=0, R^Y=0 \mid M=m, R^M=1)\textup{d}y\\&=\int_{y\in \mathcal{Y}}\bP_{+y01|m1}\frac{\bP(R^Y=0\mid Y=y,M=m,R^M=1)}{\bP(R^Y=1\mid Y=y,M=m,R^M=1)}\textup{d}y.
\end{align*}
Further define a random vector $Y^{\ast\dagger}=(Y^{\ast}\cdot R^{Y^\ast},R^{Y^\ast})$. The uniqueness of solutions $\bP(R^Y=1\mid Y=y,M=m,R^M=1)$ in $(c)$ requires that $\bP(Y^{\ast\dagger},Y,R^Y=1\mid M=m,R^M=1)$ is complete in $Y^{\ast\dagger}$ for all $m$.

So far, we have shown that the identification of $\bP(Y=y)$ and $\bP(Y=y \mid M=m)$ can be established in $(a)$ to $(c)$ under some
completeness assumptions. Subsequently, if the joint distribution $\bP(Y,M)$ is complete in $Y$, we can identify $\bP(M=m)$ by solving the following linear equations:
\begin{align}
\bP(Y=y)=\int_{m\in \mathcal{M}}\bP(Y=y \mid M=m)\bP(M=m)\textup{d}m,\nonumber
\end{align}
\text{for each} $y\in\mathcal{Y}$.

Therefore, the identification of $\bP(Y=y, M=m)$ can be achieved in $(a)$ to $(c)$ by exploiting the information on a future outcome.  Since all the probabilities and statements involved are conditioning on $T$ and $X$, the corresponding completeness conditions need to hold for all $t$ and $x$.

\section{Details for the parametric estimation}\label{sec::estimation}

For illustration, we describe the parametric methods in the scenarios considered in Theorem \ref{the1} when missingness exists only in the mediator. To simplify the notation, the likelihoods defined below are conditional on $T$ and $X$ implicitly.

Under Assumption \ref{ass1}, the log of the complete-data likelihood is 
\begin{align*}
\ell_{c}(\theta) = \sum^n_{i=1} &\log~\bP(Y_i=y_i\mid M_i=m_i,T_i=t_i, X_i = x_i )+\log~\bP(M_i=m_i \mid T_i=t_i, X_i = x_i )\\
& + \log~\bP(R^M_i=r^M_i \mid M_i=m_i,T_i=t_i, X_i = x_i ).
\end{align*}

Under Assumption \ref{ass1}, the observed-data likelihood is
\begin{align*}
L_{obs}(\theta)=\prod_{\{i:R^M_i=1\}} &\bP(Y_i=y_i\mid M_i=m_i,T_i=t_i, X_i = x_i )\bP(M_i=m_i \mid T_i=t_i, X_i = x_i )\\
&\cdot\bP(R^M_i=1 \mid M_i=m_i,T_i=t_i, X_i = x_i )\\
&\hspace{-1.4cm}\cdot\prod_{\{i:R^M_i=0\}}\int_{\mathcal{M}} \bP(Y_i=y_i\mid M_i=m,T_i=t_i, X_i = x_i )\bP(M_i=m \mid T_i=t_i, X_i = x_i )\\
&\hspace{0.8cm}\cdot\bP(R^M_i=0 \mid M_i=m,T_i=t_i, X_i = x_i )\ \textup{d} m.
\end{align*}

When $M$ is categorical, the integral involved in the above expression is reduced to summation. Since the value of $M$ is missing for some subjects, we implement the Expectation-Maximization algorithm to obtain the MLEs by treating the missing $M$ as a latent variable. Specifically, in the E-step, we find the conditional expectation of complete-data log likelihood by calculating the conditional expectation of $M_i$ for subjects with missing $M_i$. For example, if $M$ is binary,
\begin{align*}
E\{I(M_{i}=m)\mid Y_i,R^M_i=0,T_i, X_i ;\theta^{(t)}\}=\frac{\bP(Y_i,M_{i}=m,R^M_i=0\mid T_i, X_i )}{\sum_{m={0,1}}\bP(Y_i,M_{i}=m,R^M_i=0\mid T_i, X_i )}. 
\end{align*}
When $M$ is continuous, the conditional expectation of complete-data log likelihood may be complicated to calculate. Therefore, we applied fractional imputation \citepSupp{kim2011parametric} using the idea of importance sampling and weighting method to approximate the conditional expectation. Specifically, we generate the fractionally imputed data  $m^{(1)}_{i}, \ldots, m^{(S)}_{i}$ from a proposed distribution $h(M_{i}\mid T_i,  X_i )$ for subjects with missing $M_i$. Then, we compute the fractional weight for each imputed observation. The Monte Carlo approximation of the conditional expectation becomes more accurate when $S$ is large:
\begin{align*}
E\{\ell_{ci}(M_{i}=m;\theta)|Y_i,R^M_i=0,T_i, X_i ;\theta^{(t)}\}\approx\sum\limits_{j=1}^{S} \ell_{ci}(M_{i}=m^{(j)}_{i};\theta) \hat{w}(m^{(j)}_{i}),
\end{align*}
where 
\begin{align*}
\hat{w}(m^{(j)}_{i}) \propto \frac{\bP(Y_{i},M_{i}=m^{(j)}_{i},R^M_i=0\mid T_i, X_i )}{h(M_{i}=m^{(j)}_{i}\mid T_i, X_i )}
\end{align*}
is the fractional weight for $m^{(j)}_{i}$ that satisfy $\hat{w}(m^{(j)}_{i}) \geq 0$ and $\sum_{j=1}^{S} \hat{w}(m^{(j)}_{i})=1$.
We iterate between the E-step and M-step until convergence.

The same estimation methods can be applied to the situation where missingness exists in both the mediator and outcome. For subjects with both $M_i$ and $Y_i$ missing, we generate the imputed data sequentially.
For binary $M$ and binary $Y$, we generate the possible value of ($m_i$, $y_i$). For binary $M$ and continuous $Y$, we generate the possible value of $m_i$ and then the fractionally imputed data $y^{(1)}_{i}, \ldots, y^{(S)}_{i}$ for each possible value of $m_i$. For continuous $M$ and continuous $Y$, we generate the fractionally imputed data $(m^{(1)}_{i},y^{(1)}_{i})$,...,$(m^{(S)}_{i},y^{(S)}_{i})$. For continuous $M$ and binary $Y$, we generate the fractionally imputed data $m^{(1)}_{i}, \ldots, m^{(S)}_{i}$ and then the possible value of $y_i$ for each fractionally imputed $m_i$.

The outcome model is identifiable using complete cases under MNAR Assumptions \ref{ass1}, \ref{ass2}, and \ref{ass4}. Therefore, an alternative approach for those scenarios is to estimate the outcome model first using complete cases, then estimate the parameters in other models through the Expectation-Maximization algorithm by plugging in the estimated outcome model. We tried those two slightly different approaches to our simulation settings, both provided consistent results, with the alternative approach enjoying higher computation efficiency as expected. However, under MNAR Assumption \ref{ass3}, the alternative approach does not work because $\bP(Y \mid M, T,   X )$ is not identifiable using complete cases. 

\section{Details on the simulation studies}\label{sec::suppsimulation}

In this section, we show that when $M \independent  Y\mid (T,  X )$, our methods recover the underlying true values of the NIE and NDE under MNAR Assumptions \ref{ass1}, \ref{ass2}, and \ref{ass4} as expected. However, under MNAR Assumption \ref{ass3}, we observe bias when $M \independent  Y\mid (T,  X )$. In addition, we demonstrate that when $M$ has more categories than $Y$, the identifiability of the model parameters
is improved under MNAR Assumption \ref{ass4} compared to MNAR Assumption \ref{ass1} due to the one additional constraint provided by the effect of $M$ on $R^Y$. Furthermore, our results suggest that certain parametric assumptions outperform others in recovering the underlying model parameter values when the completeness assumption is violated.

\begin{figure}[H]
\begin{center}
\includegraphics[scale = 0.315]{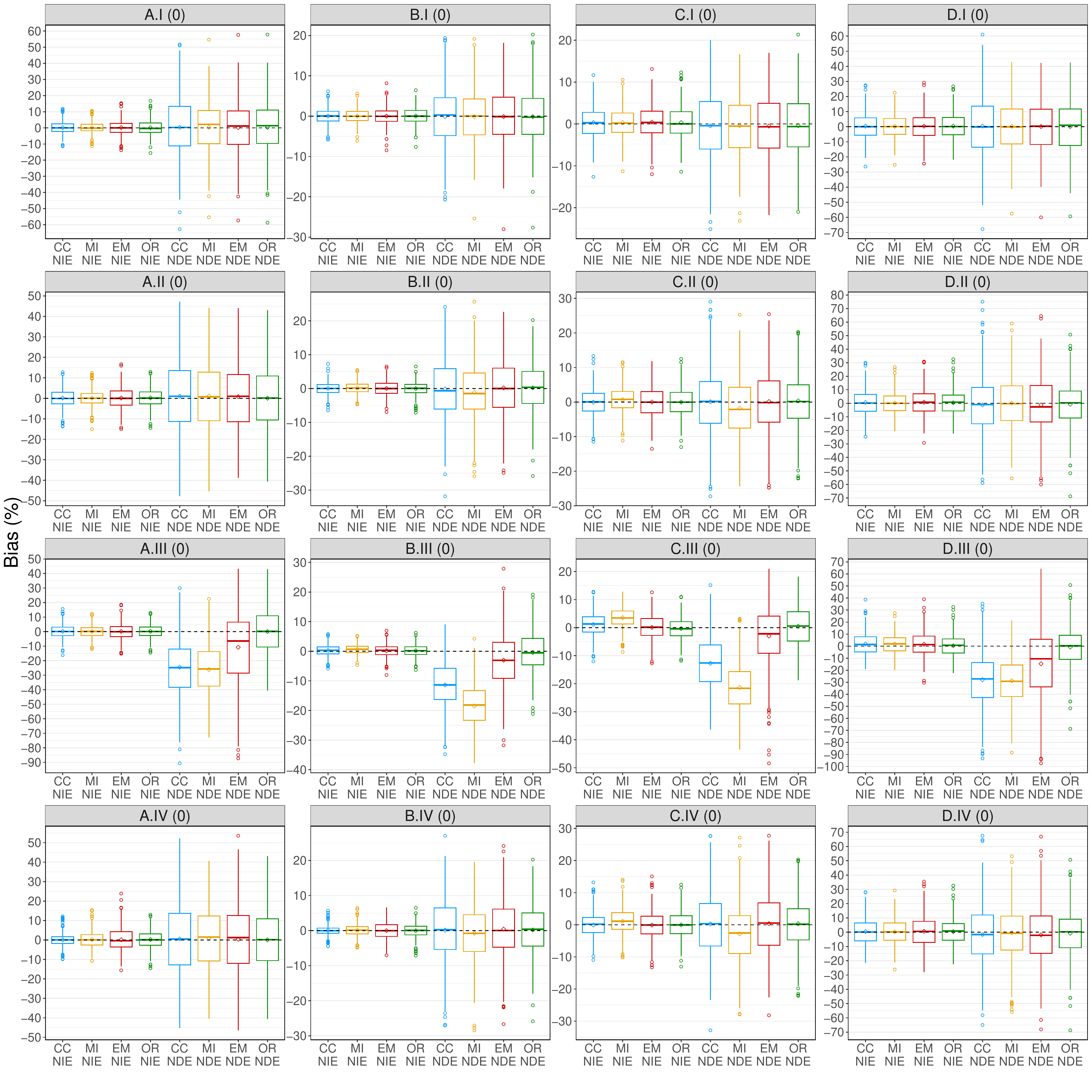}
\end{center}
\caption{Simulation results when $M \independent  Y\mid (T,   X )$. A, Binary $M$ and Binary $Y$; B, Binary $M$ and Continuous $Y$; C, Continuous $M$ and Continuous $Y$; D, Continuous $M$ and Binary $Y$; I, Assumption \ref{ass1}; II, Assumption \ref{ass2}; III, Assumption \ref{ass3}; IV, Assumption \ref{ass4}; CC, complete case analysis; MI, multiple imputation estimators; EM, our proposed Expectation-Maximization algorithm; OR, oracle estimators; (0), $M \independent  Y\mid (T,   X )$; Bias (\%), \{(estimate-truth)/truth\}*100; Bias (\%) for NIE is defined as (estimate/NDE)*100 because NIE equals 0.}
\label{fig:simulation2}
\end{figure}

Continuing the simulation studies in the main paper, Figure \ref{fig:simulation2} presents the boxplots of percentages of bias with respect to the true values for each of the simulation scenarios when $M \independent  Y\mid (T,   X )$ across the $500$ replications. Under MNAR Assumption \ref{ass1}, as shown in Figure \ref{fig:simulation2} A.I (0) to D.I (0) with $(0)$ indicating that $\mathrm{NIE}=0$, the NIE and NDE from
all three methods are consistent. This is due to the fact that $\bP(Y\mid M, T,   X )$ is identifiable using complete cases under MNAR Assumption \ref{ass1}. Under MNAR Assumption \ref{ass2} (A.II(0) to D.II(0)) and MNAR Assumption \ref{ass4} (A.IV(0) to D.IV(0)), we reached the same conclusions as those under MNAR Assumption \ref{ass1} except the fact that multiple imputation under MAR have bias in some cases (e.g. C.II(0) and C.IV(0)) where the NIE and NDE are identifiable using complete cases. Under MNAR Assumption \ref{ass3}, when $M \independent Y\mid (T,   X )$, both $\bP(Y\mid M,T,  X )$ and $\bP(M\mid T,  X )$ are not identifiable, and we observe bias using all three methods.

We also checked the performance of the proposed estimator
for a categorical $M$ with three categories and a binary $Y$ under MNAR Assumptions \ref{ass1} and \ref{ass4}, respectively, where $M$ was generated according to a multinomial logistic regression model and $Y$ was generated according to a logistic regression model. We considered a single covariate $X \sim \mathcal{N}(0,1)$ and a randomized $T\sim \mathrm{Bernoulli}(0.5)$. We generated the mediator $M$ from
\begin{align*}
\log~\frac{\bP(M=1\mid T,X)}{\bP(M=0\mid T,X)}=\alpha_{10}+\alpha_{1t} T+\alpha_{1x} X,\\
\log~\frac{\bP(M=2\mid T,X)}{\bP(M=0\mid T,X)}=\alpha_{20}+\alpha_{2t} T+\alpha_{2x} X.
\end{align*}
We generated the outcome $Y$ from
\begin{align*}
\mathrm{logit}~\bP(Y=1\mid M,T,X)&=\beta_0+\beta_{m1} I(M=1)+\beta_{m2} I(M=2)+\beta_t T\nonumber\\&+\beta_{mt1} I(M=1) \cdot T+\beta_{mt2} I(M=2) \cdot T+\beta_x X.
\end{align*}
The binary variable $R^M$ was generated from $$\mathrm{logit}~\bP(R^M=1\mid M,T,X)=\lambda_0+\lambda_{m1} I(M=1)+\lambda_{m2} I(M=2)+\lambda_t T+\lambda_x X.$$
Under (IV) Assumption \ref{ass4}, the binary variable $R^Y$ was generated from $$\mathrm{logit}~\bP(R^Y=1\mid M,T,X)=\gamma_0+\gamma_{m1} I(M=1)+\gamma_{m2} I(M=2)+\gamma_t T+\gamma_x X.$$

Table \ref{tab:simulation} (Setting E) presents the specifications of parameter values. The missing rates, sample size and number of replications are consistent with the simulation studies in the main paper.

\begin{figure}[H]
\begin{center}
\includegraphics[scale = 0.3]{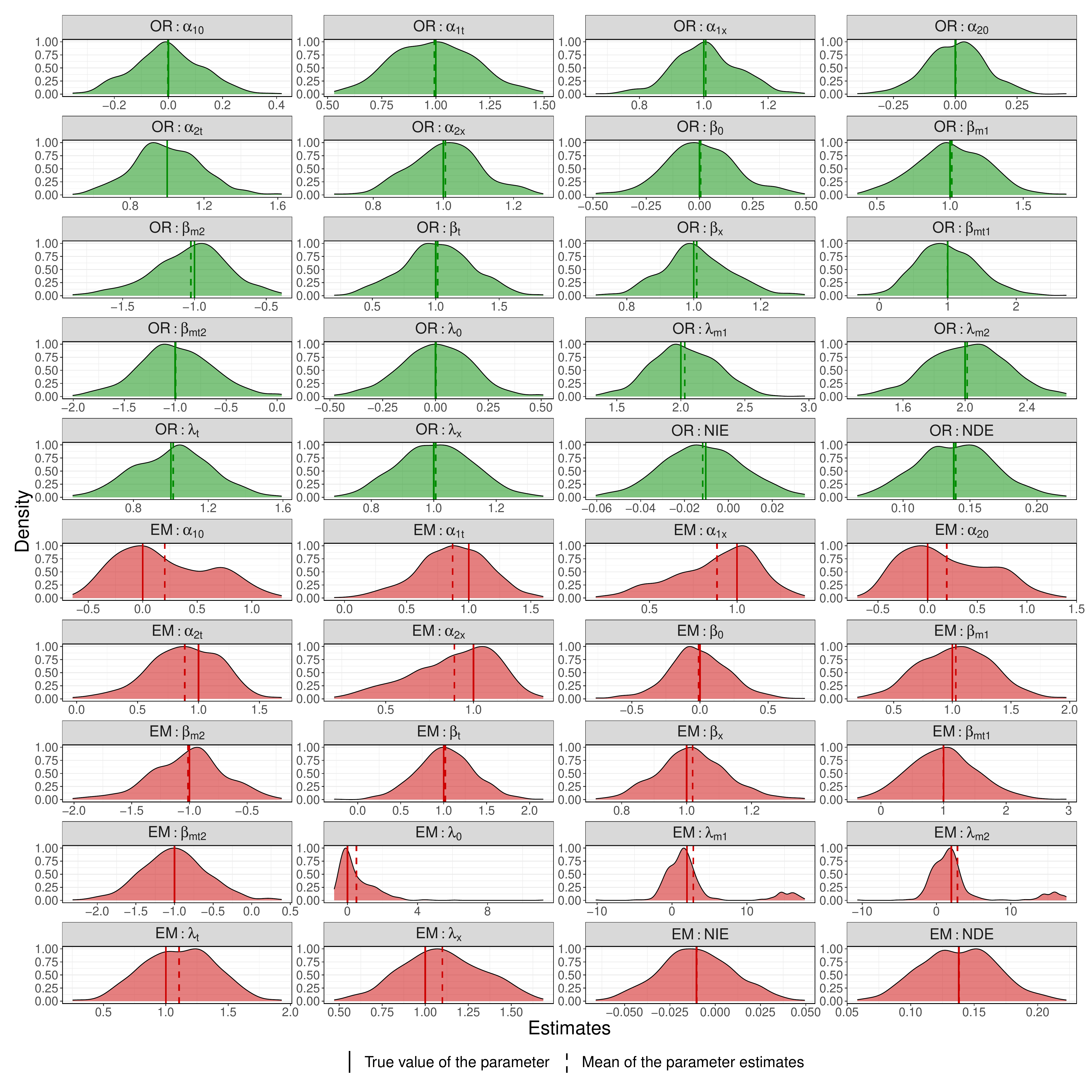}
\end{center}
\caption{Simulation results under MNAR Assumption \ref{ass1} when $M$ is under
a multinomial logistic regression model and $Y$ is under
a logistic regression model. $\alpha_{10}, \alpha_{1t}, \alpha_{1x}, \alpha_{20}, \alpha_{2t}, \alpha_{2x}$, parameters in the $M$ model; $\beta_0, \beta_{m1}, \beta_{m2}, \beta_t, \beta_x, \beta_{mt1}, \beta_{mt2}$, parameters in the $Y$ model; $\lambda_0, \lambda_{m1}, \lambda_{m2}, \lambda_t, \lambda_x$, parameters in the $R^M$ model; EM, our proposed Expectation-Maximization algorithm; OR, oracle estimators.}
\label{fig:suppfig6}
\end{figure}

\begin{figure}[H]
\begin{center}
\includegraphics[scale = 0.3]{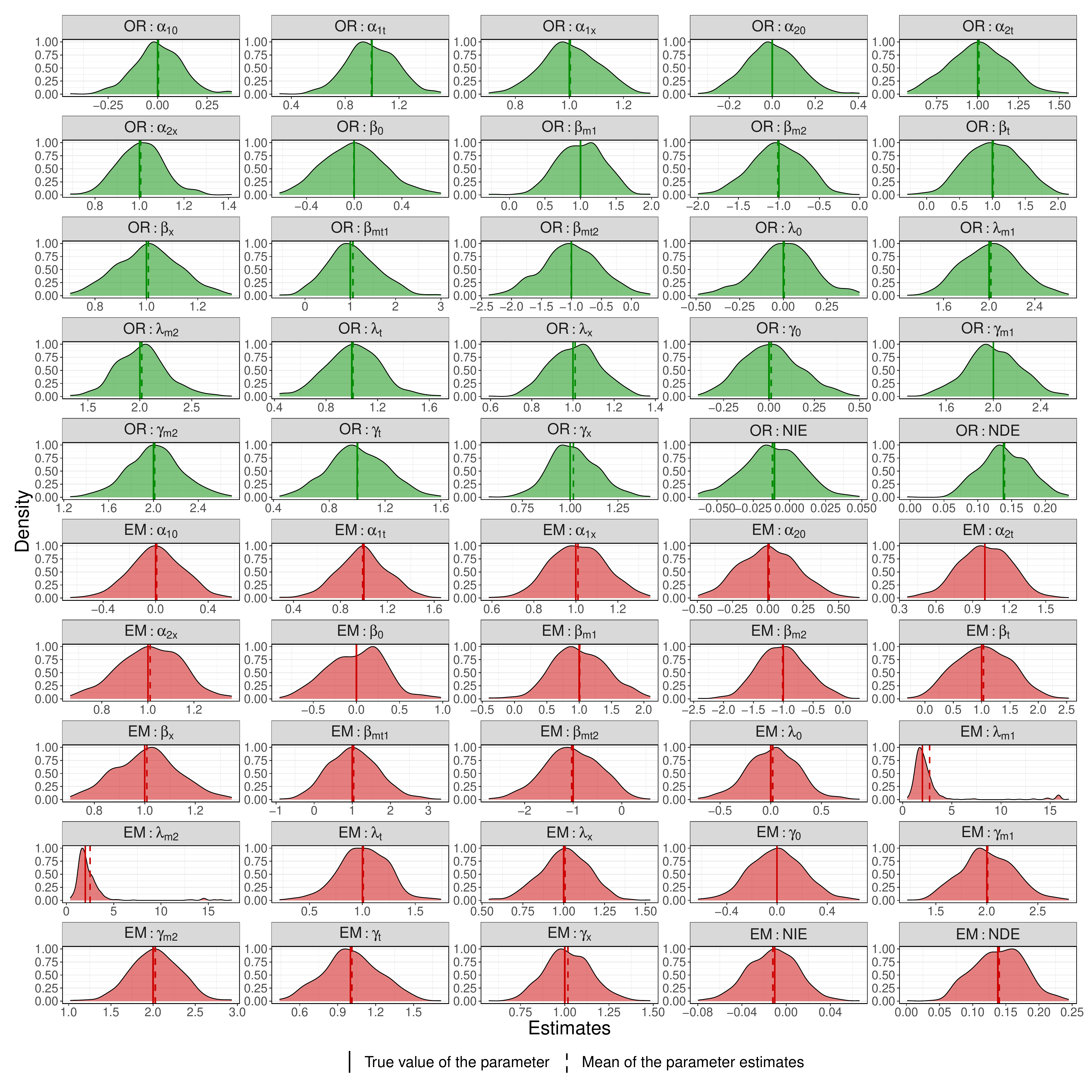}
\end{center}
\caption{Simulation results under MNAR Assumption \ref{ass4} when $M$ is under
a multinomial logistic regression model and $Y$ is under
a logistic regression model. $\alpha_{10}, \alpha_{1t}, \alpha_{1x}, \alpha_{20}, \alpha_{2t}, \alpha_{2x}$, parameters in the $M$ model; $\beta_0, \beta_{m1}, \beta_{m2}, \beta_t, \beta_x, \beta_{mt1}, \beta_{mt2}$, parameters in the $Y$ model; $\lambda_0, \lambda_{m1}, \lambda_{m2}, \lambda_t, \lambda_x$, parameters in the $R^M$ model; $\gamma_0, \gamma_{m1}, \gamma_{m2}, \gamma_t, \gamma_x$, parameters in the $R^Y$ model; EM, our proposed Expectation-Maximization algorithm; OR, oracle estimators.}
\label{fig:suppfig7}
\end{figure}

\begin{figure}[H]
\begin{center}
\includegraphics[scale = 0.3]{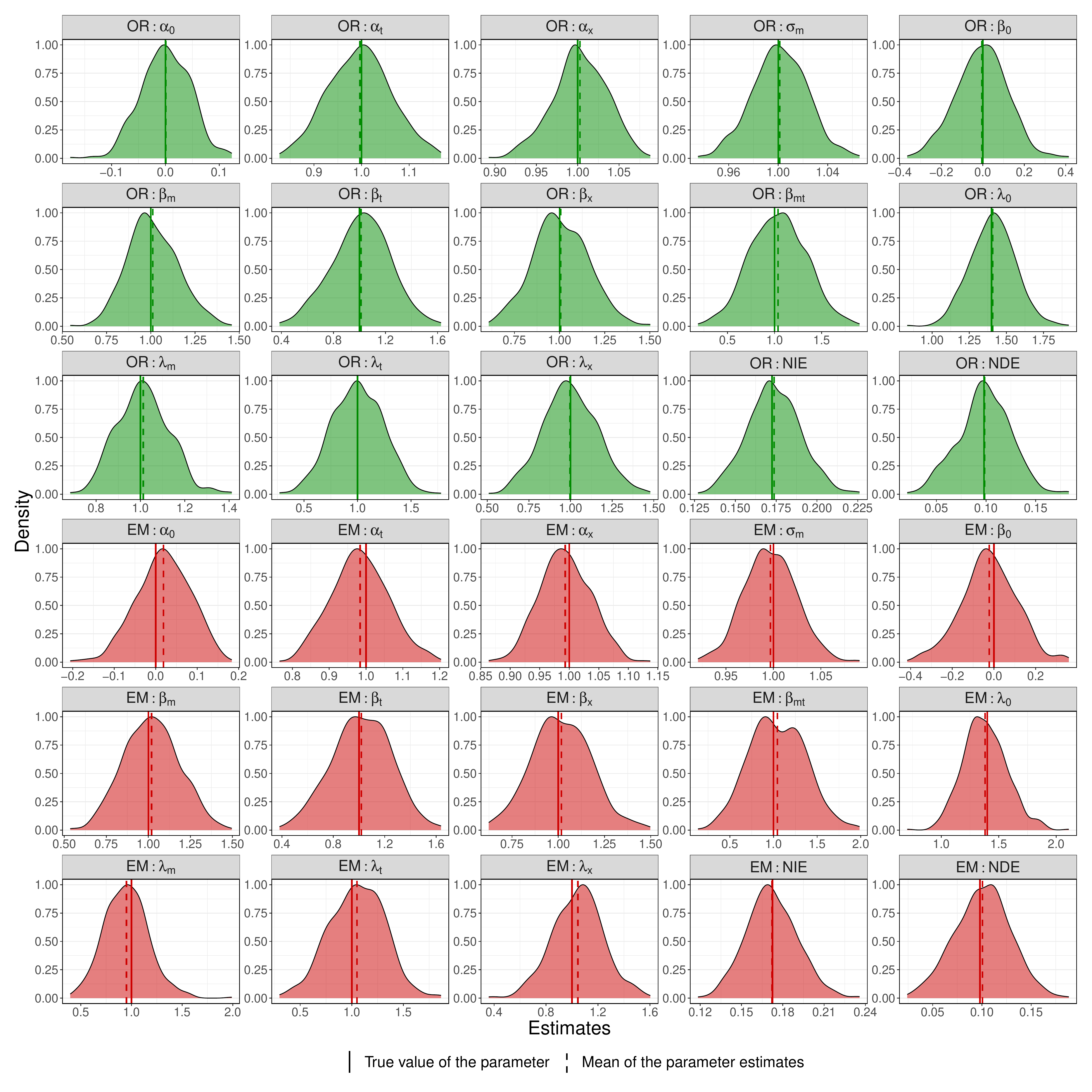}
\end{center}
\caption{Simulation results under MNAR Assumption \ref{ass1} when $M$ is under
a linear regression model and $Y$ is under
a logistic regression model. $\alpha_0, \alpha_t, \alpha_x, \sigma_m$ (residual standard error), parameters in the $M$ model; $\beta_0, \beta_m, \beta_t, \beta_x, \beta_{mt}$, parameters in the $Y$ model; $\lambda_0, \lambda_m, \lambda_t, \lambda_x$, parameters in the $R^M$ model; EM, our proposed Expectation-Maximization algorithm; OR, oracle estimators.}
\label{fig:suppfig8}
\end{figure}

Under MNAR Assumption \ref{ass1}, when $M\not\independent Y\mid (T,  X )$ but the completeness assumption does not hold, the $Y$ model is identifiable using complete cases, but the identification of both the $M$ and $R^M$ models requires the completeness assumption according to our nonparametric identification results. In Figure \ref{fig:suppfig6}, we observe that although the NIE and NDE converge to the true values, the parameters in both the $M$ and $R^M$ models exhibit more complex characteristics compared to the parameters in the $Y$ model. Specifically, $\alpha_{10}$ and $\alpha_{20}$ are concentrated around two distinct modes rather than a single point, which indicates that the parameters cannot be uniquely identified based on the available data. Also, $\alpha_{1x}$ and $\alpha_{2x}$ display an imbalance or non-symmetry in the distribution shape, that is a long tail on one side while being relatively concentrated on the other side. In addition, the irregular distribution patterns of $\lambda_{0}$, $\lambda_{m1}$ and $\lambda_{m2}$ suggest that the parameters may fail to converge to a reasonable region, and therefore, unlikely to provide a trustworthy result. Furthermore, the parameters in the $M$ model are biased. On the other hand, when data is under MNAR Assumption \ref{ass4} (E.IV) or when $M$ is under a linear regression model (D.I), the model parameters have a unique and more well-defined distribution shape, and the mean of the parameter estimates converge to the true values as shown in Figures \ref{fig:suppfig7} and \ref{fig:suppfig8}.

\begin{scriptsize}
\begin{center}
\begin{longtable}{llllllllll}
\caption{Specifications of the parameter values. Setting A, Binary $M$ and Binary $Y$; Setting B, Binary $M$ and Continuous $Y$; Setting C, Continuous $M$ and Continuous $Y$; Setting D, Continuous $M$ and Binary $Y$; Setting E, Categorical $M$ with three categories and Binary $Y$.}
\label{tab:simulation}\\
\hline 
\multicolumn{1}{l}{Setting}&\multicolumn{1}{l}{Model}&\multicolumn{1}{l}{Parameters}&\multicolumn{1}{l}{$M \not\independent  Y\mid (T,   X )$}&\multicolumn{1}{l}{$M \independent  Y\mid (T,   X )$} \\ \hline
A & $M$ & $(\alpha_0,\alpha_t,\alpha_{x})$ & $(0,1,1)$ & $(0,1,1)$\\ 
& $Y$ & $(\beta_0,\beta_m,\beta_t,\beta_{mt},\beta_{x})$ & $(0,-1,1,-1,1)$ & $(0,0,1,0,1)$\\ 
& $R^M$  & $(\lambda_0,\lambda_m,\lambda_t,\lambda_{x})$ & $(0.3,2,1,1)$ & $(0.3,2,1,1)$\\ 
& $R^Y (\mathrm{II})$ & $(\gamma_0,\gamma_{r^M},\gamma_t,\gamma_{x})$ & $(0.4,1,1,1)$ & $(0.4,1,1,1)$\\
& $R^Y  (\mathrm{III})$ & $(\gamma_0,\gamma_y,\gamma_t,\gamma_{x})$ & $(0.6,2,1,1)$ & $(0.3,2,1,1)$\\
& $R^Y (\mathrm{IV})$ & $(\gamma_0,\gamma_m,\gamma_t,\gamma_{x})$ & $(0.3,2,1,1)$ & $(0.3,2,1,1)$\\\hline
B & $M$ & $(\alpha_0,\alpha_t,\alpha_{x})$ & $(0,1,1)$ & $(0,1,1)$\\ 
& $Y$ & $(\beta_0,\beta_m,\beta_t,\beta_{mt},\beta_{x})$ & $(0,-1,1,-1,1)$ & $(0,0,1,0,1)$\\ 
& $R^M$  & $(\lambda_0,\lambda_m,\lambda_t,\lambda_{x})$ & $(0.3,2,1,1)$ & $(0.3,2,1,1)$\\ 
& $R^Y (\mathrm{II})$ & $(\gamma_0,\gamma_{r^M},\gamma_t,\gamma_{x})$ & $(0.4,1,1,1)$ & $(0.4,1,1,1)$\\
& $R^Y  (\mathrm{III})$ & $(\gamma_0,\gamma_y,\gamma_t,\gamma_{x})$ & $(0.8,-1,1,1)$ & $(1.4,1,1,1)$\\
& $R^Y (\mathrm{IV})$ & $(\gamma_0,\gamma_m,\gamma_t,\gamma_{x})$ & $(0.3,2,1,1)$ & $(0.3,2,1,1)$\\\hline
C & $M$ & $(\alpha_0,\alpha_t,\alpha_{x})$ & $(0,1,1)$ & $(0,1,1)$\\ 
& $Y$ & $(\beta_0,\beta_m,\beta_t,\beta_{mt},\beta_{x})$ & $(0,1,1,1,1)$ & $(0,0,1,0,1)$\\ 
& $R^M$  & $(\lambda_0,\lambda_m,\lambda_t,\lambda_{x})$ & $(1.4,1,1,1)$ & $(1.4,1,1,1)$\\ 
& $R^Y (\mathrm{II})$ & $(\gamma_0,\gamma_{r^M},\gamma_t,\gamma_{x})$ & $(0.4,1,1,1)$ & $(0.4,1,1,1)$\\
& $R^Y  (\mathrm{III})$ & $(\gamma_0,\gamma_y,\gamma_t,\gamma_{x})$ & $(1.8,1,1,1)$ & $(1.4,1,1,1)$\\
& $R^Y (\mathrm{IV})$ & $(\gamma_0,\gamma_m,\gamma_t,\gamma_{x})$ & $(1.4,1,1,1)$ & $(1.4,1,1,1)$\\\hline
D & $M$ & $(\alpha_0,\alpha_t,\alpha_{x})$ & $(0,1,1)$ & $(0,1,1)$\\ 
& $Y$ & $(\beta_0,\beta_m,\beta_t,\beta_{mt},\beta_{x})$ & $(0,1,1,1,1)$ & $(0,0,1,0,1)$\\ 
& $R^M$  & $(\lambda_0,\lambda_m,\lambda_t,\lambda_{x})$ & $(1.4,1,1,1)$ & $(1.4,1,1,1)$\\ 
& $R^Y (\mathrm{II})$ & $(\gamma_0,\gamma_{r^M},\gamma_t,\gamma_{x})$ & $(0.4,1,1,1)$ & $(0.4,1,1,1)$\\
& $R^Y  (\mathrm{III})$ & $(\gamma_0,\gamma_y,\gamma_t,\gamma_{x})$ & $(0.4,2,1,1)$ & $(0.3,2,1,1)$\\
& $R^Y (\mathrm{IV})$ & $(\gamma_0,\gamma_m,\gamma_t,\gamma_{x})$ & $(1.4,1,1,1)$ & $(1.4,1,1,1)$\\\hline
E & $M$ & $(\alpha_{10},\alpha_{1t},\alpha_{1x},\alpha_{20},\alpha_{2t},\alpha_{2x})$ & $(0,1,1,0,1,1)$\\ 
& $Y$ & $(\beta_0,\beta_{m1},\beta_{m2},\beta_t,\beta_{mt1},\beta_{mt2},\beta_{x})$ & $(0,1,-1,1,1,-1,1)$\\ 
& $R^M$  & $(\lambda_0,\lambda_{m1},\lambda_{m2},\lambda_t,\lambda_{x})$ & $(0, 2,2,1,1)$\\
& $R^Y (\mathrm{IV})$ & $(\gamma_0,\gamma_{m1},\gamma_{m2},\gamma_t,\gamma_{x})$ & $(0, 2,2,1,1)$\\\hline
\end{longtable}
\end{center}
\end{scriptsize}

\section{Details on the distribution of the covariates}\label{sec::xtable}

\begin{table}[H]
\begin{tabular}{|lc|lc|}
\hline
Characteristics X& Prevalence&Characteristics X& Prevalence \\
\hline
Female& 41.61\% &Education: no high school diploma / GED& 76.72\% \\
Male& 58.39\%&Education: GED certificates& 4.67\%\\
\cline{1-2}
Age: 16 - 17& 40.21\%&Education: high school diploma& 17.95\%\\
Age: 18 - 19& 31.64\%&Education: missing &0.65\%\\
\cline{3-4}
Age: 20 - 24& 28.15\%& Earnings (past year): 0& 35.32\%\\
\cline{1-2}
Race: white& 22.10\%&Earnings (past year): 0 - 1000&10.20\%\\
Race: black& 52.74\% &Earnings (past year): 1000 - 5000&25.49\%\\
Race: hispanic&18.03\%&Earnings (past year): 5000 - 10000&12.94\%\\
Race: others & 7.13\%&Earnings (past year): $\geq 10000$ &6.47\%\\
\cline{1-2}
Had child: no &79.76\%&Earnings (past year): missing&9.59\% \\
Had child: yes &20.24\%&& \\
\cline{1-2}
Ever arrested: no &72.23\%&& \\
Ever arrested: yes &21.03\%&& \\
Ever arrested: missing &6.74\%&& \\
\hline
\end{tabular}
\end{table}

\section{Sensitivity analysis}\label{sec::suppsens}

\begin{figure}[H]
\centering
\begin{tikzpicture}
    \node (t)  at (0,0) {$T$};
    \node (x)  at (2,0) {$M$};
    \node (rm) at (2,2) {$R^M$};
    \node (y)  at (4,0) {$Y(H)$};
    \node (ry) at (4,2) {$R^Y$};

    \path[-latex] (t) edge (x);
    \path[-latex] (t) edge (rm);
    \path[-latex] (t) edge (ry);
    \path[-latex] (x) edge (y);
    \path[-latex] (x) edge (rm);
    \path[-latex] (rm) edge (ry);
    \path[-latex] (t) edge [bend right] (y);
    \path[-latex] (y) edge (ry) [color=red];
    \path[-latex] (x) edge (ry) [color=red];
\end{tikzpicture}
\caption{The DAG describing the missing mechanism for the sensitivity analysis (the DAG conditions on $X$ and allow $X$ to have directed arrows to all variables in the DAG).}
\label{fig:Sensitivity Analysis}
\end{figure}
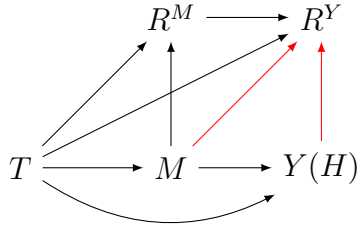

We consider the Gamma model under MNAR Assumption \ref{ass2} from the data analysis as a starting model for building the sensitivity analysis. It is possible that missingness of earnings also depends on the earnings itself and/or the educational and vocational attainment, in addition to missingness of the educational and vocational attainment, as described in Figure \ref{fig:Sensitivity Analysis}. The goal is to assess the sensitivity of our causal conclusions to the additional impacts on $R^Y$ from $H$ and/or $M$. The revised model for $R^Y$ is as follows: 
\begin{eqnarray*}
    &&\mathrm{logit}~\bP(R^Y_i=1\mid R^M_i=r^M, H_i=h, M_i=m, T_i=t,    X _i=  x   )\\
    &=&\gamma_0+\gamma_{r^M} r^M +\gamma_h h+\gamma_m m +\gamma_t t +\gamma_x^ \textsc{t}  x,
\end{eqnarray*}
where $\gamma_h$ and $\gamma_m$ are the sensitivity parameters. We consider a large effect in the log odds ratio \citepSupp{chen2010big} and let both sensitivity parameters vary among $-2$, 0 and 2. When $\gamma_h=0$ and $\gamma_m=0$, it is the same as the MNAR mechanism under Assumption \ref{ass2} that stands out in the data analysis.

The sensitivity analysis results are presented in Table \ref{tab:Sensitivity Analysis}. The NIE estimate increases more than 10\% in the case where $\gamma_m=-2$ and $\gamma_h=2$, and the case where $\gamma_m=0$ and $\gamma_h=2$. The NDE estimate decreases more than 10\% in the case where $\gamma_m=0$ and $\gamma_h=2$, and increases more than 10\% in the case where $\gamma_m=2$ and $\gamma_h=2$. However, the NIEs are estimated to be positive and significant at the $0.05$ level and the NDEs are estimated to be positive but not significant at the $0.05$ level, for all pairs of values $(\gamma_m, \gamma_h)$ considered. In summary, the causal conclusions on the NIE and NDE are not sensitive to a strong impact of $H$ on $R^Y$ and/or $M$ on $R^Y$ in addition to the impact of $R^M$ on $R^Y$.

\begin{table}[H]
\centering
\caption{Sensitivity analysis results from the Gamma model under MNAR Assumption \ref{ass2}. Est, estimate; CI, confidence interval based on $500$ bootstrap samples; $\gamma_h$ (sensitivity parameter), coefficient of $H$ in the $R^Y$ model; $\gamma_m$ (sensitivity parameter), coefficient of $M$ in the $R^Y$ model.} 
\label{tab:Sensitivity Analysis}
\begin{tabular}{lllllllllll}
\hline 
\multicolumn{2}{l}{}&\multicolumn{2}{l}{$\gamma_h=-2$}&\multicolumn{2}{l}{$\gamma_h=0$}&\multicolumn{2}{l}{$\gamma_h=2$}\\
Parameters & $\gamma_m$ & Est & $95\%$ CI & Est & $95\%$ CI & Est & $95\%$ CI \\ \hline
NIE & $-2$ & $11.15$ & $(7.97, ~14.49)$ & $11.49$ & $(8.24, ~14.83)$ & $14.33$ & $(11.02, ~17.89)$\\ 
& $0$ & $11.30$ & $(8.12, ~14.58)$ & $10.94$ & $(7.94, ~14.29)$ & $13.40$ & $(10.22, ~16.78)$\\ 
& $2$ & $11.39$ & $(8.19, ~14.63)$ & $10.83$ & $(7.98, ~14.25)$ & $10.48$ & $(7.15, ~14.81)$\\ 
NDE & $-2$ & $13.18$ & $(-1.53, ~27.88)$ & $13.90$ & $(-1.00, ~28.57)$ & $11.72$ & $(-2.33, ~26.34)$\\
& $0$ & $12.82$ & $(-1.90, ~27.52)$ & $12.93$ & $(-1.95, ~27.64)$ & $11.27$ & $(-3.12, ~25.57)$\\
& $2$ & $12.50$ & $(-2.24, ~27.16)$ & $12.25$ & $(-2.38, ~27.33)$ & $15.43$ & $(-0.18,~29.47)$\\\hline
\end{tabular}
\end{table}

\section{Instrumental variable analysis with the treatment and outcome MNAR}\label{sec::suppiv}

We focus on the MNAR problem in mediation analysis in the main paper, and analogous results apply to the instrumental variable setting with the treatment and outcome MNAR. When evaluating a treatment's effect on an outcome of interest, many studies are challenged by the concern of the unmeasured confounding in the treatment and outcome relationship. To control for the  unmeasured confounding, the method of instrumental variable is often adopted. An instrument is a variable that (i) is associated with the treatment, (ii) has no direct effect on the outcome that is not through the treatment, and (iii) is independent of the unmeasured confounding conditional on the measured confounders. \citeSupp{angrist1996identification} showed that under SUTVA, along with (iv) the monotonicity assumption, the two-stage least squares estimator identifies the complier average causal effect (CACE), where compliers are the subjects who would take the treatment only when being encouraged by the instrument. In this section, we explain that when there are missing data in the treatment and the outcome that are potentially MNAR, the parallel results extend naturally to identify the CACE.

Consider a sample of size $n$ that are independent and identically distributed samples drawn from an infinite superpopulation. Let $Z$ denote a binary instrumental variable, with $z=1$ encouraging the receipt of the treatment and $z=0$ otherwise. We use $D$ to denote the treatment received, with $d=1$ and $d=0$ representing the treatment condition and the control condition, respectively. We use Y to denote the outcome. We continue to adopt the potential outcomes framework to define the causal effect. Under SUTVA, we use $D(z)$ to denote the individual's potential treatment value under instrument value $z$ for $z=0,1$, and use $Y(z)$ to denote the individual's potential outcome value under instrument value $z$ for $z=0, 1$. We use $U$ to denote the unobserved confounding in the relationship of $D$ and $Y$, and use $X$ to represent the measured covariates. Further, let $R^D$ be the missingness indicator for $D$ such that $R^D=1$ if $D$ is observed and $R^D=0$ otherwise, and let $R^Y$ be the missingness indicator for $Y$ such that $R^Y=1$ if $Y$ is observed and $R^Y=0$ otherwise. 

We invoke the standard assumptions on the instrumental variable, i.e., assumptions (i) to (iv). For the mathematical formulation of each assumption, we refer readers to \citeSupp{angrist1996identification} and \citeSupp{baiocchi2014instrumental}. When there is no missing data, the two-stage least squares estimator identifies the CACE:
$$\mathrm{CACE} = \mathbb{E}\{Y(1)-Y(0)\mid D(0)=0, D(1)=1\}  = \mathbb{E}[\mathbb{E}\{Y(1)-Y(0)\mid D(0)=0, D(1)=1, X\}]$$
where 
$$\mathbb{E}\{Y(1)-Y(0)\mid D(0)=0, D(1)=1, X=x\} = \frac{\mathbb{E}(Y\mid Z=1,X=x)-\mathbb{E}(Y\mid Z=0,X=x)}{\mathbb{E}(D\mid Z=1, X=x)-\mathbb{E}(D\mid Z=0,X=x)}.$$

When there is missing data, the identification of $\bP(Y, D \mid Z, X)$ in Figure \ref{fig:iv missingness mechanisms} $(a)$ to $(d)$ can be established on the basis of the theoretical results presented in the main paper under the corresponding completeness assumptions. Specifically, let $D$ and $R^D$ play the roles as $M$ and $R^M$, respectively, and let $Z$ play the role as $T$. The identification of $\bP(Y, D \mid Z, X)$ can be achieved following the identification results of $\bP(Y, M \mid T, X)$ in the proofs of Theorems \ref{the1} to \ref{the4}. Then, we can subsequently identify $\bP(D \mid Z, X)$ and $\bP(Y \mid Z, X)$, which are the components for identifying the CACE in the instrumental variable analysis. 

\begin{figure}[ht]
\centering
\begin{tikzpicture}

    \node (z)  at (0,-5) {$Z$};
    \node (d)  at (2,-5) {$D$};
    \node (rd) at (2,-3) {$R^D$};
    \node (y)  at (4,-5) {$Y$};
    \node (u)  at (3,-6) {$U$};
    \node (c)  at (2,-7) {$(a)$ MNAR mechanism I};

    \path[-latex] (z) edge (d);
    \path[-latex] (d) edge (rd);
    \path[-latex] (z) edge (rd);
    \path[-latex] (d) edge (y);
    \path[-latex] (u) edge (d);
    \path[-latex] (u) edge (y);

    \node (z)  at (6,-5) {$Z$};
    \node (d)  at (8,-5) {$D$};
    \node (rd) at (8,-3) {$R^D$};
    \node (ry) at (10,-3) {$R^Y$};
    \node (y)  at (10,-5) {$Y$};
    \node (u)  at (9,-6) {$U$};
    \node (c)  at (8,-7) {$(b)$ MNAR mechanism II};

    \path[-latex] (z) edge (d);
    \path[-latex] (d) edge (rd);
    \path[-latex] (z) edge (rd);
    \path[-latex] (z) edge (ry);
    \path[-latex] (d) edge (y);
    \path[-latex] (u) edge (d);
    \path[-latex] (u) edge (y);
    \path[-latex] (rd) edge (ry);

    \node (z)  at (0,-10) {$Z$};
    \node (d)  at (2,-10) {$D$};
    \node (rd) at (2,-8) {$R^D$};
    \node (ry) at (4,-8) {$R^Y$};
    \node (y)  at (4,-10) {$Y$};
    \node (u)  at (3,-11) {$U$};
    \node (c)  at (2,-12) {$(c)$ MNAR mechanism III};

    \path[-latex] (z) edge (d);
    \path[-latex] (d) edge (rd);
    \path[-latex] (z) edge (rd);
    \path[-latex] (d) edge (y);
    \path[-latex] (u) edge (d);
    \path[-latex] (u) edge (y);
    \path[-latex] (y) edge (ry);
    \path[-latex] (z) edge (ry);

    \node (z)  at (6,-10) {$Z$};
    \node (d)  at (8,-10) {$D$};
    \node (rd) at (8,-8) {$R^D$};
    \node (ry) at (10,-8) {$R^Y$};
    \node (y)  at (10,-10) {$Y$};
    \node (u)  at (9,-11) {$U$};
    \node (c)  at (8,-12) {$(d)$ MNAR mechanism IV};

    \path[-latex] (z) edge (d);
    \path[-latex] (d) edge (rd);
    \path[-latex] (z) edge (rd);
    \path[-latex] (d) edge (y);
    \path[-latex] (u) edge (d);
    \path[-latex] (u) edge (y);
    \path[-latex] (d) edge (ry);
    \path[-latex] (z) edge (ry);

\end{tikzpicture}
\caption{DAGs describing the MNAR mechanisms in the instrumental variable setting (all DAGs condition on $X$ and allow $X$ to have directed arrows to all variables in the DAGs).}
\label{fig:iv missingness mechanisms}
\end{figure}

\bibliographystyleSupp{apalike}
\bibliographySupp{ref}

\end{document}